\documentclass[11pt]{article}%

\usepackage{amssymb}
\usepackage{url}
\usepackage{bbm}
\usepackage{fullpage}
\usepackage{amsmath}
\usepackage[normalem]{ulem}

\usepackage{bm}
\usepackage{amsmath}

\newcommand{\mb}[1]{\mbox{\boldmath$#1$}}
\newcommand{\bs}[1]{\mathbf{#1}}
\usepackage{adjustbox}
\newcommand{\datasetFmt}[1]{\emph{#1}}

\usepackage{color}
\usepackage{hyperref}
\usepackage[capitalise]{cleveref}
\title{Core--Periphery Structure in Directed Networks}

\author{%
Andrew Elliott$^{1,2}$,
Angus Chiu$^{2}$,
Marya Bazzi$^{1,3,4}$,
\\ 
Gesine Reinert$^{1,2}$ 
and  
Mihai Cucuringu$^{1,2,4}$
\\
{\small
$^{1}$
The Alan Turing Institute, London, UK}
{\small
$^{2}$ Department of Statistics, University of Oxford, %
UK 
}
\\
{\small
$^{3}$ Mathematical Institute, University of Warwick, %
UK
}
{\small
$^{4}$ Mathematical Institute,
University of Oxford, UK
}
}
\date{ }

\begin{document}

\maketitle
\begin{abstract} 
While studies of meso-scale structures in networks often focus on community structure, core--periphery structures can reveal new insights. This structure
typically consists of a well-connected core and a periphery that is well connected to the core but sparsely connected internally. Most studies of core--periphery structure focus on undirected networks.
We propose a 
generalisation 
of core--periphery structure 
to directed networks. Our approach yields a %
family of core--periphery block model formulations in which core and periphery sets are edge-direction dependent. We mainly focus on a particular core--periphery structure consisting of two core sets and two periphery sets which we motivate empirically.

To detect this directed core--periphery structure we propose 
four different 
methods, with different trade-offs between computational complexity and accuracy.  
We assess these methods 
 on three benchmarks and compare to four standard methods. On simulated data, the proposed methods match or outperform the standard methods. 
Applying our methods to three empirical networks --  a political blogs networks, a faculty hiring network, and a trade network --  illustrates  that this directed core--periphery structure can offer novel insights about the underlying dataset.
\end{abstract}

\section{Introduction}
\label{intro}
Networks provide useful representations of complex systems across many applications~\cite{Newman2018}, such as   physical, technological, information, biological, financial, 
and social systems. 
A network in its simplest form is a graph in which vertices represent entities of interest and edges represent pairwise interactions of interest. In  weighted 
graphs, each edge has an associated edge 
weight;
in an  unweighted graph, the edge weights 
are 0 (absent) or 1 (present). Edges can also incorporate directions to represent asymmetric interactions; here, we consider directed unweighted networks (with some methods adaptable to weighted networks).

Given a network representation of a system, it can be useful to apply coarse-graining techniques to investigate so-called meso-scale features that lie between the micro-scale  (local vertex properties e.g. subgraph counts) and the macro-scale (global network properties e.g., total edge weight, degree distribution, average local clustering coefficient). Typical meso-scale structures are community structure, core--periphery structure, role structure, and hierarchical structure~\cite{Newman2018,journals/corr/abs-1202-2684, peixoto2013hierarchical,beguerisse2014interest}; 
often, more than one of these is present in a network, see for example~\cite{peixoto2013hierarchical}.
Communities are the most commonly studied type of meso-scale structure. A community is loosely defined as a set of vertices that are more densely connected to each other than they are to vertices in the rest of the network. 
 Many algorithms have been developed to  detect communities, which   have led to insights in a wide variety of applications, see for example ~\cite{masonams,fortunato,Newman2018}.
In the present paper, we focus on core--periphery structure. The concept of core--periphery stems from studies of economic and social networks~\cite{journals/corr/abs-1202-2684}, and was first formalised by Borgatti and Everett~\cite{borgatti1999models}. Typically, core--periphery structure is a partition of an undirected network into two sets, a core and a periphery, such that there are dense connections within the core and sparse connections within the periphery~\cite{borgatti1999models}. Furthermore, core vertices are reasonably well-connected to the periphery vertices. 
Extensions allow for multiple core--periphery pairs
and nested core--periphery structures~\cite{Borgatti2000, Masuda2017, peixoto2013hierarchical}. 
Many algorithms have been developed for detecting (different variants) of core--periphery structure. These include approaches based on the optimisation of a quality function~\cite{borgatti1999models, journals/corr/abs-1202-2684, Holme2005, journals/corr/abs-1205-6228, zhang2014identification, peixoto2013hierarchical}, spectral methods~\cite{journals/corr/CucuringuRLP14,  Tudisco2019, Mondragon2016}, and notions of core--periphery based on transport (e.g., core vertices are likely to be on many shortest paths between other vertices in the network) rather than edge densities and weights~\cite{journals/corr/CucuringuRLP14, SangHoonOxford}.
Core--periphery detection has been applied 
to various fields 
such as economics, sociology, international relations, journal-to-journal networks, and networks of interactions between scientists; see \cite{tang2019recent} for a recent survey.

Many  methods for detecting core--periphery were developed for undirected networks, and although they can be
(and some have been) generalised to directed graphs,
they do not also generalise the definition of a discrete core and periphery to be edge-direction dependent,
but rather, either disregard the edge-direction or  consider the edge in each direction as an independent observation ~\cite{peixoto2013hierarchical,Tafreshi2013, LidthdeJeude2019a, borgatti1999models} or use a continuous structure~\cite{BOYD2010125}.

 To the best of our knowledge, the only structure that can
be interpreted as a form of generalisation of core--periphery structure to
directed networks where the definition of  core and periphery is
edge-direction dependent, is the bow-tie structure~\cite{journals/compnet/CsermelyLWU13, Broder2000}. Bow-tie structure consists of a core (defined as the largest connected component), an in-periphery (all vertices with a directed path to a vertex in the core), an out-periphery (all vertices with a directed path from a vertex in the core), and other sets containing any remaining vertices~\cite{Broder2000,Lu2016, Yang2011}. 

In this paper, we introduce a generalisation of the block-model core--periphery structure introduced in~\cite{borgatti1999models} to directed networks, in which the definition of both core and periphery are edge-direction dependent. We propose four methods to detect the proposed directed core--periphery structure, and illustrate their performance on synthetic and empirical networks. We include comparisons to bow-tie structure in our synthetic experiments and illustrate that the structure we propose yields additional insights about empirical networks. Moreover, we suggest a
 framework for defining cores and peripheries  in way that accounts for edge direction, 
 which yields as special cases a bow-tie-like structure and the structure we focus on in the present paper. 
This paper is organised as follows. In~\cref{structure} we introduce a novel block-model for directed core--periphery structure that consists of four sets (two periphery sets and two core sets) and a two-parameter synthetic model that can generate the proposed structure.  %
In \cref{directeddetected} we introduce four methods for detecting the proposed
directed core--periphery structure. The first one is a generalisation of the
low-rank approximation approach introduced
in~\cite{journals/corr/CucuringuRLP14}
(see~\cref{LowRank}), 
the second and third are adaptations and extensions of the HITS
algorithm~\cite{kleinberg_1998} (see~\cref{HITS_sec}),
and the fourth is a likelihood-maximisation approach
(see~\cref{DDCSBM}).  \cref{syntheticResults}
illustrates the performance of our methods on synthetic benchmark networks.
In~\cref{realworld}, we apply the methods to three real-world data sets.
\cref{conc} summarises   our main results and offers directions for future
work. 

The code for our proposed methods and the implementation for bow-tie structure is    available at   
\url{https://github.com/alan-turing-institute/directedCorePeripheryPaper}.

\section{Core--periphery structure}
\label{structure}

\paragraph{Core--periphery structure in undirected networks}
\noindent
The most well-known quantitative formulation of core--periphery structure was introduced by Borgatti and Everett~\cite{borgatti1999models}.
They propose both a discrete and a continuous model for detecting core--periphery structure in undirected networks. We describe below the (discrete) block model formulation introduced in~\cite{borgatti1999models}, which we generalise to directed networks later in this section. 

In the discrete notion of core--periphery structure,~\cite{borgatti1999models} suggests that a core--periphery model should consist of two sets: a densely connected core and a loosely connected periphery, with dense connections between the core and the periphery. 
Formally, let $n_c$ denote the number of vertices in the core and $n_p$ the number of vertices in the periphery, with $n_c + n_p = n$ (i.e., the core and the periphery form a partition of the vertices in the graph). The idealised probability matrix and  network representation
that match the undirected core--periphery structure is given by 

\begin{flalign} 
\bs{M_0} = 
\begin{array}{ccc}
            &  \ \ \ \  \text{Core} & \text{Periphery}\ \ \  \\
\text{Core}\! \! \! \!\! \! \! \! \! \!& \vline \ \ \ 1 & 1 \ \ \ \vline  \\
\text{Periphery}\! \! \! \!\! \! \! \! \! \!& \vline \ \ \ 1 & 0 \ \ \  \vline \\ 
\end{array}
\qquad
\begin{minipage}{0.3\textwidth}
\resizebox{0.95\textwidth}{!}
{\includegraphics{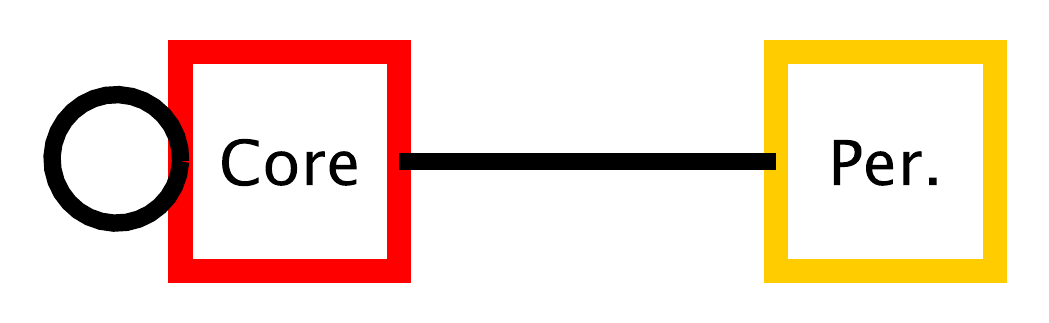}
}
\end{minipage}
\label{UBM}
\end{flalign} 
with corresponding adjacency matrix
\begin{equation}\label{nullmodel} 
	\bs{A}_0=
\begin{vmatrix}
	\mb{1}_{n_c \times n_c} & \mb{1}_{n_c \times n_p} \\
	 \mb{1}_{n_p \times n_c} & \mb{0}_{n_p \times n_p} \\
\end{vmatrix}, 
\end{equation}
where 
$\mb{1}_{n_1 \times n_2}$
(respectively, $\mb{0}_{n_1 \times n_2}$)
 denotes an $n_1 \times n_2$
matrix in which every entry takes the value 1 (respectively, 0).
In adjacency matrices of real-world data sets, any
structure of the form~\cref{nullmodel}, if present, is likely observed with
random noise.  With 
 $sgn$ denoting the signum function that maps a
real number to $1$ if the number is positive, to $-1$ if the number if
negative, and to $0$ if the number is zero, the block matrix
$\bs{A}_0$ is equivalent (up to
row and column permutations) to the matrix 
with $(i,j)^{\text{th}}$ entry $sgn(x_i + x_j)$, with
$x_i=1$ if $i$ is in the core and $0$ otherwise~\cite{Tudisco2019}.
To detect core--periphery structure in a graph with adjacency matrix
$\bs{A}\in\{0,1\}^{n\times n}$, one approach introduced
in~\cite{borgatti1999models}, see also \cite{Tudisco2019}, is to optimise the  quality function
\begin{equation}
 \sum_{i,j=1}^{n} A_{ij}sgn(x_i + x_j)\,,
\label{BE}
\end{equation}
over $\bs{x} = (x_1, \ldots, x_n) \in \Omega = \{\bs{x} \in \{0,1\}^n \vert \sum_{i=1}^n x_i= n_c\}$ for a
fixed $n_c \in\mathbb{N}$. The intuition is that the larger the
double summation in~\cref{BE}, the greater the extent to which the adjacency
matrix matches the idealised block matrix in~\cref{nullmodel}. One can use~\eqref{BE} to detect core--periphery structure in weighted 
and directed 
networks~\cite{borgatti1999models}, and different detection methods for undirected networks build on this formulation by changing the signum function in~\cref{BE}~\cite{borgatti1999models, Tudisco2019, journals/corr/abs-1202-2684}. %
\paragraph{Core--periphery structure in directed networks}
Now we introduce a block model for directed core--periphery in which the definitions of the core and periphery sets are edge-direction-dependent. 
To incorporate information about edge direction in the formulation of the block model itself, we  split each of the
sets in~\cref{nullmodel} into one that has incoming edges and another one 
that has outgoing edges. This yields four sets in total, which we denote 
$\mathcal{C}_{in}$ ({\it core-in}\/), 
$\mathcal{C}_{out}$ ({\it core-out}\/), 
$\mathcal{P}_{in}$ ({\it periphery-in}\/)
and 
$\mathcal{P}_{out}$ ({\it periphery-out}\/). 
Within each of the two core sets ($\mathcal{C}_{in}$ and
$\mathcal{C}_{out}$) and periphery sets ($\mathcal{P}_{in}$ and
$\mathcal{P}_{out}$), we adopt the same convention as in~\cref{nullmodel}: the
two core sets are fully internally connected and the two periphery sets have no
internal edges.  In line with the intuition behind core--periphery structure,
we also assume that edges do not exist between the periphery sets. There are no multiple edges, but there are self-loops.
Specifically, we assume the following block probability matrix and corresponding network representation
\newcommand{\highlight}[1]{%
  \colorbox{red!50}{$\displaystyle#1$}}
\vspace{-0.25cm}
\begin{flalign} 
\bs{M} = 
\begin{array}{rccccccr}
 & \ \ \ \  \mathcal{P}_{out} 
 & \mathcal{C}_{in}  
 & \mathcal{C}_{out} 
 & \mathcal{P}_{in} \ \ \ \\
 \mathcal{P}_{out} \! \! \! \!\! \! \! \! \! \! & \vline \ \ \ 0  &  
 \color{red}{\textbf{1}} &  0  &  0 \ \ \ \vline \\
 \mathcal{C}_{in} \! \! \! \!\! \! \! \! \! \! & \vline \ \ \ 0  & 
  \color{red}{\textbf{1}}  &  0  &  0 \ \ \ \vline \\
 \mathcal{C}_{out}\! \! \! \!\! \! \! \! \! \! & \vline \ \ \ 0  &  
 \color{red}{\textbf{1}}  &   \color{red}{\textbf{1}} &  { \color{red}{\textbf{1}}} \ \ \  \vline \\
 \mathcal{P}_{in}\! \! \! \!\! \! \! \! \! \!  & \vline \ \ \ 0  &  0  &  0  &  0 \ \ \ \vline
\end{array}\,\qquad  
\begin{minipage}{0.3\textwidth}
\resizebox{0.8\textwidth}{!}
{\includegraphics{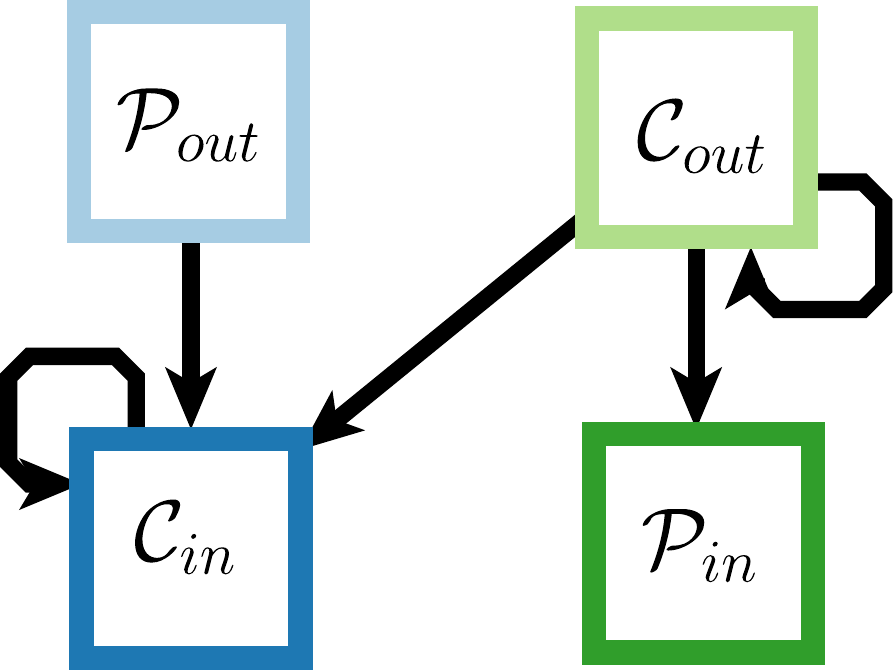}
}
\end{minipage}
\label{DBM}
\end{flalign} 
To gain some empirical  intuition for the structure in~\cref{DBM}, we motivate it with a few real-world examples. Consider networks that represent a type of information flow, with two sets that receive information 
($\mathcal{C}_{in}$ and $\mathcal{P}_{in}$)
and two sets that transmit/send information 
($\mathcal{C}_{out}$ and $\mathcal{P}_{out}$).
Furthermore, within each of these categories, there is one set with core-like properties and another one with periphery-like properties. Inspired by \cite{beguerisse2014interest},
in a Twitter network for example, $\mathcal{C}_{in}$ and $\mathcal{P}_{in}$ could correspond to
consumers of information, with $\mathcal{C}_{in}$ having the added property of being a
closely-knit community that has internal discussions (e.g., 
interest groups) rather than individuals collecting information independently 
(e.g., an average user). On the other hand, $\mathcal{C}_{out}$ and $\mathcal{P}_{out}$ could 
correspond to transmitters of information, with $\mathcal{C}_{out}$ having the added property 
of being a well-known closely-knit community (e.g., broadcasters) rather than
individuals spreading information independently (e.g., celebrities). 
Another class of examples is
networks that represent a type of social flux, when   there are two sets that entities move out of, and two sets that 
entities move towards. Furthermore, within each of these categories, there is
one with core-like properties and one with periphery-like properties.
In a faculty hiring network of institutions 
for example,
$\mathcal{C}_{out}$ may correspond to highly-ranked institutions with sought-after alumni, 
while 
$\mathcal{C}_{in}$
 may 
correspond to good institutions which take in more faculty than they let go. For the periphery sets, $\mathcal{P}_{out}$ may correspond to lower-ranked institutions which do not attract faculty from higher-ranked institutions,  and
$\mathcal{P}_{in}$ may correspond to a set of institutions which attract many alumni from highly-ranked ones. 
 These ideas will be illustrated on real-world data in~\cref{realworld}.
While the extension of~\cref{UBM} in~\cref{DBM} can be regarded as a natural one to consider
in that the edge-direction dependence is defined in the same way for both core and periphery sets, %
there are other choices that one can pursue. 
In Supplementary Information (SI)~A, %
a framework is provided from which one can derive a set of directed core--periphery structures of which~\cref{DBM} is one example; they 
also  include a %
block model formulation of bow-tie structure.%
\paragraph{Synthetic model for directed core--periphery structure}
Here we describe a 
stochastic block model that we will use as a synthetic graph model to benchmark our methods.
Consider a partition with four sets, namely 
${\mathcal{P}_{out}}$,
${\mathcal{C}_{in}}$,
${\mathcal{C}_{out}}$
and 
${\mathcal{P}_{in}}$
with respective sizes 
$n_{{\mathcal{P}}_{out}}$,
$n_{\mathcal{C}_{in}}$,
$n_{\mathcal{C}_{out}}$
and 
$n_{\mathcal{P}_{in}}$.
For any two vertices $u,v$ let $X(u,v)$ be the random variable which equals 1 if there is an edge from $u$ to $v$, and 0 otherwise. We refer to $X(u,v)$ as an {\it{edge indicator}}. For an edge indicator which should equal 1 according to the perfect structure (\cref{DBM}) let $p_1$ be the probability that an edge is observed. Similarly for an edge indicator which should be 0 according to the perfect structure (\cref{DBM}) let $p_2$ be the probability that an edge is observed. Thinking of $p_2$ as noise and of $p_1$ as signal, we assume that $p_1 > p_2$ so that the noise does not overwhelm the structure~\cref{DBM}.  
We represent this model as a stochastic block model with probability matrix
\begin{equation}
p_1{\bf M}  + p_2({\bf 1} - {\bf M}) = \begin{vmatrix}
	p_2 & \color{red}{p_1}& p_2 & p_2  \\
	p_2 &  \color{red}{p_1}& p_2 & p_2  \\
	p_2 &  \color{red}{p_1}&  \color{red}{p_1} &  \color{red}{p_1}  \\
	p_2 & p_2& p_2 & p_2  \\
\end{vmatrix}\,,
\label{2parm_model}
\end{equation}
where ${\bf 1}={\bf 1}_{4\times 4}$.
Setting $p_1=1$ and $p_2=0$ recovers the idealised block structure in~\cref{DBM}. This two-parameter model allows  to increase the difficulty of the detection by reducing the difference between $p_1$ and $p_2$, and to independently modify the expected density of edges matching (respectively, not matching) the planted structure by varying $p_1$ (respectively, $p_2$).

A special case of~\cref{2parm_model} with $p_1=0.5+p$ and $p_2 = 0.5-p$, $p \in [0,0.5]$, yields the stochastic block model with probability matrix 
\begin{equation}
(0.5+p)\bs{M} + (0.5-p)({\bf 1} - \bs{M})\,.
\label{1parm_model}
\end{equation}
This model yields the idealised block structure in~\cref{DBM} when $p=0.5$ and an  Erd\H{o}s-R\'enyi random graph when $p=0$. By varying $p$ between these bounds, one simultaneously varies the expected density of edges that match
the planted structure.

\begin{figure}[htp] 
\begin{center}
\includegraphics[height=5.5cm]{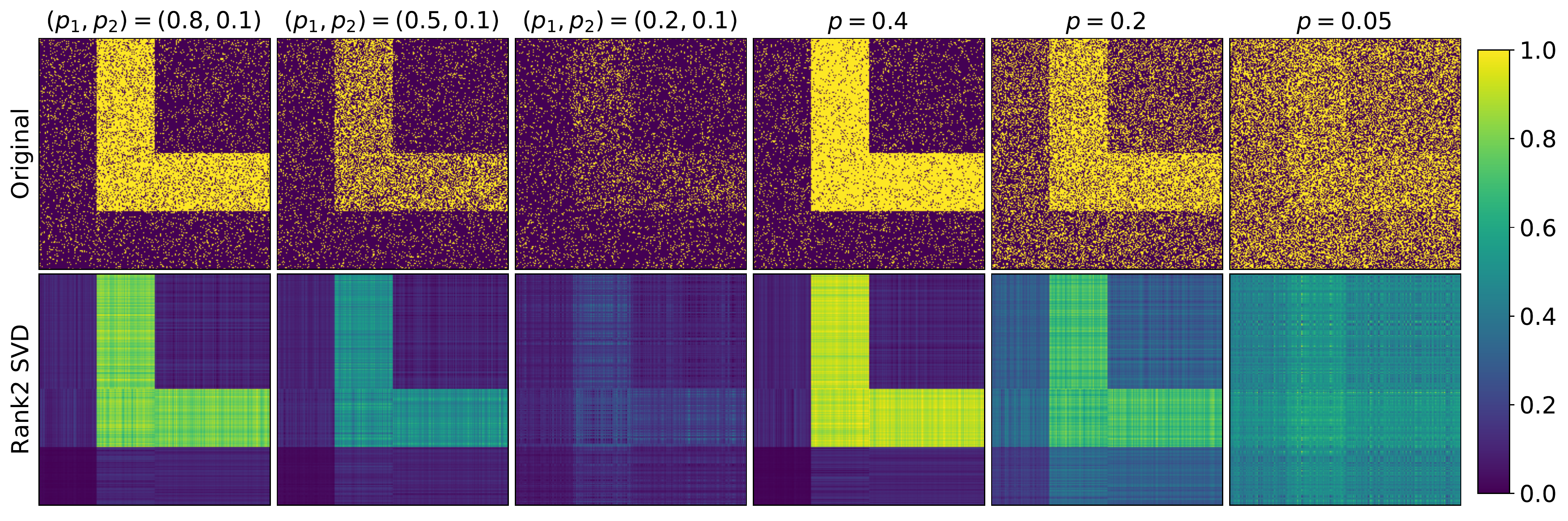}
\end{center}
\caption[Short Caption]{Heatmaps illustrating our model and the  rank-2 SVD approximation. Top row: heatmaps of the original adjacency matrix, with $n=400$ vertices.  We generate the first three examples with ~\eqref{2parm_model} and the second three examples with~\eqref{1parm_model}, for which $p_1 = 0.5+p$ and $p_2=0.5-p$,  Blocks are  equally-sized  in both cases. Bottom row: heatmaps of the low-rank
SVD reconstruction for the
adjacency matrix above it.
}
\label{fig:adjPlotCombinNew}
\end{figure}

Examples of both model instances are shown in the first row of \cref{fig:adjPlotCombinNew}. The first three panels of the first row of \cref{fig:adjPlotCombinNew} show example adjacency matrices obtained with \cref{2parm_model}, $n=400$, and equally-sized 
$n_{{ \mathcal{P}}_{out}}
=
n_{\mathcal{C}_{in}}
=
n_{\mathcal{C}_{out}}
=
n_{\mathcal{P}_{in}} = 100$.
We fix $p_2=0.1$ and vary $p_1$. As $p_1$ decreases with fixed $p_2$, the `L'-shaped pattern starts to fade away and the sparseness of the network increases. The last three panels show realisations of adjacency matrices obtained with~\cref{1parm_model} for $p \in \{0.1,0.3,0.45\}$, $n=400$, and four equally-sized blocks.
As anticipated, the `L'-shaped pattern is less clear for larger values of $p$. We shall return to  \cref{fig:adjPlotCombinNew} in the next section.

\section{Core--periphery detection in directed networks} 
\label{directeddetected}

The
directed core--periphery block structure in~\cref{DBM}
differs from the original undirected core--periphery
structure in~\cref{UBM} in two important ways: the structure in~\cref{DBM} is asymmetric,  and it consists of two core sets and two periphery sets, each with their own distinct characteristics. %
In this section, we describe three method classes  for detecting this novel structure, ordered by run time, from fast to slow. From these three method classes we shall select four methods for directed core--periphery detection.
The first is based on a generalisation of the low-rank approximation approach
in~\cite{journals/corr/CucuringuRLP14}, the second and third are based on an adaptation of
the popular HITS algorithm~\cite{kleinberg_1998}, and the fourth is based on
likelihood-maximisation.

\subsection{Method class 1: Low-rank approximation} \label{LRA}
\label{LowRank}

Our first approach generalises the low-rank  approximation method which was developed for undirected networks  in~\cite{journals/corr/CucuringuRLP14}.
For a low-rank approximation of an asymmetric matrix, 
 the singular value decomposition
(SVD) of the $n \times n$ adjacency matrix $\bs{A}$
is given by
$
\bs{A} = \bs{U} \bs{\Sigma}\bs{V}^T,
$
where $\bs{U}$ (left-singular vectors) and $\bs{V}$ (right singular vectors) are orthogonal $n \times n$ matrices
and $\bs{\Sigma}$ is a $n\times n$ diagonal matrix with the so-called singular values on the diagonal. %
Singular values
are real and non-negative, and by convention, sorted in descending order. For a rank-$r$ approximation, we keep the top $r$ singular vectors of $\bs{U}$ and $\bs{V}$.
Formally, let $\bs{U}_r$ (respectively, $\bs{V}_r$) be a $n \times r$ (respectively, $r \times n$) matrix storing the first $r$ columns of $\bs{U}$ (respectively, $\bs{V}$), and let $\bs{\Sigma}_r$ be a $r \times r$ matrix constructed from the first $r$ rows and columns of $\bs{\Sigma}$.
A rank-$r$ approximation of $\bs{A}$ is

\begin{equation}
\widehat{\bs{A}} = \bs{U}_r \bs{\Sigma}_r \bs{V}^T_r\,,
\label{low_rank_approx}
\end{equation}

\noindent the matrix of rank $r$ with the lowest possible Frobenius error, by  the Eckart-Young theorem. %

Similarly to the undirected case in~\cite{journals/corr/CucuringuRLP14}, one can interpret a low-rank approximation as a `denoising' of the original adjacency matrix. 
The appropriate rank of approximation for detecting a given structure is dictated by the rank of the idealised block model. The rank-2 block matrix in~\cref{DBM} would suggest a rank-2 approximation. 
The second row of~\cref{fig:adjPlotCombinNew} shows the rank-2 approximations for the  synthetic adjacency matrices in the first row of~\cref{fig:adjPlotCombinNew}. In all of these cases, a rank-2 approximation of the adjacency matrix is able to  highlight the planted `L' structure.  %

To partition the network vertex set into $\mathcal{P}_{out}$, $\mathcal{C}_{in}$, $\mathcal{C}_{out}$ and $\mathcal{ P}_{in}$ based on the low-rank approximation in~\cref{low_rank_approx} with $r=2$, we use a score which is analogous to the degree-based score of~\cite{journals/corr/CucuringuRLP14} for the undirected case. Let the in- or out-degree of a set be the sum of the in- or out-degrees of its vertices. In a network which perfectly matches the directed core-periphery structure~\eqref{DBM}, the in-degree of  $\mathcal{ P}_{out}$ would be 0, the out-degree of $\mathcal{ P}_{in}$ would be 0, the in-degree of  $\mathcal{ C}_{in}$ would equal 
($n_{\mathcal{ P}_{out}}$
+
$n_{\mathcal{ C}_{out}}$
+
$n_{\mathcal{ C}_{in}}$
)
$n_{\mathcal{ C}_{in}}$
, and also the 
  out-degree of  $\mathcal{ C}_{out}$ would equal 
($n_{\mathcal{ P}_{in}}$
+
$n_{\mathcal{ C}_{out}}$
+
$n_{\mathcal{ C}_{in}}$
)
$n_{\mathcal{ C}_{out}}$. %
We denote by
${{ C}^{LR}_{in}}(i)$ the $\mathcal{C}_{in}$ score of vertex $i$ with respect to our low rank method, 
with equivalent formulations for the remaining sets. 
Based on the idealised block structure in~\cref{DBM}, $\mathcal{C}_{in}$
should have the highest number of incoming edges and  
$\mathcal{C}_{out}$ 
the highest number of out-going edges. Accordingly,  one would expect a vertex $i$ to contribute more to the ${{\bf C}^{LR}_{in}}$ score if it has high in-degree, and to contribute more to the ${{\bf C}^{LR}_{out}}$ score if it has high out-degree. 
We thus set
\begin{equation}
    \text{\textbf{${{ C}^{LR}_{in}}$}}(i)= \sum\limits_{j=1}^{n} \widehat{A}_{ji},
    \hspace{2cm}
    \text{\textbf{${ { C}^{LR}_{out}}$}}(i)= \sum\limits_{j=1}^{n} \widehat{A}_{ij},
\label{Cin}
\end{equation}	
which yield the 
in-degree and 
out-degree, respectively, of vertex $i$ in the network with adjacency matrix $\widehat{\bs{A}}$ from~\cref{low_rank_approx}. 
Similarly, based on the idealised  structure in~\cref{DBM}, a vertex should have a high ${ {\bf P}^{LR}_{in}}$ score if it has low out-degree compared to the maximal out-degree, and
a vertex %
should have a high ${ {\bf P}^{LR}_{out}}$ score 
if it has low in-degree compared to the maximal in-degree. 
We set
\begin{flalign} 
    \label{pin2}
	 { P}^{LR}_{in}(i) = & 
    \text{max}_j\left(\sum_{a=1}^{n}\widehat{A}_{ja}\right)
    -
    \sum_{a=1}^{n}\widehat{A}_{ia}
    =
    \text{max}_j\left(\textbf{${ { C}^{LR}_{out}(j)}$}\right)
    - \text{\textbf{${ { C}^{LR}_{out}(i)}$}},  
    \\
    \label{pout2LR}
	{{ P}^{LR}_{out}(i)} = & 
    \text{max}_j\left(\sum_{a=1}^{n}\widehat{A}_{aj}\right)
    -
    \sum_{a=1}^{n}\widehat{A}_{ai}
    =
    \text{max}_j\left(\textbf{${ { C}^{LR}_{in}(j)}$}\right)
    - \text{\textbf{${ { C}^{LR}_{in}(i)}$}}.  
\end{flalign}
In SI~C %
we consider alternative variants of these scores and note that the  above scores 
achieve the best performance on the synthetic benchmarks.%
To convert the scores into a partition of a network,
we consider the $n \times 4$ score matrix $\mathbf{S}^{LR} = 
[\mathbf{P}^{LR}_{out}, \mathbf{C}^{LR}_{in},  \mathbf{C}^{LR}_{out}, \mathbf{P}^{LR}_{in}]$.
We normalise each row of $\mathbf{S}^{LR}$ such that it
has an $L_2$-norm of 1. Heuristically  the normalisation  allows the rows of
$\mathbf{S}^{LR}$ (vectors in $4$-dimensional space) 
not only
to concentrate in four different directions, but to do so while also having a small within-set Euclidean distance~\cite{journals/corr/CucuringuKCMP16, journals/corr/abs-1111-1055}. 
We then partition the vertices by clustering $\mathbf{S}^{LR}$ using 
k-means$++$ \cite{1283494}, a variant of the popular k-means clustering algorithm that 
alleviates the issues of unstable clusterings retrieved by k-means \cite{MacQueen1967}. 
We refer to this method as {\sc LowRank}.

\subsection{Method class 2: The Hyperlink-Induced Topic Search (HITS) algorithm}
\label{HITS_sec}

\noindent
Our second method builds on a well-known algorithm in link analysis known as 
Hyperlink-Induced Topic Search (HITS)~\cite{kleinberg_1998}.
Similarly to PageRank, the HITS algorithm was originally designed to measure the importance of webpages or other documents 
using the structure of the directed links between the webpages~\cite{Borodin2005}. 
The 
underlying intuition is that authoritative webpages on a topic should not only have large in-degrees (i.e., they constitute hyperlinks on many webpages) but should also considerably overlap in the sets of pages that point to them. There exists a mutually reinforcing relationship between authoritative webpages for a topic, referred to as ``authorities'', and pages that link to many related authorities, referred to as ``hubs''.
A good hub is a page that points to many good authorities, and a good authority is a page that is pointed to by many good hubs. Consequently, the HITS algorithm assigns to each webpage  two vertex-based scores the  ``authority scores'' $\bs{a}$
and the ``hub scores'' $\bs{h}$, with the following recursive relationship 
$ \bs{a} = \bs{A}^T\bs{h}$ and $\bs{h} = \bs{A}\bs{a}\,$.
The HITS algorithm determines the scores iteratively.
Initially, all authority and hub %
scores 
are set to $\bs{1}_n$. At each iteration of the HITS algorithm, one sequentially updates the authority and hub %
scores. 
A normalisation step is then applied, so that the vectors $\bs{a}$ and $\bs{h}$ become unit vectors in some norm~\cite{Borodin2005}. Kleinberg~\cite{kleinberg_1998} proves that the algorithm converges to the principal left and right singular vectors of the adjacency matrix $\bs{A}$, provided that the initial authority and hub vectors are not orthogonal to the principal eigenvectors of  $\bs{A}^T\bs{A}$ and $\bs{A}\bs{A}^T$.

We use this method in two ways: (i) we construct a score based directly on
the hub and authority scores, (ii) we generalise this method from the case of
two scores (hub and authority) to the case of four scores, motivated by the four sets in our 
directed core-periphery structure. 
\paragraph[(i)] {Direct use of HITS scores.}
To construct the 
${\bf C}^{HITS}_{in}$,
${\bf C}^{HITS}_{out}$,
${\bf P}^{HITS}_{in}$,
${\bf P}^{HITS}_{out}$ scores,
we begin by noting that a vertex would have a high
authority score if it has many incoming edges, whereas it would have a high
hub score if it has many outgoing edges. Based on the idealised block structure in~\cref{DBM},
vertices with the highest authority scores
should also have a high ${{\bf C}^{HITS}_{in}}$ score, and vertices with the highest hub
scores should also have a high ${{\bf C}^{HITS}_{out}}$ score. We let
$$  {{ C}^{HITS}_{in}(i)}  = { h}(i)  \quad \quad  \text{ and } \quad \quad    {{ C}^{HITS}_{out}(i)}  ={ a}(i).  $$ 
To set the remaining scores we use the same intuition as for the low-rank setting, and let
	\begin{flalign} 
    \label{pin2hits}
	{{ P}^{HITS}_{in}(i)} = & 
    \text{max}_j(\textbf{${{ C}^{HITS}_{out}(j)}$})
    - {{ C}^{HITS}_{out}(i)}\,,
    \\
    \label{pout2HITS}
	{{ P}^{HITS}_{out}(i)} = & 
    \text{max}_j(\textbf{${{ C}^{HITS}_{in}(j)}$})
    - {{ C}^{HITS}_{in}(i)}\,.
	\end{flalign}
We then cluster the resulting scores using the same procedure used for {\sc LowRank} (see Section~\ref*{directeddetected}\ref{LowRank}), 
and following {\sc LowRank} we refer to this method as {\sc HITS}.  Recalling the relation between the HITS algorithm and SVD from Kleinberg~\cite{KleinberHITS}, our score based on the HITS algorithm can be construed as a variant of {\sc LowRank}, in which we only consider a rank-$1$ approximation and use the SVD components directly.

\paragraph[(ii)]{An alternative $4$ score scheme {\sc AdvHits} (Advanced Hits).}  
Instead of using hub and authority scores, in each set, edge indicators are rewarded when they match the structure in~\cref{DBM} and penalised otherwise,  %
through the reward-penalty matrix associated to  $\bs{M}$ 
given by
\begin{equation*}
{\bf D} = 2{\bf M} - 1 = 
\begin{array}{|rrrr|}
-1 & \ {\color{red}{1}} & -1 & -1 \\
-1 & \ {\color{red}{1}} & -1 & -1 \\
-1 & \ {\color{red}{1}} & \  {\color{red}{1}} &\ {\color{red}{1}} \\
-1 & -1 & -1 & -1 
\end{array}
= 
\begin{array}{|rrrr|}
{\bf d}_1 & 
{\bf d}_2 & 
{\bf d}_3 & 
{\bf d}_4  
\end{array}
= 
\begin{array}{|c|}
{\bf e}_1 \\ 
{\bf e}_2 \\ 
{\bf e}_3 \\ 
{\bf e}_4  
\end{array}
\,,
\end{equation*}
where ${\bf d_i}$ is the $i^{\text{th}}$ column vector of ${\bf D}$, and ${\bf e_i}$ is the $i^{\text{th}}$
row vector of ${\bf D}$. 
The
first column/row corresponds to $\mathcal{P}_{out}$, the second column/row to
$\mathcal{C}_{in}$, and so on.
We shall iteratively update two sets of scores:
(1) raw scores ${\bf S}^{Raw} =  [{\bf{S}}^{Raw} _1, {\bf{S}}^{Raw}_2, {\bf{S}}^{Raw}_3, {\bf{S}}^{Raw} _4] = [{\bf P}_{out}^{Raw},{\bf C}_{in}^{Raw},{\bf C}_{out}^{Raw},{\bf P}_{in}^{Raw}]$
which are  sums of the connection-based rewards and penalties for each vertex; and (2) normalised scores 
${\bf S}^{{Nrm}} =  [{\bf{S}}^{{Nrm}} _1, {\bf{S}}^{Nrm}_2, {\bf{S}}^{{Nrm}}_3, {\bf{S}}^{{Nrm}} _4]= [{\bf P}_{out}^{Nrm},{\bf C}_{in}^{Nrm},{\bf C}_{out}^{Nrm},{\bf P}_{in}^{Nrm}]$
with 
\begin{flalign}\label{advhits} 
S^{Raw}_i = 
& 
\left( 1-\frac{m}{n^2} \right)  \bs{A} 
\bs{S}^{Nrm}
{\bf e}_i^T
+
\frac{m}{n^2} (1-\bs{A}) 
\bs{S}^{Nrm}
(-{\bf e}_i^T) \nonumber \\
& 
+
\left( 1-\frac{m}{n^2} \right) \bs{A}^T 
\bs{S}^{Nrm}
{\bf d}_i
+
\frac{m}{n^2}(1-\bs{A}^T) 
\bs{S}^{Nrm}
(-{\bf d}_i)\,,
\end{flalign}
for $i\in\{1, \ldots, 4\}$.
The first two terms score the outgoing edge indicators and the last two terms score the incoming edge indicators.  The multiplicative constants are chosen to weigh edges in each direction evenly, and to fix the contribution of non-edges to be equal to that of edges.

As the scores in \eqref{advhits}
can be negative and the scores for a given vertex may not sum to $1$, we normalise as follows. 
For $j \in \{1, \ldots, n\}$ let  $B(j) = \min \{{ P}_{out}^{Raw}(j),{ C}_{in}^{Raw}(j),{ C}_{out}^{Raw}(j),{ P}_{in}^{Raw}(j) \}$, and for $i\in\{1, \ldots, 4\}$ we set
\begin{equation}
\label{eq:hitsNorm}
{ S}_{i}^{Nrm}(j)
=
\frac{{ S}_{i}^{Raw}(j)
-B(j)
}
{\sum_{k=1}^4 \left({ S}^{Raw}_k(j) 
- B(j)\right)
}\,.
\end{equation}
Thus for each vertex the scores for the four groups sum to 1. If for a given vertex, the raw scores in each of the sets are equal up to floating point error (defined as the denominator 
of 
\cref{eq:hitsNorm}
being less than an 
arbitrarily fixed constant ($10^{-10}$)), we set the normed score in each set to $0.25$.
We initialise and update the scores as follows. First,  we assign a raw score for each vertex set combination uniformly at random from 0 to 1, which we then use to calculate the normalised scores. %
We then update each of the scores in turn, first updating ${\bf P}_{out}^{Raw}$ using ${\bf S}^{Nrm}$, next updating ${\bf S}^{Nrm}$ using the new value of ${\bf S}^{Raw}$, then repeating the same procedure for each of the remaining scores in the order ${\bf C}_{in}$, ${\bf C}_{out}$ and ${\bf P}_{in}$. 
We repeat the procedure until convergence, which we
measure by computing the largest change observed in each normalised score when we update its
raw counterpart; convergence is determined if the largest change over each of the sets is less than
$10^{-8}$.
 The general iteration can fail to converge within $1000$ iterations. 
If the scheme has not converged after $1000$ steps, we fall back to a scheme which updates the scores on each vertex in turn which often empirically removes the convergence problem with the cost of additional computational complexity. 
To obtain the set labels, we cluster the normalised score matrix ${\bf S}^{Nrm}$ using k-means$++$, and 
allocate each of the clusters to a named set by  mapping the  clusters to sets which maximises the log-likelihood as will be explained in the next subsection,  Section~\ref*{directeddetected}\ref{DDCSBM}. 
We call this method 
{\sc AdvHits}. 

The {\sc AdvHits} method may lead to large sets having a higher contribution to the score than small sets. Thus, 
as an alternative scoring system 
with score matrix ${\bf{S}}^{Nrm}$, we introduce a $4 \times 4$ diagonal  matrix ${\bf F_Y}$ with diagonal entries the sum of the 
scores for the vertices, so that   $ ({ F_Y})_{ii} = 1/\sum_{j=1}^{n} { S}^{Nrm}_{ji}$ for $i\in\{1, \ldots, 4\}$, and use the iteration scheme 
\begin{flalign} 
\label{advhitsGrp} 
{\bf S}_{i}^{Raw} = 
& 
 {\bf  A} {\bf S}^{Nrm}{\bf F_Y} 
{\bf e}^T_{i} 
+ 
{\bf A}^T {\bf S}^{Nrm}{\bf F_Y} 
{\bf d}_{i} 
- \frac{m}{n^2} {\bf S}^{Nrm}{\bf F_Y} 
(
{\bf d}_{i} 
+ 
{\bf e}^T_{i} 
)\,.
\end{flalign}
This iteration scheme makes the contribution from each set to a vertex score 
identical even for 
very differently sized sets. The same normalisation scheme and updating scheme as for {\sc AdvHits} is then used; we call this method  {\sc AdvHitsGrp}.

\subsection{Method class 3: Likelihood maximisation}
\label{DDCSBM}

As third method we 
 maximise the likelihood of the
directed core--periphery model ~\cref{2parm_model}, which is a 
stochastic block model with $4$
blocks and our particular connection structure. For maximising the likelihood numerically we use, first, a greedy approach from \cite{snijdersSBM} which we call {\sc HillClimb}, and the second, a slower but potentially more accurate approach  from \cite{karrer2011} which we call {\sc MaxLike}. 
{\sc HillClimb} iteratively updates the values of $p_1$ and $p_2$, and  
the set assignments. Our implementation deviates slightly from that in~\cite{snijdersSBM} for example 
in having a fixed number of iterations; see SI~B. %
Unlike  {\sc HillClimb}, {\sc MaxLike}
does not consider a random order of vertices, but instead updates the vertices in an
order which increases the likelihood the most at each step, thus giving a
potentially more accurate (albeit slower) algorithm.
The complete algorithm can be found in SI~B.%
For multimodal or shallow likelihood surfaces the maximum likelihood algorithms may fail to detect the maximum and instead find a local optimum.
Hence we  use a range of starting points for the algorithms.

There is a 
considerable difference in computational cost between {\sc HillClimb} and {\sc MaxLike}, with one iterate of the {\sc HillClimb} taking $O(m)$ time where $m$ is the number of edges, while for {\sc MaxLike} in sparse networks the computational complexity is $O(n^2+m)$.
A detailed discussion around the computational complexity can be found in SI~B(c). %

\section{Numerical Experiments on Synthetic Data}
\label{syntheticResults}

To compare the performance of the 
methods from~\Cref{directeddetected}, we create three benchmarks using the
synthetic model in~\cref{structure}
and measure accuracy by computing the  Adjusted Rand Index
(ARI)~\cite{huberta85} 
between the  output
partition of a network and the ground truth, using the
implementation from~\cite{sklearn}.
ARI takes values in $[-1,1]$, with $1$ indicating a perfect match, and an expected score of approximately $0$ under a given model of randomness; %
a 
detailed description  %
is given in
SI~D(a). %
{For completeness, we also
compute the similarity using VOI (Variation of Information
\cite{Meila01arandom}) and NMI
(Normalised Mutual Information~\cite{kvalseth1987entropy}), and achieve qualitatively similar
results, shown in SI~D. 
}  As each of our new methods is 
specifically designed to output the
structure in~\cref{DBM}, 
all of these comparisons can be seen as assessing their 
performance gain over generic methods 
in detecting the directed core periphery
structure.

We first compare our method against 
a na\"ive classifier ({\sc Deg.}), which performs k-means$++$~\cite{1283494} clustering  solely on the in- and out- degree of each vertex. 
To compare against other approaches, 
we divide our methods into three classes: fast methods ({\sc LowRank} and {\sc
HITS} 
slow methods ({\sc HillClimb}
and {\sc MaxLike}), and variable methods, namely {\sc AdvHits}. We use a
separate category for {\sc AdvHits} as it can be fast, but in networks with
non-existent or weak structure (e.g., Erd\H{o}s-R\'enyi random graphs) it can fall back on the single vertex
update scheme described at the end of~\cref{HITS_sec}, which reduces the speed
considerably.
The methods in each of these three classes are compared to 
methods from the literature
that have a similar 
speed--quality trade-off 
as the method class it is being compared against.
Our second comparison assesses  
 fast methods against  other methods with a similar run time.
For completeness, we also compare the {\sc AdvHits} method in this comparison.
We assess the performance against 
two well-known fast approaches for directed networks, namely 
{\sc SaPa} from~\cite{satuluri2011sym} 
and 
{\sc DiSum} from~\cite{Rohe12679}. 
For full details of our implementation and the variants we consider see 
SI~D(e).
For brevity we only include the best performing {\sc SaPa} and {\sc DiSum} variant 
namely {\sc SaPa2}, using degree-discounted symmetrisation, and {\sc {DiSum3}}, a combined row and column clustering into four sets, using the concatenation of the left and right singular vectors.

The final comparison assesses  full likelihood methods.
We compare to the block
modelling fitting approach ({\sc GraphTool}) from the graph-tool software
package~\cite{graphtool} based on the approach from
\cite{peixoto2014efficient,peixoto2013hierarchical},
by using a procedure that minimises the the minimum description length of the observed data
directly. 
To make this a fair comparison we use the non-degree corrected block
modelling fitting approach, and we fix the number of sets at
$4$, which makes this approach to equivalent to maximising the likelihood. %
For completeness, we also compare against {\sc AdvHits}, as it has a time
profile between the fast methods and the slow methods. 

\paragraph{Benchmark 1}

We test our approaches using the $1$-parameter model from \cref{1parm_model}, with equally-sized sets  
and varying   the parameter 
$p \in \{0.5,0.49,0.48,\ldots,0.21\}\cup \{0.195,0.19,0.185,\ldots,0.005\}$. 
We choose a finer discretisation step in the $p$-range which corresponds to a weaker planted structure, 
and 
we average over $50$ 
network samples for each value of $p$. 
Recall that for $p=0.5$ the planted structure corresponds to the idealised block structure in ~\cref{DBM} and for $p=0$ the planted structure corresponds to an Erd\H{o}s-R\'enyi random graph with edge probability $0.5$. 

The performance results  are shown{
{in Table~\ref{table:model1andeari}, giving  the ARI for  $p=0.4$ and $p=0.1$, and then for values of $p$ between $0.05 $ and $0.015$ with  step size $0.005$, in decreasing order.}} 

\begin{table}[h!]
\begin{center}
\begin{tabular}{|c| |c |c |c |c |c |c |c |c |c |c |c |}
%\begin{tabular}{|c|
%|>{\hspace{-0.25mm}}c<{\hspace{-0.25mm}}
%|>{\hspace{-0.25mm}}c<{\hspace{-0.25mm}}
%|>{\hspace{-0.25mm}}c<{\hspace{-0.25mm}}
%|>{\hspace{-0.25mm}}c<{\hspace{-0.25mm}}
%|>{\hspace{-0.25mm}}c<{\hspace{-0.25mm}}
%|>{\hspace{-0.25mm}}c<{\hspace{-0.25mm}}
%|>{\hspace{-0.25mm}}c<{\hspace{-0.25mm}}
%|>{\hspace{-0.25mm}}c<{\hspace{-0.25mm}}
%|>{\hspace{-0.25mm}}c<{\hspace{-0.25mm}}
%|>{\hspace{-0.25mm}}c<{\hspace{-0.25mm}}
%|>{\hspace{-0.25mm}}c<{\hspace{-0.25mm}}
%|}
\hline
$p$:
& $0.4$ & $0.1$  & $0.05$ & $0.045$ & $0.04$ & $0.035$ & $0.03$ &
$0.025$ & $0.02$ 
&
$0.015$ %
\\ \hline 
{\sc Deg.} & $1.0$ & $0.985$ & $0.745$ & $0.670$ & $0.592$ & $0.492$ & $0.384$ & $0.279$ & $0.179$ & $0.104$%
\\
{\sc Disum3}   & $1.0$ & $0.919$ & $0.180$ & $0.115$ & $0.057$ & $0.020$ & $0.007$ & $0.002$ & $0.001$ & $0.000$ %
\\
{\sc SaPa2} & $1.0$ & $0.954$ & $0.186$ & $0.137$ & $0.089$ & $0.044$ & $0.015$ & $0.005$ & $0.002$ & $0.001$ %
\\ 
{\sc HITS} & $1.0$ & $0.995$ & $0.766$ & $0.687$ & $0.605$ & $0.507$ & $0.393$ & $0.255$ & $0.160$ & $0.094$ %
\\ 
{\sc LowRank} & $1.0$ & $0.992$ & $0.764$ & $0.687$ & $0.605$ & $0.506$ & $0.393$ & $0.255$ & $0.160$ & $0.094$ %
\\
{\sc AdvHits} & $1.0$ & $1.0$ & $0.888$ & $0.825$ & $0.733$ & $0.613$ & $0.479$ & $0.335$ & $0.196$ & $0.103$ %
\\
{\sc GraphTool} & $1.0$ & $1.0$ & $0.965$ & $0.928$ & $0.797$ & $0.013$ & $0$ & $0$ & $0$  & $0$  %
\\
{\sc HillClimb} & $1.0$ & $1.0$ & $0.967$ & $0.932$ & $0.874$ & $0.763$ & $0.611$ & $0.412$ & $0.216$ & $0.096$ %
\\
{\sc MaxLike} & $1.0$ & $1.0$ & $0.967$ & $0.932$ & $0.874$ & $0.751$ & $0.611$ & $0.409$ & $0.215$ & $0.093$ %
\\ \hline \end{tabular} 
\end{center}
\caption{ARI of the methods under comparison on Benchmark 1 for different values of $p$ and with network size $n=1000$. 
}
\label{table:model1andeari}
\end{table}

\begin{figure}
\centering
\includegraphics[height=7cm]{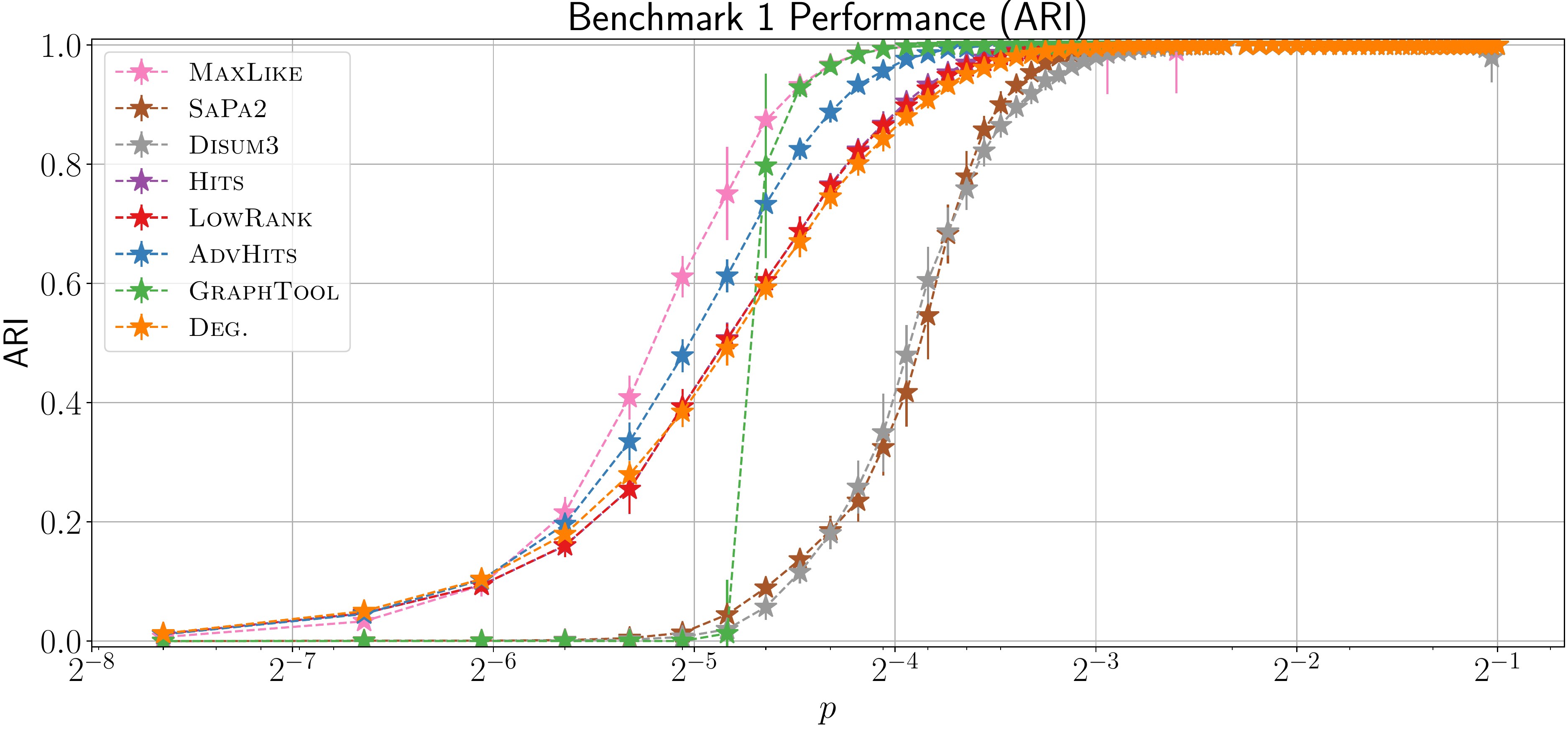} 
\caption{Performance on Benchmark $1$ (see~\cref{1parm_model}). The ARI between the planted
partition of the graph and the partition detected by each method
for networks of size 
$n=1000$. On the $x$ axis, we vary the parameter $p$
on a log scale.  
Error bars are one sample standard
deviation.
}
\label{fig:model1ande}
\end{figure}

We graphically compare the performance of 8 of these methods 
in Figure~\ref{fig:model1ande} 
for $n=1000$, and we further present a comparison with the case where $n=400$
in SI~D(b). %
For ease of readability {\sc HillClimb} is not included in
Figure~\ref{fig:model1ande} as Table~\ref{table:model1andeari} illustrates
that its performance is very similar to {\sc MaxLike}.

We observe that in all cases, Method class 3 (likelihood maximisation)
outperforms the other approaches for a large range of $p$. 
The {\sc AdvHits} algorithm also achieves good performance, while
taking 
less run time than the  likelihood approaches
(see SI~D(c) %
for further details on run time).
For  NMI and VOI  we observe similar qualitative results, see SI~D. %

Above a certain threshold of $p$ (roughly around $p=0.2$), {many approaches including the} degree-based approach {\sc Deg.} achieve 
optimal performance, indicating that in this region of the 
networks
obtained with Benchmark 1, the degrees alone are sufficient to uncover the
structure. %
Below this threshold,  the {\sc AdvHits} and likelihood
approaches  strongly outperform the degree-based approach,  while the {\sc LowRank} and
{\sc HITS} methods equal the performance of the degree method in some regions and
outperform it in others. 
{\sc SaPa2} from \cite{satuluri2011sym} and {\sc DiSum3} 
from
\cite{Rohe12679} are outperformed by our methods
{\sc LowRank}, {\sc HITS}, and {\sc AdvHits}.
The slow likelihood-based methods {\sc MaxLike}  and {\sc{HillClimb}} 
outperform even 
{\sc GraphTool} which fits a general block
model to the data,  
although the difference is less pronounced than the difference between the
fast methods.
Finally, we note that the performance of {\sc GraphTool} collapses
as $p$ gets close to $0$ (similar behaviour is observed for $n=400$
see SI~D(b)).
Further investigation indicated that  for low values of $p$, {\sc GraphTool} often places most vertices in a single set (see SI~D
for further details). 
\paragraph{Benchmark $2$}

\begin{figure}
\centering
\includegraphics[width=1.0\textwidth]{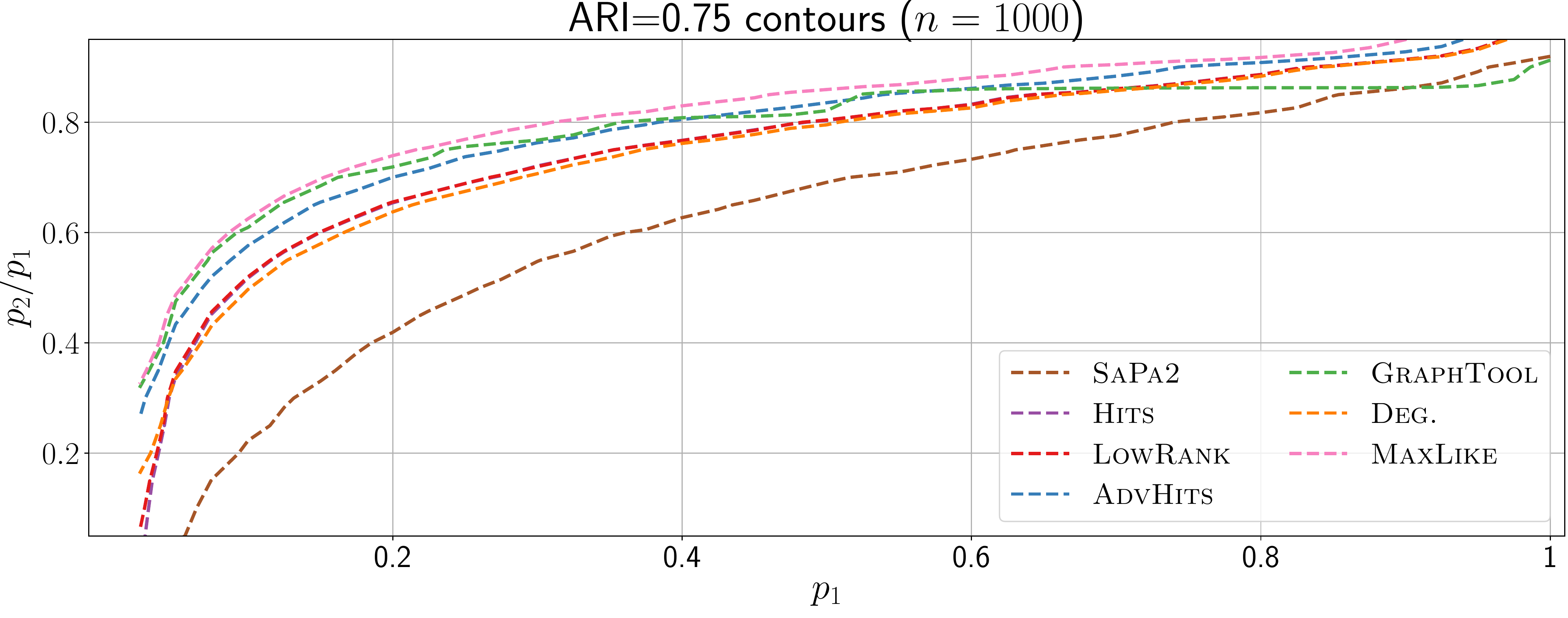}
\caption{Contour plot of {ARI=0.75} 
on Benchmark $2$ (see~\cref{2parm_model}). 
For this plot, we calculate the average ARI over 10 networks of size 
$n=1000$ 
varying $p_1 \in
\{0.025,0.05,\ldots,1.0\}$ and $\frac{p_2}{p_1} \in \{0,0.05,0.1,\ldots,0.95\}$, and
then display the contours corresponding to an average ARI of $0.75$. %
We note
that the fast methods {\sc HITS} and {\sc LowRank} outperform the comparisons
{\sc Deg.} and {\sc SaPa2}, although they are hard to distinguish due to similar performance. 
}
\label{fig:model2_mk2}
\end{figure}

We use the 
two-parameter model from~\cref{2parm_model}, 
again with all four sets of the same size 
$\frac{n}{4}$, 
now the edge probabilities $(p_1,p_2)$   vary  
the density 
and   the strength of the core
periphery structure.  
To this end, we vary $p_1$, and the ratio $0 \le \frac{p_2}{p_1}<1$. For a given $p_1$,
$\frac{p_2}{p_1}=0$ corresponds to the strongest structure and   
$\frac{p_2}{p_1}=1$ to the weakest structure.
We generate 10 networks each with  { $p_1 \in
\{0.025,0.05,\ldots,1.0\}$} and $\frac{p_2}{p_1} \in \{0, \allowbreak
0.05,\ldots, \allowbreak 0.95\}$, resulting in $820$ parameter {instances of $(p_1,\frac{p_2}{p_1})$}. 
The results for $n=1000$ are in
\cref{fig:model2_mk2} with a comparison to $n=400$ presented in
SI~D(b). 
We present the contours corresponding to an average ARI of $0.75$. 
Due to space constraints, we only display {\sc HITS}, {\sc LowRank}, {\sc
AdvHits} and {\sc MaxLike} as well as the best-performing other methods {\sc
SaPa2}, {\sc GraphTool} and {\sc Deg}. 
The performance of {\sc HillClimb} is again very similar to that of {\sc
MaxLike} and not shown.
The  same comparisons performed using VOI and NMI can be found in
SI~D. %

As observed for  Benchmark $1$,  the full likelihood
approaches 
outperform all other methods, with the
performance of  {\sc AdvHits} coming
close and outperforming  {\sc GraphTool}  in certain regions. 
Comparing 
the fast methods only,
again, the slower {\sc AdvHits} approach outperforms all of the other fast
approaches, performing almost as well as the full likelihood {\sc MaxLike}
method. Within the slow methods,  {\sc MaxLike}  outperforms 
 {\sc GraphTool}, although occasionally it only finds a local maximum.

The $0.9$ contour (
Fig.~SI~2
of SI~D)
shows that the performance of {\sc GraphTool} is  closer to that
of the likelihood methods compared to the 0.75 contour.   
This could again be related to 
{\sc GraphTool} often 
placing all vertices in the same community when the structure is weak.%

\paragraph{Benchmark 3} 

Benchmark 3  assesses 
the sensitivity of our methods to different set sizes.
In  the 1-parameter model from
\cref{1parm_model}, fix $p=0.1$ as this value is
sufficiently large to see variation in performance between our approaches, but
sufficiently small that most of the methods can detect the underlying
structure. 
We then fix the size of three of the sets and vary the size of the final set.
For example, to vary $\mathcal{P}_{out}$, we fix  $n_{\mathcal{C}_{in}}=
n_{\mathcal{C}_{out}}=n_{\mathcal{P}_{in}}=\frac{n}{4}$ and test 
performance when we let $n_{\mathcal{P}_{out}} \in
\{2^{-3}\frac{n}{4},2^{-2}\frac{n}{4},\ldots,2^{3}\frac{n}{4}\}$, with
equivalent formulations for the other sets.

\begin{figure}[htp]
\centering
\includegraphics[height=9.25cm]{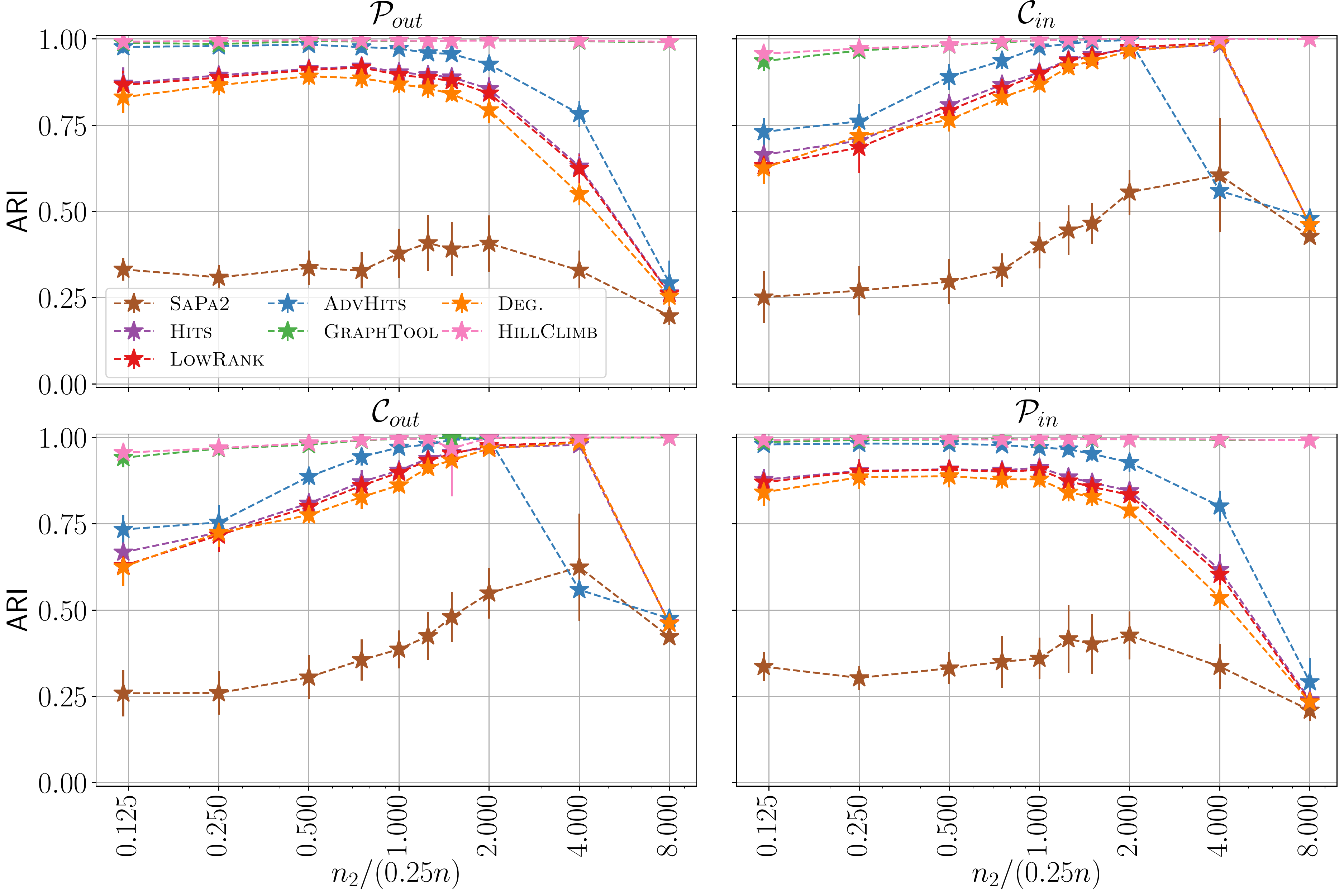}
\caption{ARI of all methods on Benchmark $3$ with
sets of different sizes (see~\cref{2parm_model}). We fix $p=0.1$, the size of 3 sets at
$n/4$ with $n=400$, and vary the size (denoted as $n_2$) of the fourth set. Error bars are one sample standard deviation.
}
\label{fig:model3ande}
\end{figure}

Results for $n=400$ are shown in \cref{fig:model3ande}. The relative
performance of the methods matches that of the first two
synthetic experiments, with the likelihood methods performing best, followed by
{\sc AdvHits},  
then by the faster {\sc LowRank} and {\sc
HITS} methods, and finally by the other spectral approaches.
The likelihood methods appear to be robust to set size
changes. The performance of the faster methods depends on core or periphery set changes.  
In the periphery, smaller set
sizes seem to have a minimal impact on performance, while large set sizes
lead to a large decay in performance. In the core, both smaller and
larger sets appear to have a significant impact. For a large core set, the fast methods outperform the {\sc AdvHits}
methods, as the performance of {\sc AdvHits} decays much faster, and in some
cases, {\sc AdvHits} is even outperformed by {\sc Deg}. 
\section{Application to real world data}
\label{realworld}

Next, we test our methods on three real-world data sets, namely Political Blogs 
(\datasetFmt{Blogs})
from \cite{politBlogs} 
(Section~\ref*{realworld}\ref{politicalBlogs}), 
Faculty hiring data 
(\datasetFmt{Faculty})
from \cite{clauset2015systematic}
(Section~\ref*{realworld}\ref{facultyHiring}), 
and Trade data 
(\datasetFmt{Trade})
from \cite{wtf} 
(Section~\ref*{realworld}\ref{worldTrade}). 
In each case our methods find a division into four sets
and we explore the identified
structure using any known underlying attributes. 
For each data set, we first assess the consistency of the partitions both within and between each
of the approaches.  We compute $11$ runs for each of the fast methods, 
$1$ run for the slow methods, and
then compute the within-method ARI between the resultant partitions and
the ARI between methods of different types. We also compare with the
structure uncovered by bow-tie~\cite{Broder2000},
as discussed in \cref{intro}. Bow-tie can
allocate vertices to several sets -- there is a core set, an incoming periphery set, an
outgoing periphery set and $4$ additional sets corresponding to other connection
patterns. 
Thus, we consider the ARI between the partition into $7$ sets ({\sc BowTie}),
and the partition induced only by the core set and the in- and out-periphery sets ({\sc BowTieAdj}).
{When computing the ARI between the partition induced by the core and periphery sets of {\sc BowTie} (the partition  we call {\sc BowTieAdj}) with another partition $\mathcal{S}$, we consider the partition induced by $\mathcal{S}$ on the vertex-set in {\sc BowTieAdj} (and thus, by construction, the ARI between {\sc BowTieAdj} and {\sc BowTie} is always $1$). }
As there is no ``ground truth'' core-periphery structure available for these networks, we carry out a Monte Carlo test to assess whether the detected structure could plausibly be explained as arising from random chance. 
To this end, we take the difference
between the probability of connection within the `L' shape ($d_{in}$), with
that outside of the `L' shape ($d_{out}$),
maximising 
the likelihood
over the $4!=24$  ways of rearranging the sets.
This statistic is used in two Monte Carlo tests,  with $250$ repeats each, against the following null models:
(1) a directed Erd\H{o}s-R\'{e}nyi (ER) model without self-loops (using the standard NetworkX implementation~\cite{networkx}) in which we set the number of vertices and the
connection probability equal to those of the observed network, and (2), 
 inspired in part by~\cite{BARUCCA2016244}, 
a directed configuration model with self-loops (again using the standard
NetworkX implementation~\cite{networkx}), which adjusts for the in- and out-degree of the
vertices, collapsing multi-edges into a single edges. 
In {\sc AdvHits}, these Monte Carlo tests 
often fall back to the single vertex update scheme and thus can be slow.

\subsection{Political Blogs}
\label{politicalBlogs}

\begin{figure}
\begin{center}
\includegraphics[height=5.50cm]{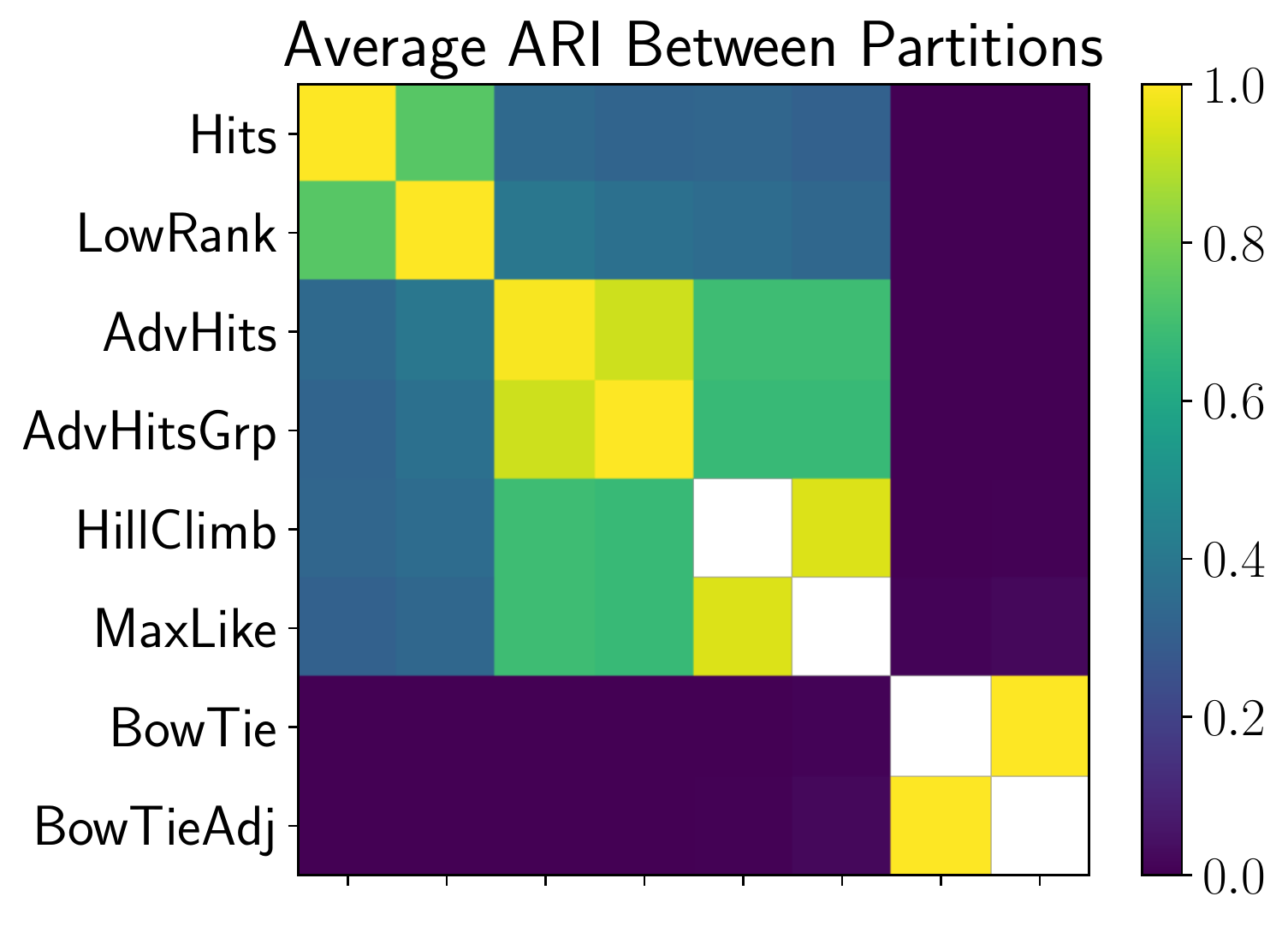}
\includegraphics[height=5.50cm]{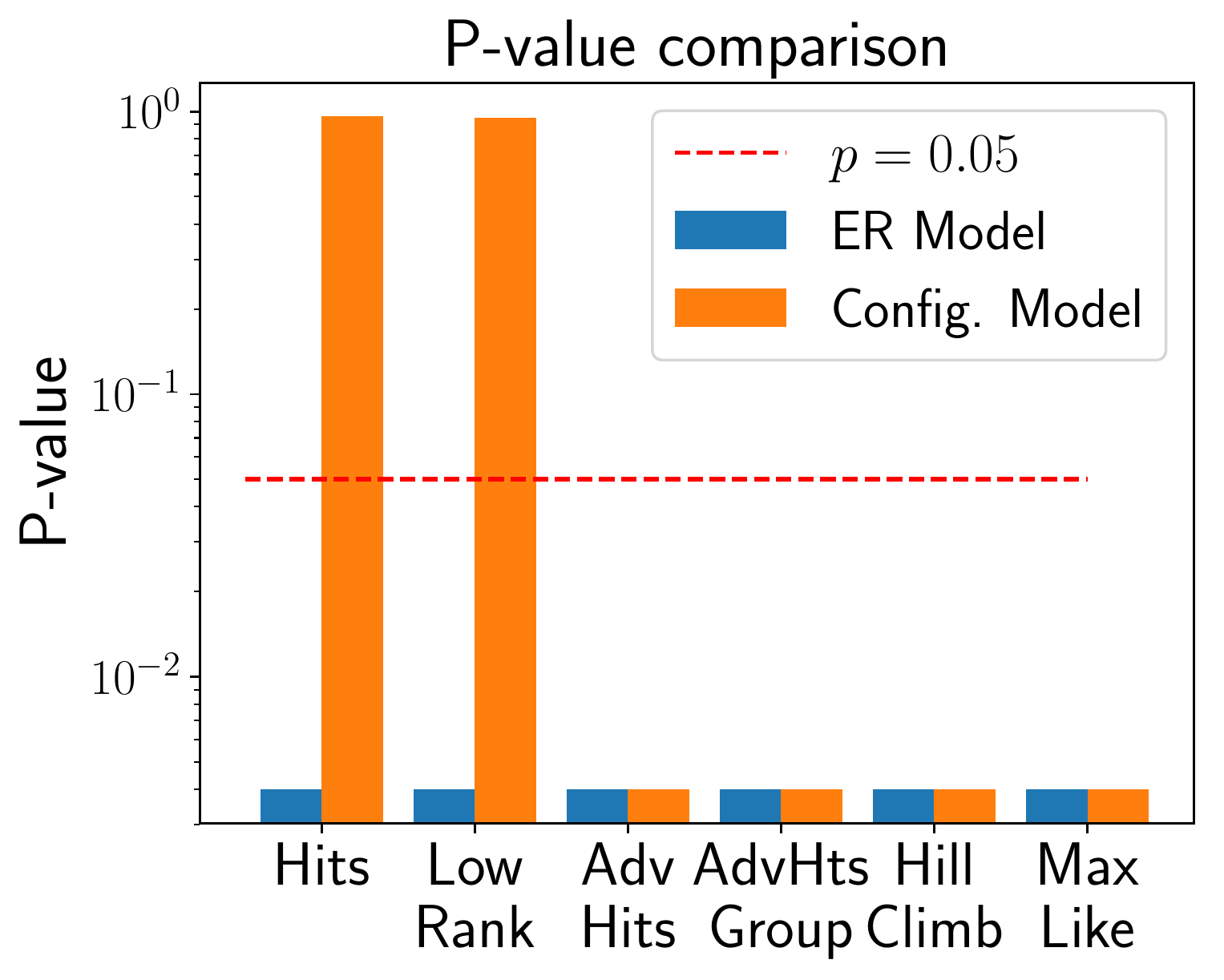}
\end{center}
\caption{\datasetFmt{PolBlog} results 
{\bf Left panel} The ARI between the partitions uncovered by each method. For ease of comparison, negative values are set to $0$. 
For the faster methods we compare with $11$ runs %
and show the average similarity between all pairs of partitions
whereas for the slow methods, 
we use a single run and thus display a blank (white) square on the corresponding diagonal blocks. 
To compare to bow-tie, 
we compare both to the partition into $7$ sets and a partition formed by a subset of the vertices corresponding to the 
main three sets.
{\bf Right panel} The $p$-values of the Monte Carlo test 
on each partition against a directed ER and a directed configuration model.
}
\label{politBlogResults}
\end{figure}

The data set, \datasetFmt{PolBlogs}  from \cite{politBlogs} 
consists of political blogs as vertices,  and  directed edges  from one Blog to another denoting that the first blog contains at least one link to the second blog. 
The data set was collected on a single day during the 2004 US presidential election.  After collapsing 65 multi-edges, it 
contains $1490$ vertices and $19090$ edges. The set of blogs is divided into 758 liberal blogs and 732 conservative blogs. %
For the analysis in this section, we focus on the largest weakly connected component of the network, which has 
$n=1222$ vertices and  $19024$ edges.

The first panel of \cref{politBlogResults} shows the ARI between the different approaches.
We note a high within-method ARI, and a high ARI between {\sc HITS} and {\sc LowRank} (perhaps unsurprisingly, as both use centrality-based scores), 
and also a high ARI  between {\sc AdvHits} and {\sc AdvHitsGrp} and between the likelihood methods. %
Furthermore, there is a low ARI between our methods and bow-tie,  indicating that our methods uncover a different structure. 

The Monte Carlo $p$-values of the tests described at the beginning of this Section 
 can be seen in 
\cref{politBlogResults}. The {\sc HITS} and {\sc LowRank} methods are 
significant at $5\%$ level  against the directed  ER null model but not against  the directed configuration
model, {potentially indicating that the structures found are a function
of the in- and out-degree distribution.}  It is of course possible that 
{\sc LowRank} and {\sc HITS}  uncover significantly non-random 
structures, just that these structures  do not correspond to
an `L'-shape structure. 
The other structures are significant at a $5\%$ level on both tests.
Noting the similarity between 
the likelihood methods, for brevity we restrict our discussion to  {\sc AdvHits}
method and {\sc MaxLike}.

\begin{figure}
\centering
\includegraphics[height=5.5cm]{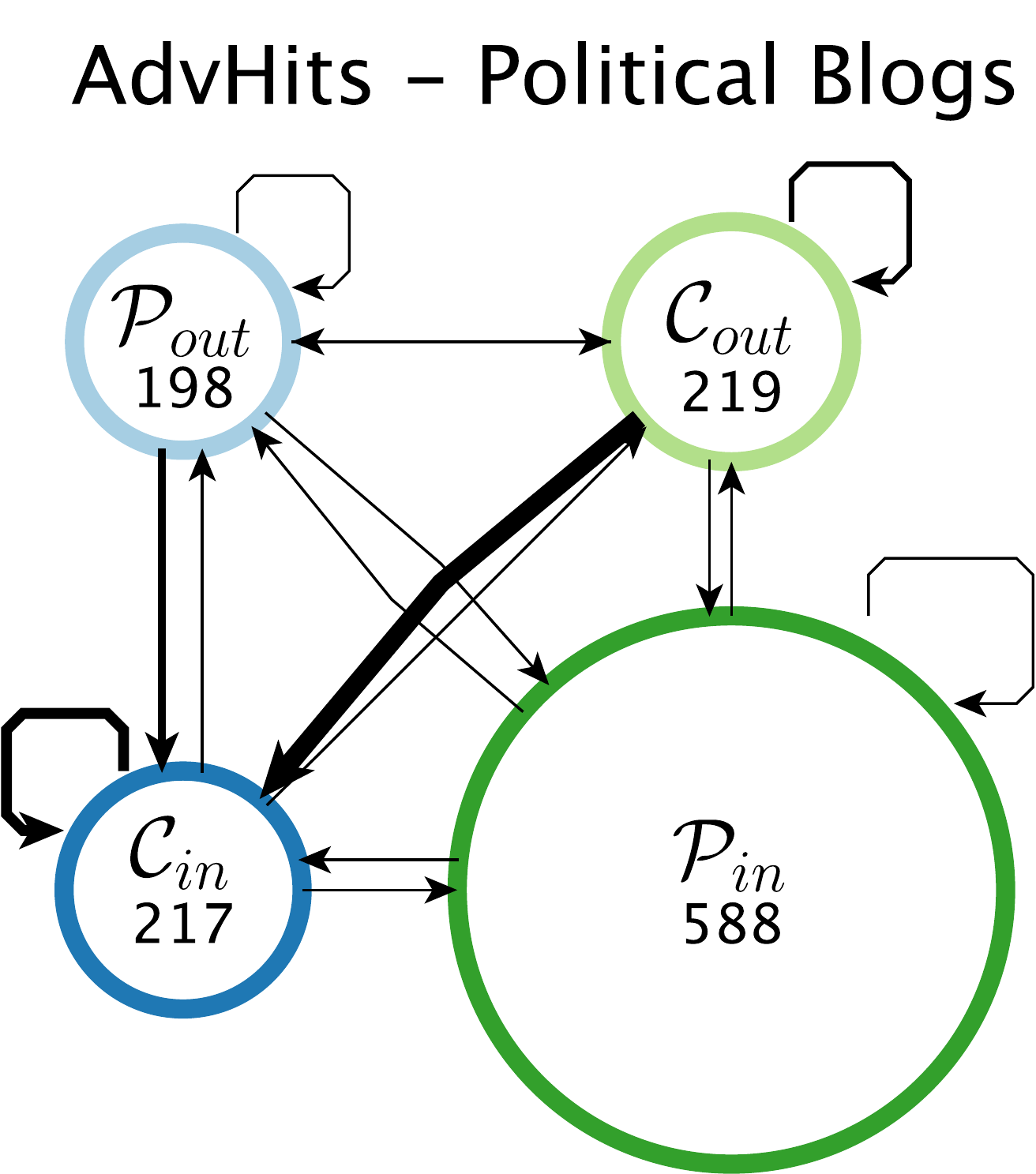}
\includegraphics[height=5.5cm]{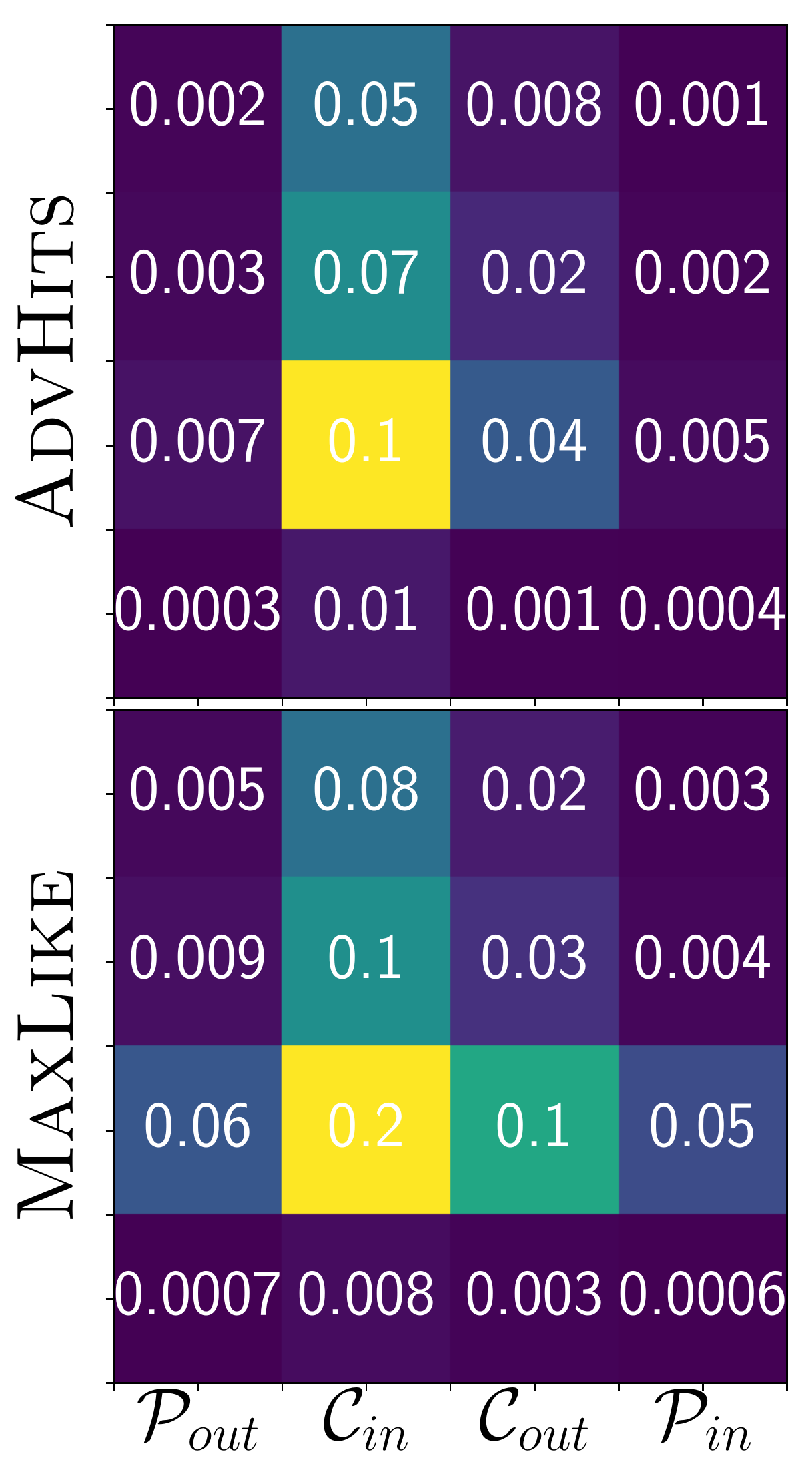}
\includegraphics[height=5.5cm]{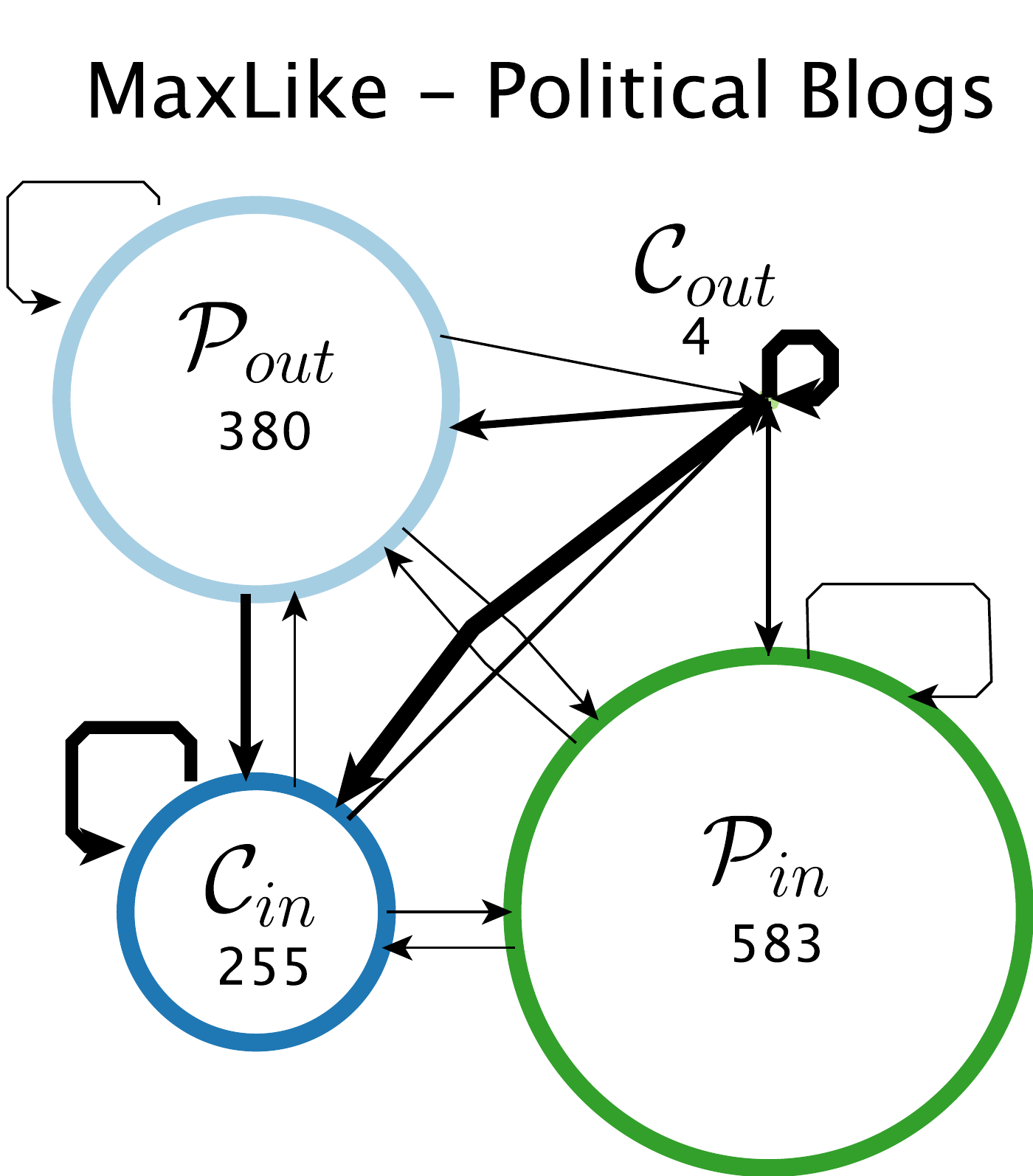}
\caption{
Structures in \datasetFmt{PolBlogs}. 
In the {\bf left} ({\sc AdvHits}) and {\bf right} ({\sc MaxLike}) panels, we show summary network diagrams associated with the uncovered structures. In these summaries the size of each of the vertices is proportional to the number of vertices in each set, and  the width of the lines is the percentage of edges that are present between the
sets. The width of lines is designed to highlight the differences and is not comparable between plots. The centre panel 
displays the percentage of edges between each pair of blocks, 
allowing for easy visualisation of the `L'-structure. 
}
\label{politBlogGraphPlot}
\end{figure}

Beginning with the {\sc AdvHits} method, %
\cref{politBlogGraphPlot} presents both a network visualisation and a matrix showing the density of edges between each pair of sets.
The underlying density structure reveals
an `L'-shape, albeit 
a weaker connection than anticipated in $\mathcal{P}_{in}$. 

\begin{figure}%
\centering
\begin{center}
\includegraphics[height = 5.50cm]{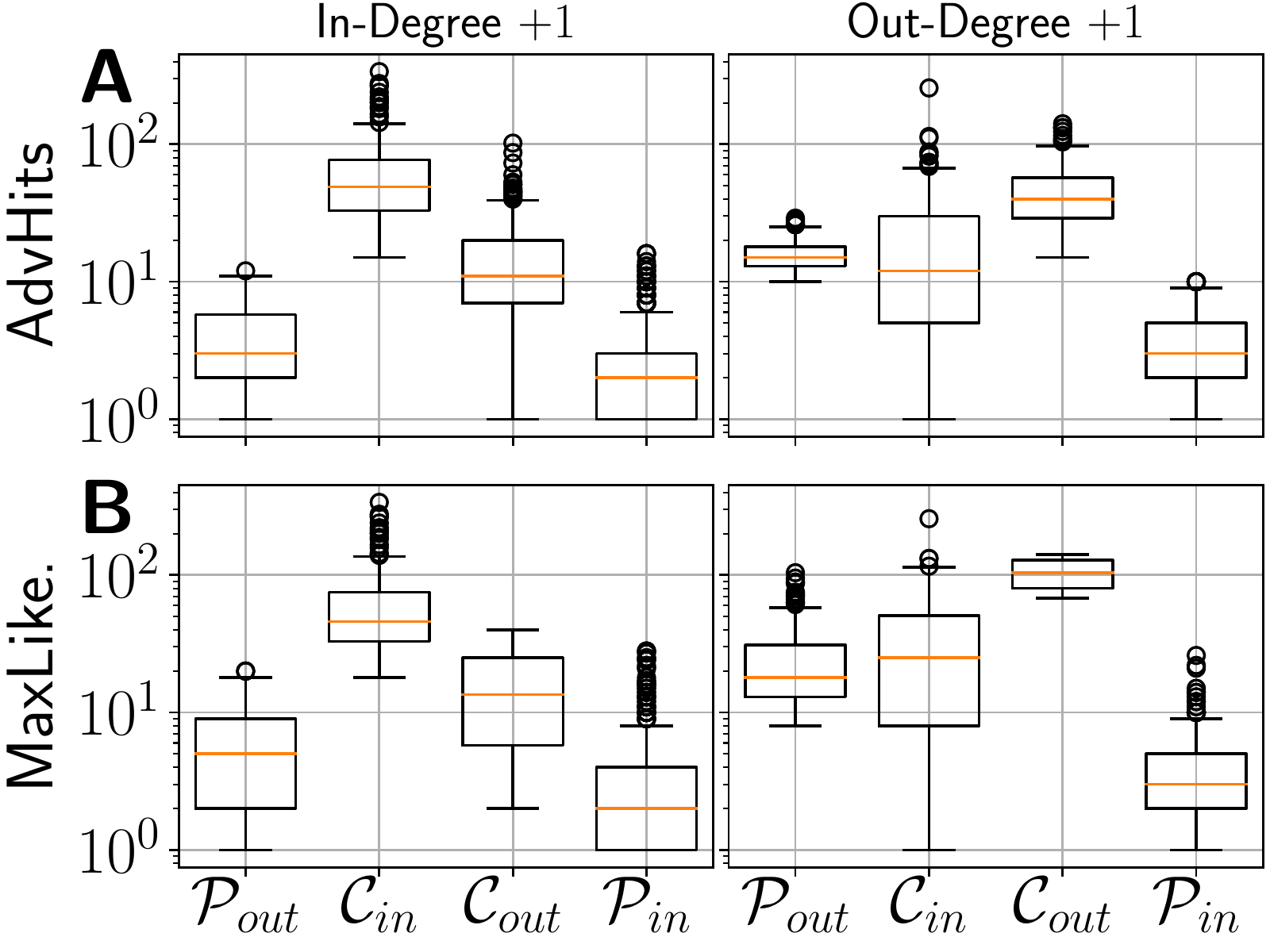}
\includegraphics[height=5.50cm]{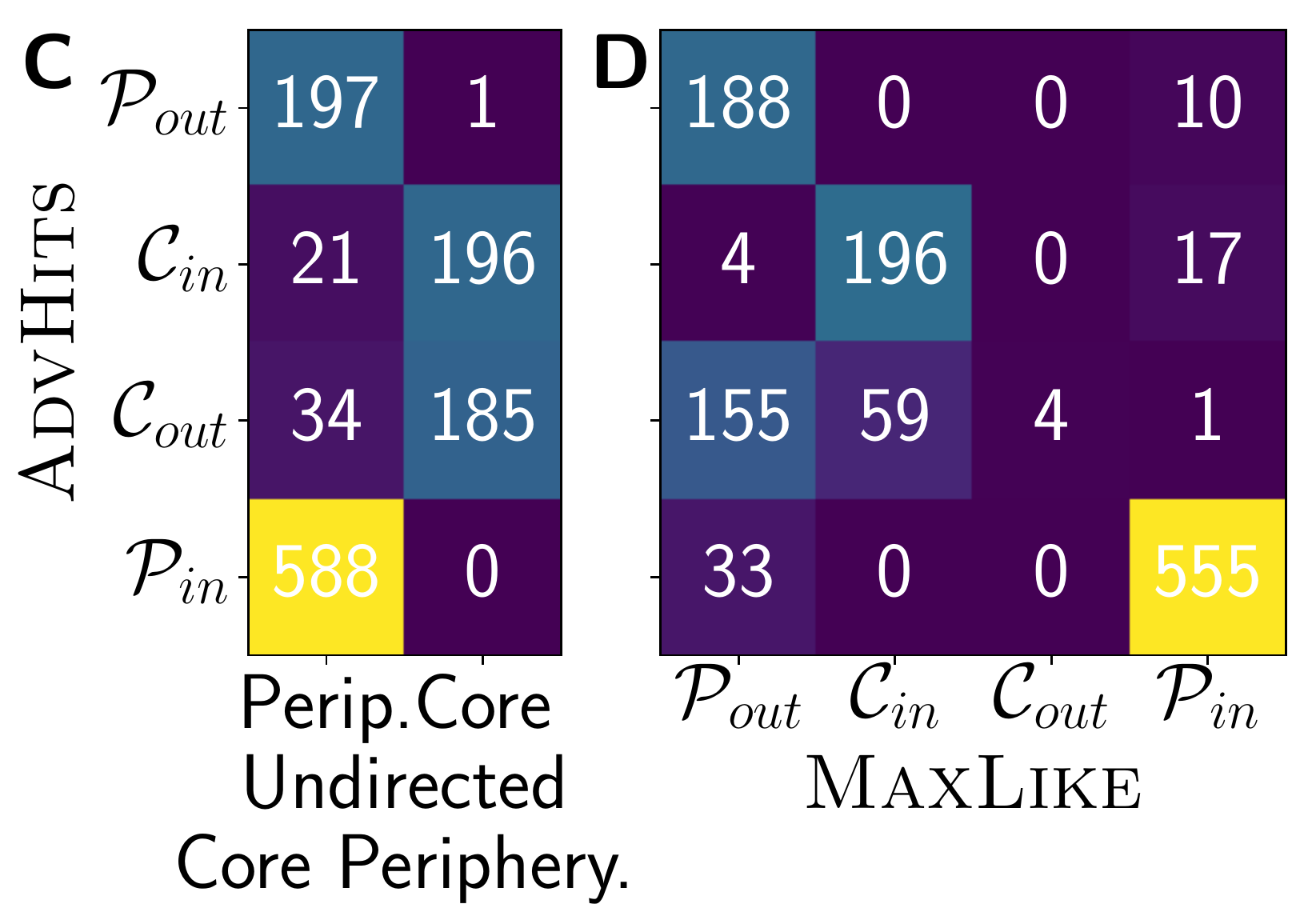}
\end{center}
\caption{\textbf{A} -  Boxplot of in- and out-degree in each of the sets in {\sc AdvHits}.
\textbf{B} -  Boxplot of in- and out-degrees in each of the sets in {\sc MaxLike}.
To visualise the in- and out-degrees on a log scale, we add $1$ to the degrees.
\textbf{C} -  Confusion table between the {\sc AdvHits} partition and the division into a standard core-periphery structure.
\textbf{D} -  Confusion table between the partition given by {\sc AdvHits} and the partition by {\sc MaxLike}. 
} 
\label{fig:politBlogsDegBoxplot}
\end{figure}

Assessing whether the partitions which are found 
by {\sc AdvHits} and {\sc MaxLike} methods relate to the classification of a blog as liberal or conservative, we observe
that the partitions 
are not just a function of
political affiliation; see 
Fig.~SI~9
in SI~E %
for full details.
Instead, similar to the division in \cite{zhang2014identification}, %
the sets may relate to the role they play in the network. Indeed, 
the in- and out-degree distributions in each of the sets, as shown in \cref{fig:politBlogsDegBoxplot}A,
indicate that the sets $\mathcal{P}_{out}$, $\mathcal{C}_{in}$ and $\mathcal{C}_{out}$ have in-and out-degrees which correspond  to a core--periphery structure.

Directly comparing the {\sc AdvHits} partition with the core-periphery partition, using the stochastic block model approach from \cite{zhang2014identification}, as implemented in the cpalgorithm package \cite{cpalgorithm}, we uncover the confusion table in \cref{fig:politBlogsDegBoxplot}C.
Our method splits the previously detected core into two distinct sets which fulfil different roles, and to a lesser extent, splits the periphery in a similar way.
In the case of the periphery sets, this could be the propensity of the individual site owners to link out to other content, as both periphery sets have a similar average in-degree, but differ on out-degree, a property controlled by the authors of the blog.
A possible explanation for the split within the cores is 
that 
the $\mathcal{C}_{in}$ set represents blogs that are seen as authorities, and thus do not link to many other blogs (but
are linked to themselves), whereas the opposite is true of those in
$\mathcal{C}_{out}$. 
A  potentially related structure was observed in $15$-block SBM from~\cite{peixoto2013hierarchical}, 
in which they identify blocks that cite but are not cited, and 
other blocks which are highly cited either in general or by specific groups.
To assess this hypothesis, an inspection of the
web-address corresponding to the vertices in each set suggests that certain
sets are enriched for  `blogspot' sites. A `blogspot' site is  a free blogging site that requires less
expertise to set up than a full website.  
Indeed, while  the fractions of `blogspot' sites in $\mathcal{P}_{out}$ ($0.566$),
$\mathcal{C}_{out}$ ($0.411$), and  $\mathcal{P}_{in}$ ($0.454$) are relatively
similar, the percentage in $\mathcal{C}_{in}$ is distinctly smaller at
$0.180$, supporting our hypothesis.
Moreover,~\cite{politBlogs} provides a list of 
the top $20$ conservative and the top $20$ liberal blogs, by filtering the top $\approx 100$ liberal and conservative blogs by degree, and ranking the top blogs by the number of citations of blog posts in an external index (see~\cite{politBlogs} for details).
In {\sc AdvHits}, 100\% of these top blogs are in $\mathcal{C}_{in}$, again supporting our hypothesis. 

The structure uncovered with {\sc MaxLike}, shown in
\cref{politBlogGraphPlot} again has  an `L'-shape structure, indicating a
split into the $4$ roles; compared to {\sc AdvHits},  the size of
$\mathcal{C}_{out}$ is greatly reduced with a much larger $\mathcal{P}_{out}$
and a slightly enlarged $\mathcal{C}_{in}$. {
We first note that the
likelihood of this structure is higher than that of the {\sc AdvHits} method.}  
The confusion table between this structure and the {\sc AdvHits} structure in
\cref{fig:politBlogsDegBoxplot}D 
suggests that most of the vertices in $\mathcal{C}_{out}$ in the {\sc AdvHits}
structure have moved to $\mathcal{P}_{out}$, with a reasonably sized subset also
moving to $\mathcal{C}_{in}$. 
The confusion table has a small number of large values, indicating that many pairs of vertices in the same set in one partition are also in the same set in the other.   
This is consistent with our previous assertion that the partitions are similar (as measured by ARI).
The degree patterns of each of the sets are still preserved, see  \cref{fig:politBlogsDegBoxplot}B, 
although with a much larger out-degree in $\mathcal{C}_{out}$. This split indicates  a
different partition of the sets into a large out-linking set consisting of $4$
vertices with large out-degree, with the remaining vertices mostly moving to
$\mathcal{P}_{out}$.  Furthermore, while  the undirected core--periphery  structure is not visible here (see 
Fig.~SI~10
in
SI~E), %
the confusion table indicates that the  $\mathcal{C}_{in}$ set is
relatively
unchanged,. The `blogspot' statistic ($0.21$ in $\mathcal{C}_{in}$
versus $>0.43$ in all other sets) and
$39$ out of $40$ top blogs are in $\mathcal{C}_{in}$ in this division,  also support the interpretation that $\mathcal{C}_{in}$ is enriched for  authorities.

\subsection{Faculty Hiring}
\label{facultyHiring}

In the faculty hiring network from~\cite{clauset2015systematic}, vertices are academic institutions, and  a directed edge from
institution $i$ to $j$ indicates that an academic received their PhD at $i$ and
then became faculty at $j$. The data set is divided by gender, faculty
position, and into three fields (Business, Computer Science, and History). For
brevity, we only consider the overall connection pattern in Computer Science. 
In \cite{clauset2015systematic} it is found that 
a large percentage of the faculty is
trained by a small number of institutions, and it is suggested  that there
exists a core--periphery-like structure.

We apply our methodology to this data set, and find that the 
results from the {\sc
AdvHits} variants and the likelihood methods are significant at $5 \%$  under both random null models. The partitions within method class are again very similar (data not shown). 
For brevity, due to the similarity between the partitions, we focus on the
results from the {\sc MaxLike} and  {\sc AdvHits}. 

\begin{figure} 
\begin{center}
\includegraphics[height=5.5cm]{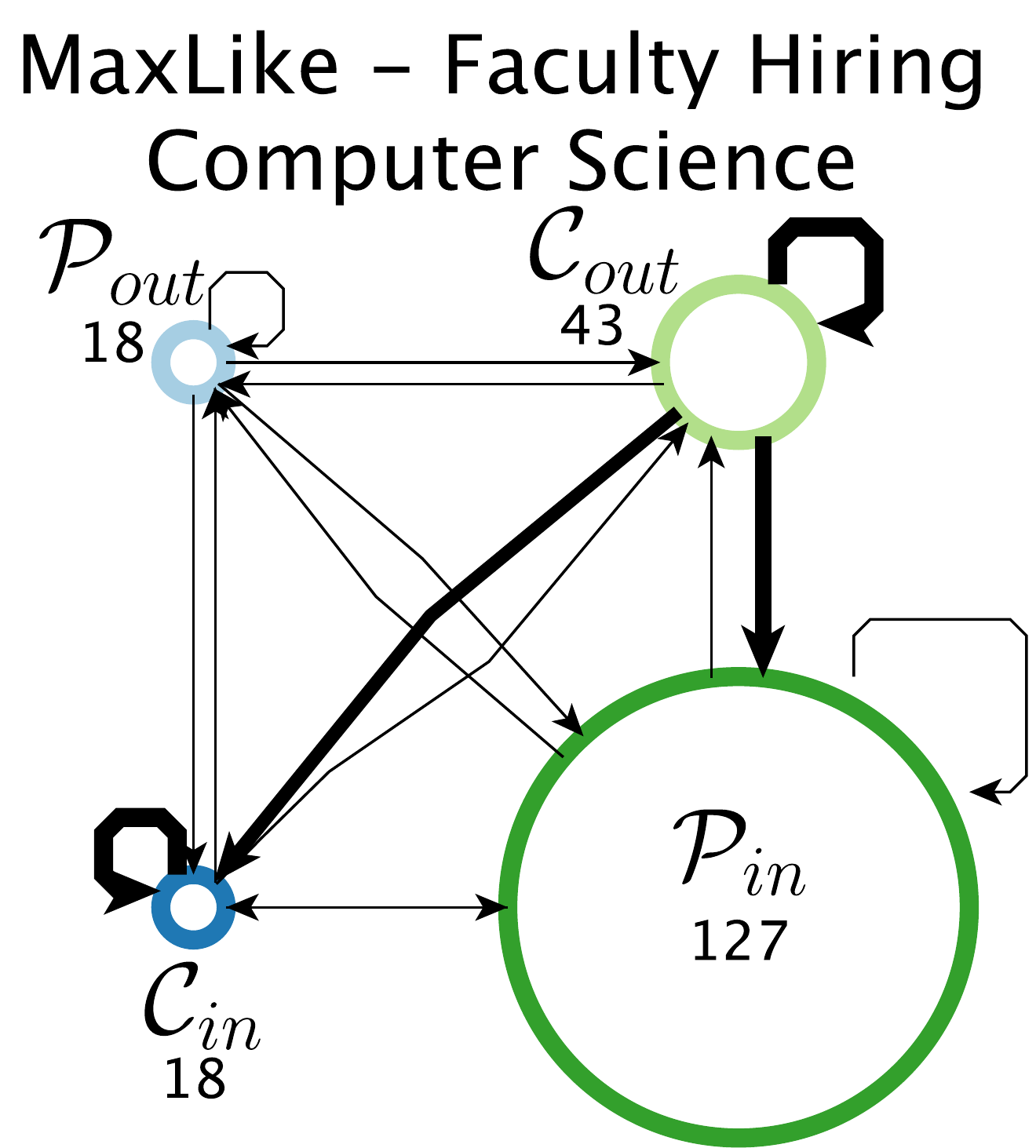}
\includegraphics[height=5.5cm]{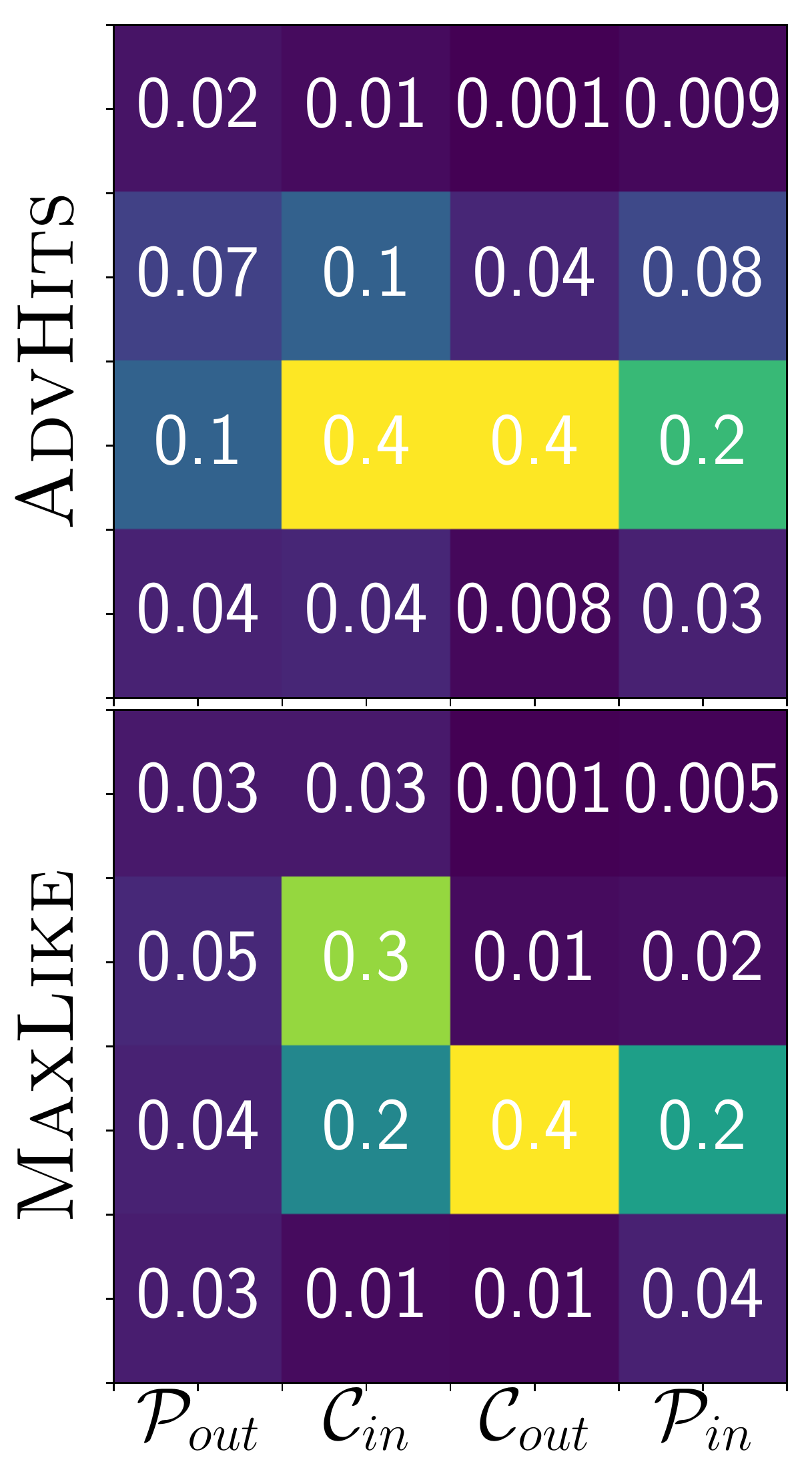}
\includegraphics[height=5.5cm]{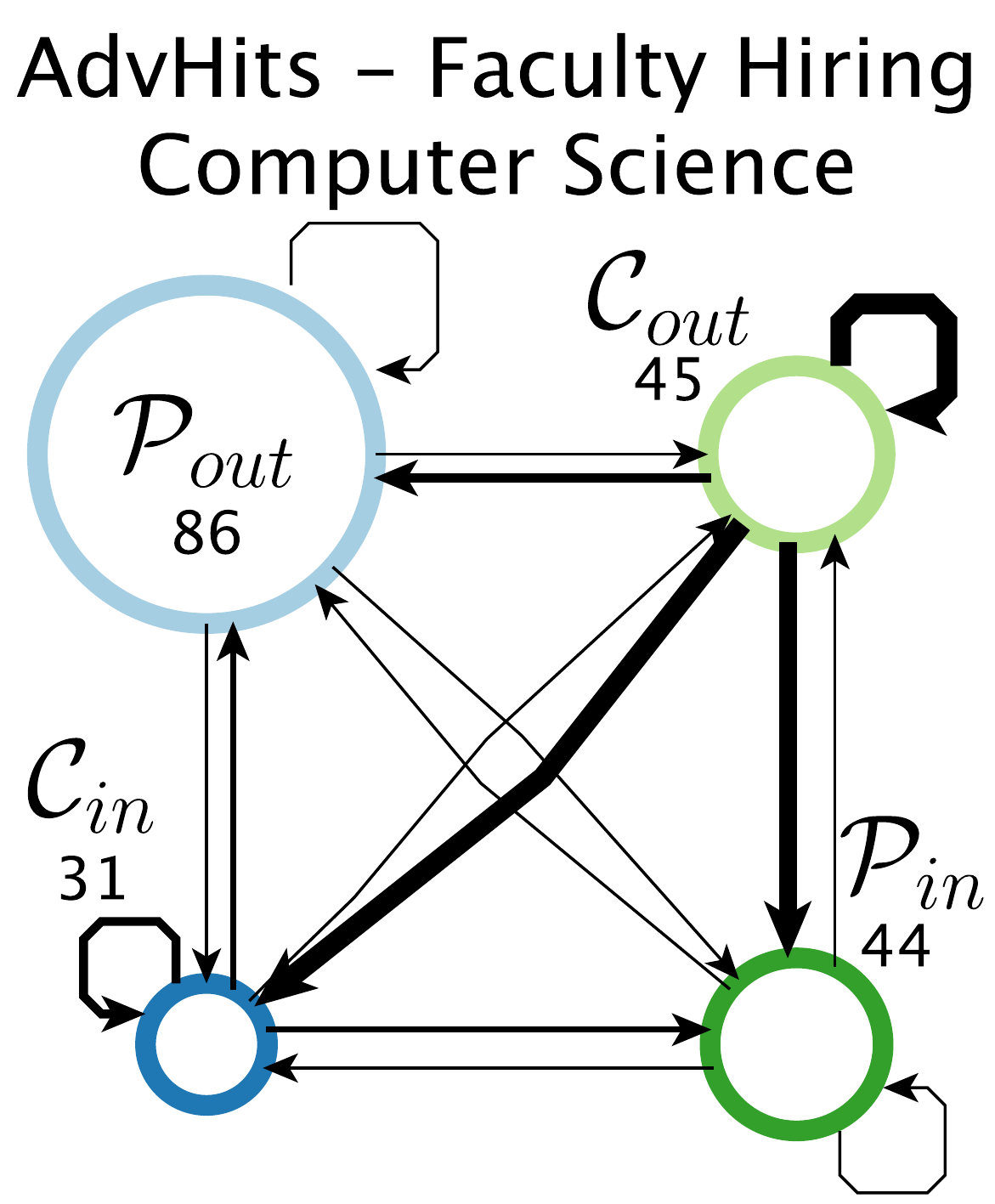}
\end{center}
\caption{Structures in \datasetFmt{Faculty} uncovered by 
{\sc MaxLike} and the {\sc AdvHits}. In the {\bf left} ({\sc MaxLike})
and {\bf right} ({\sc AdvHits}) panels we show summary network diagrams associated with the
uncovered structures.
In these summaries the size of each of
the vertices is proportional to the number of vertices in each set, the width of
the lines is given by the percentage of edges that are present between the
sets. The width of lines is designed to highlight the
differences and is therefore not comparable between plots. The centre panel
displays the percentage of edges between each pair of blocks which
allowing easily visualisation of the
`L'-structure. }
\label{fig:facultyHiringNetworks}
\end{figure}

\begin{figure}%
\begin{center}
\includegraphics[height = 5.50cm]{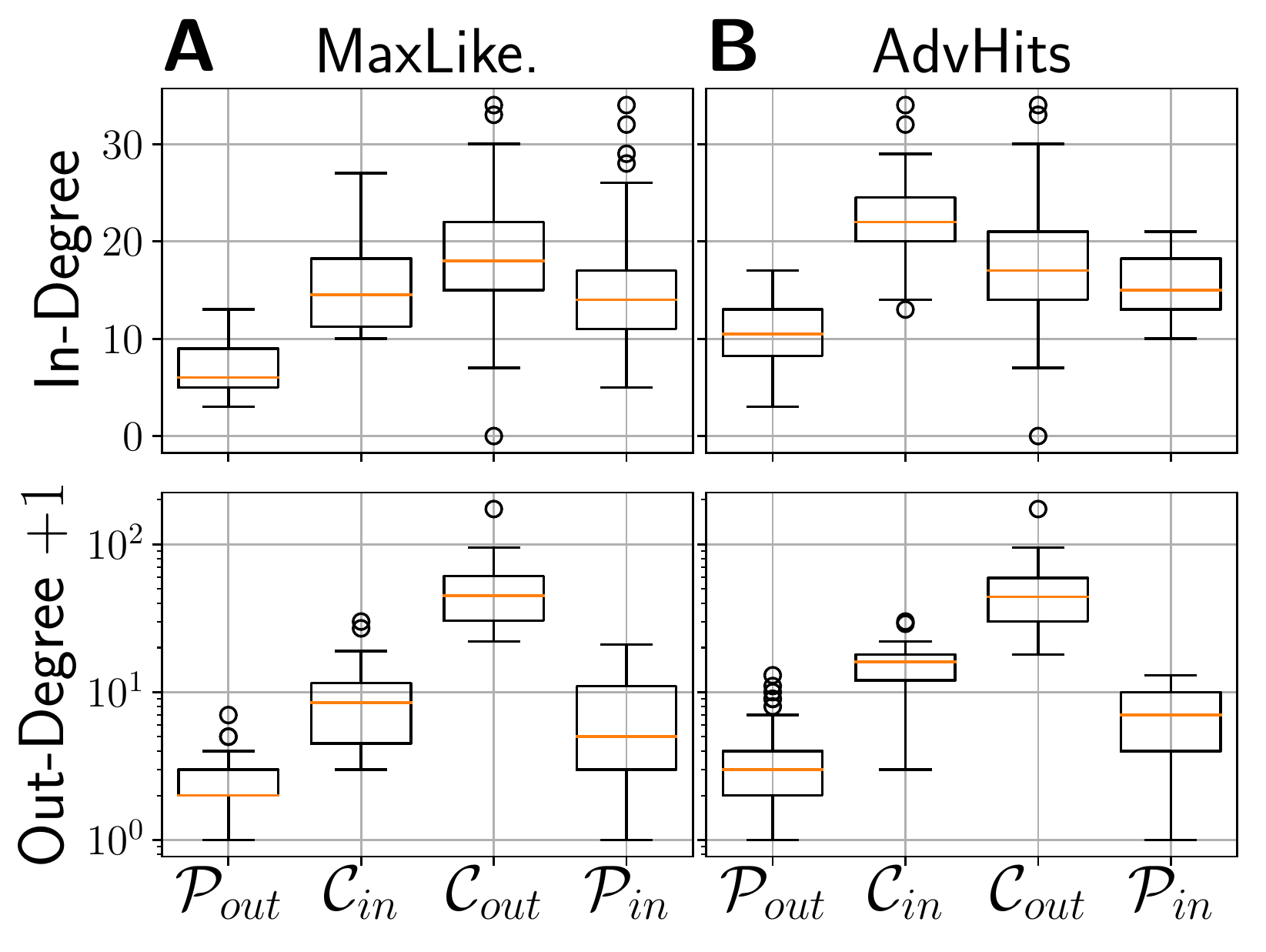}
\includegraphics[height=5.50cm]{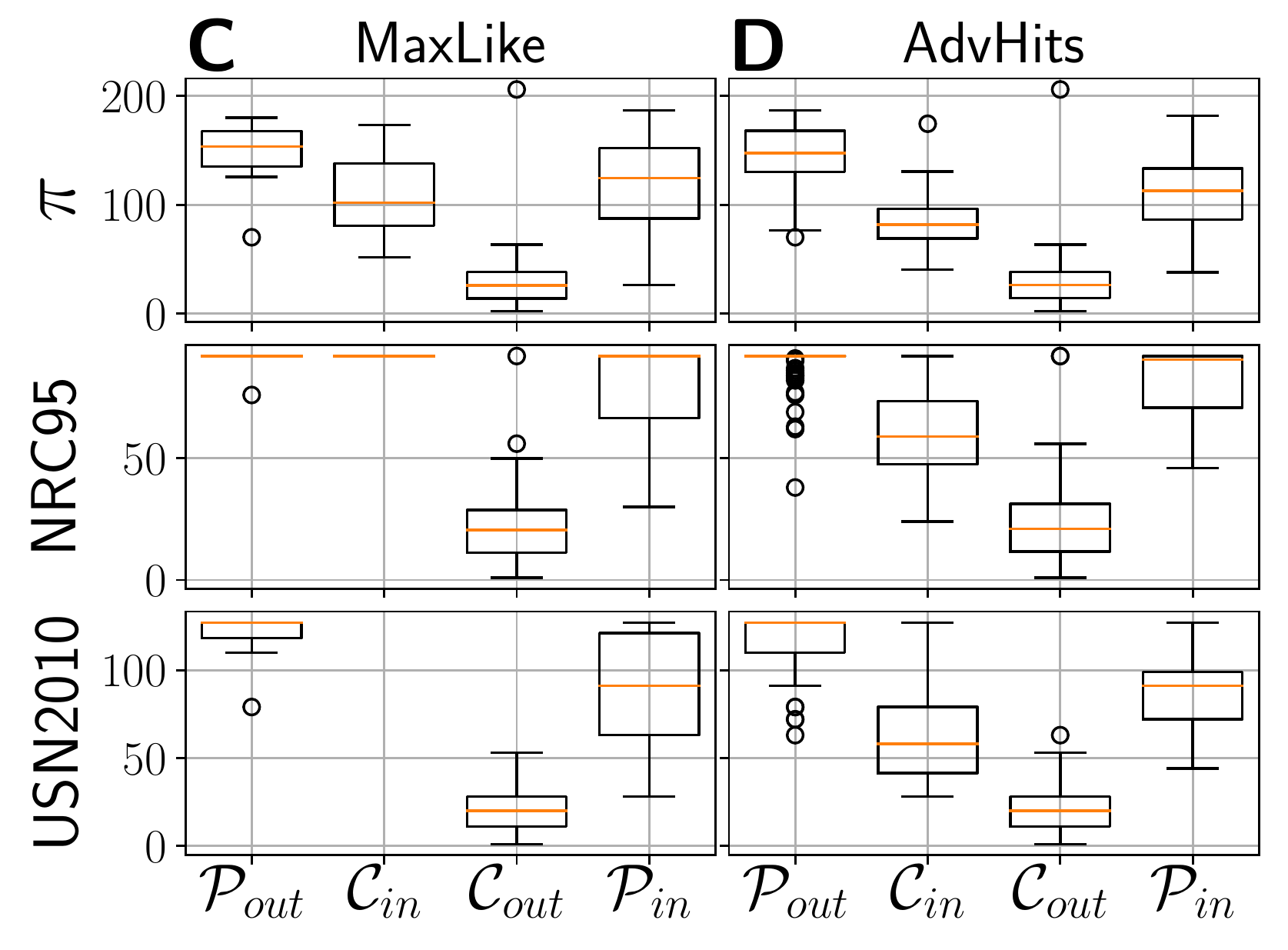}
\end{center}
\caption{\textbf{A} -  Boxplot of in- and out-degrees in each of the sets in {\sc MaxLike}.
\textbf{B} -  Boxplot of in- and out-degrees in each of the sets in {\sc AdvHits}.
To visualise the out-degrees on a log scale, we add $1$ to the degrees.
\textbf{C} -  Boxplot of the ranking in \cite{clauset2015systematic}, denoted $\pi$, ranking in NRC95 and the ranking in USN2010 in each of the sets in {\sc
MaxLike}. 
\textbf{D} -  Boxplot of the ranking in \cite{clauset2015systematic}, the
ranking in NRC95 and the ranking in USN2010 in each of the sets in {\sc
AdvHits}.
} 
\label{fig:facultyBoxplot}
\end{figure}

The {\sc MaxLike} results in \cref{fig:facultyHiringNetworks} show a clear
`L'-shape structure, albeit with a weakly defined $\mathcal{P}_{out}$. 
Inspecting the sets, $\mathcal{C}_{out}$ consists of highly-ranked
institutions, including Harvard, Stanford, MIT and also a vertex that represents
institutions outside of
the data set. 
\cref{fig:facultyBoxplot}C shows the University ranking  $\pi$ obtained by
\cite{clauset2015systematic}, and the two other University rankings used in the
\cite{clauset2015systematic}, abbreviated NRC95 and USN2010, 
in each of the sets which are found using {\sc MaxLike}. Here, the NRC95 ranking from 1995 was used because the computer science community rejected the 2010 NRC ranking for computer science as inaccurate. The NRC ranked only a subset of the institutions; all other institutions were assigned the highest NRC rank $+ 1 = 92$.
The set $\mathcal{C}_{out}$ has considerably smaller ranks than the other
sets, indicating that $\mathcal{C}_{out}$ is enriched for  highly ranked institutions.  
The set $\mathcal{P}_{in}$ from {\sc MaxLike} appears to represent a second
tier of institutions who take academics from the schools in $\mathcal{C}_{out}$
(\cref{fig:facultyHiringNetworks}) but do not return them to the job market.  
This observation can again be validated by considering the rankings 
in \cite{clauset2015systematic} (\cref{fig:facultyBoxplot}C).
The $\mathcal{C}_{in}$ set loosely fits the expected structure with a strong
incoming link from $\mathcal{C}_{out}$ and a strong internal connection
(\cref{fig:facultyHiringNetworks}), suggesting a different role to that of  the
institutions in $\mathcal{P}_{in}$.  A visual inspection of the vertices in
$\mathcal{C}_{in}$ reveals that 
$100$\% of the institutions in $\mathcal{C}_{in}$ are Canadian (also explaining
the lack of ranking in USN2010 (\cref{fig:facultyBoxplot}C)). In contrast, the proportion of Canadian universities in 
$\mathcal{P}_{out}$ is $11.1$\%; in $\mathcal{C}_{out}$ it is $2.3$\%, and in 
$\mathcal{P}_{in}$ it is  $0.79$\%. 
This finding suggests that Canadian universities tend to play a structurally different role to US universities,
tending to recruit faculty from other Canadian universities, as well as from
the top US schools. 
In \cite{clauset2015systematic} the insularity of Canada was already noted, but without a core-periphery interpretation. 

Finally, $\mathcal{P}_{out}$ is weakly connected both internally and to the remainder of the network (\cref{fig:facultyHiringNetworks}). In each of the
rankings
(\cref{fig:facultyBoxplot}C),
$\mathcal{P}_{out}$
has slightly lower average ranks than the other sets (with the exception 
$\mathcal{C}_{in}$, due to the default/missing rankings of 
Canadian institutions). This could indicate that
$\mathcal{P}_{out}$ consists of lower ranked institutions which are not
strong enough to attract faculty from the larger set of
institutions. 
The in- and out-degree distributions, (\cref{fig:facultyBoxplot}A),
show that $\mathcal{P}_{out}$ has lower in and out-degree distributions than
the other sets. Thus, an alternative hypothesis is  that $\mathcal{P}_{out}$
consists of universities with smaller Computer Science departments which do not interact with the wider network. We leave addressing this interpretation to future work.
In either case,
the institutions in $\mathcal{P}_{out}$ do not appear
to match the pattern observed in the remainder of the network and hence it is plausible to delegate them into one set.

While in the  {\sc AdvHits} result (\cref{fig:facultyHiringNetworks}),
$\mathcal{C}_{out}$ plays the same role as in the {\sc MaxLike} partition, with an overlap of $42$ vertices out of sets of size 
$43$ and $45$, the remaining sets appear to have a different structure.
The distribution of in- and out-degrees, shown in 
\cref{fig:facultyBoxplot}B, indicates 
that the out-degrees of the sets tend to be   much smaller
than the sets uncovered by   {\sc MaxLike}. 
Further, unlike in the {\sc MaxLike} case, there is a stronger ordering
with respect to 
the academic performance of the sets (\cref{fig:facultyBoxplot}D).
This suggests that $\mathcal{C}_{in}$ could represent a second tier of
institutions which take faculty members from $\mathcal{C}_{out}$ but also take
faculty members from within (the density of self-loops in $\mathcal{C}_{in}$ is
$0.11$, in contrast to $0.011$ in $\mathcal{P}_{out}$ and $0.039$ in
$\mathcal{P}_{in}$), $\mathcal{P}_{in}$ may represent  a third tier of institutions which contains institutions that  
attract a large number of academics from $\mathcal{C}_{out}$, while
$\mathcal{P}_{out}$ may represent a fourth tier set which contains institutions that do not tend to take
academics from the highest-ranked universities.

\subsection{World Trade data}
\label{worldTrade}

The World Trade  data set  from~\cite{wtf} has countries as vertices and directed edges between countries representing trade. 
For simplicity, we focus on data from 2000 and restrict our attention to the trade in fresh, chilled and frozen bovine meats. We remove trades that do not correspond to a specific country, resulting in 720 trades for bovine meats  involving 125 countries, 
which leads to a network density of roughly $ 0.046$. 

\begin{figure}[htp]
\includegraphics[width=0.5cm]{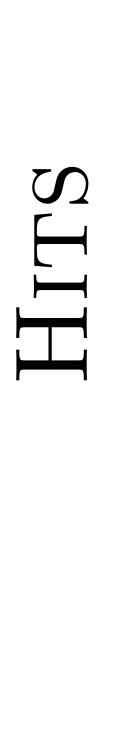}
\hspace{-0.25cm}
\includegraphics[width=0.25\linewidth]{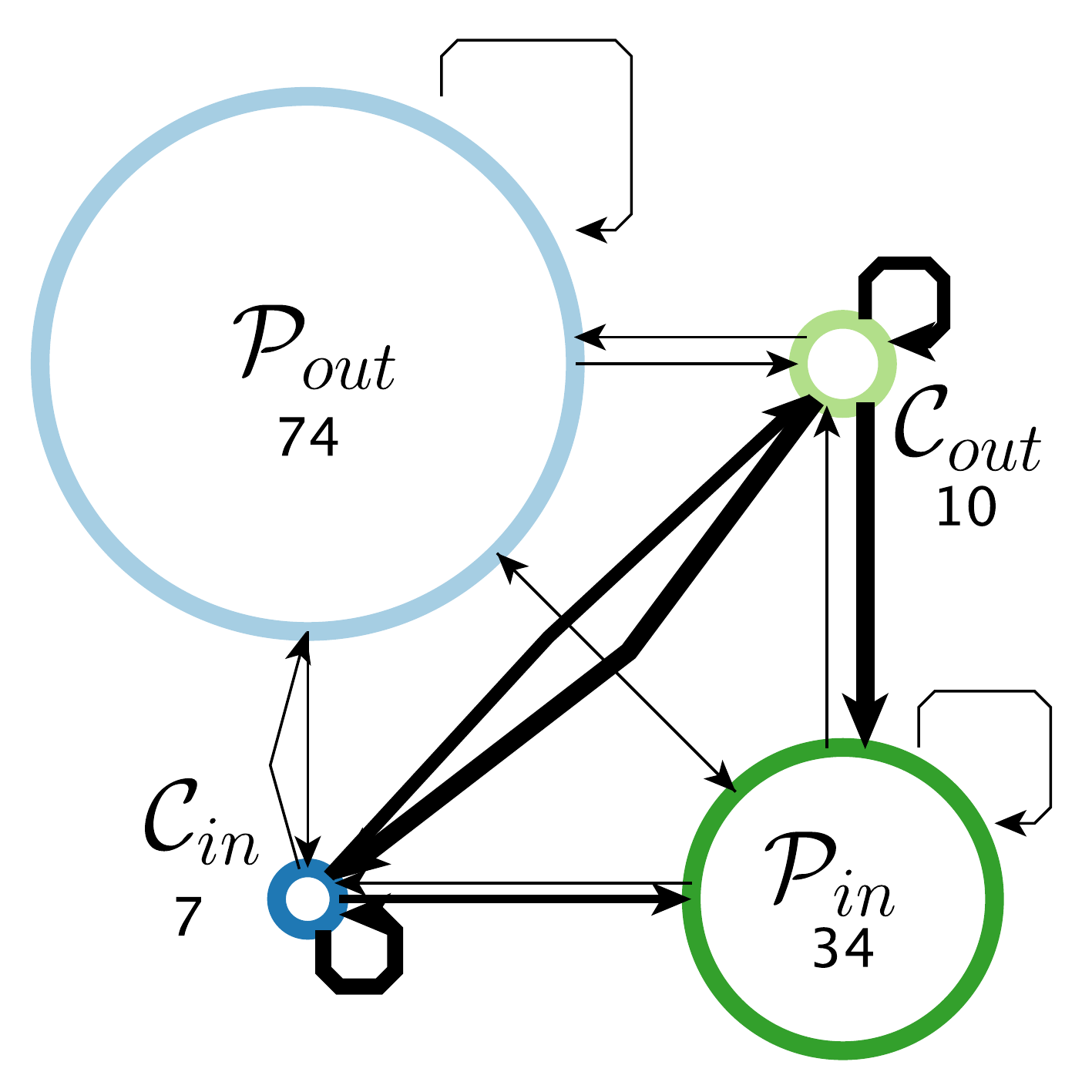}
\hspace{-0.25cm}
\includegraphics[width=0.23\linewidth]{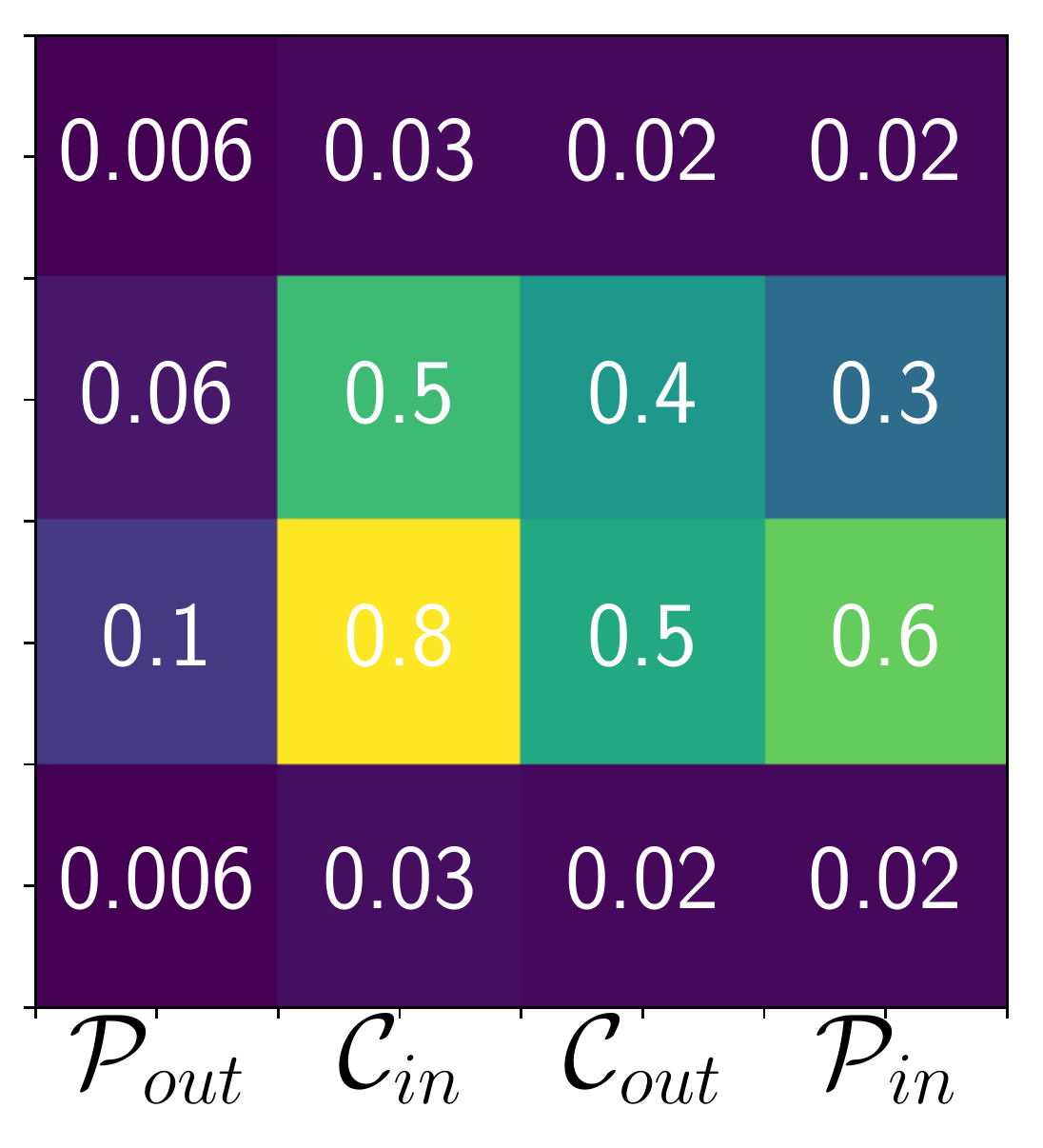}
\hspace{-0.2cm}
\includegraphics[width=0.5\linewidth]{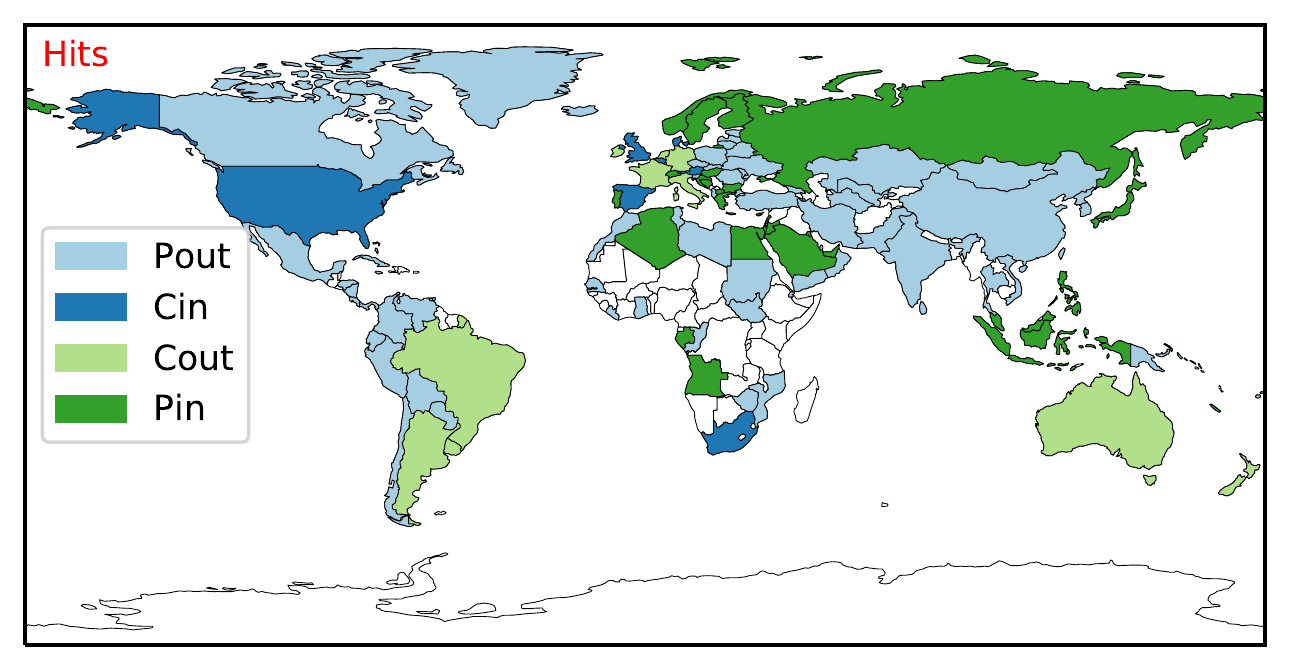}

\vspace{-0.20cm}
\includegraphics[width=0.5cm]{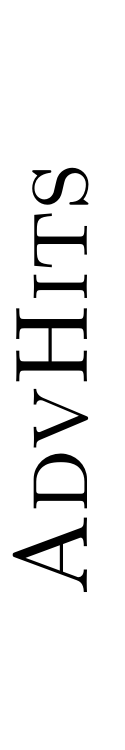}
\hspace{-0.25cm}
\includegraphics[width=0.25\linewidth]{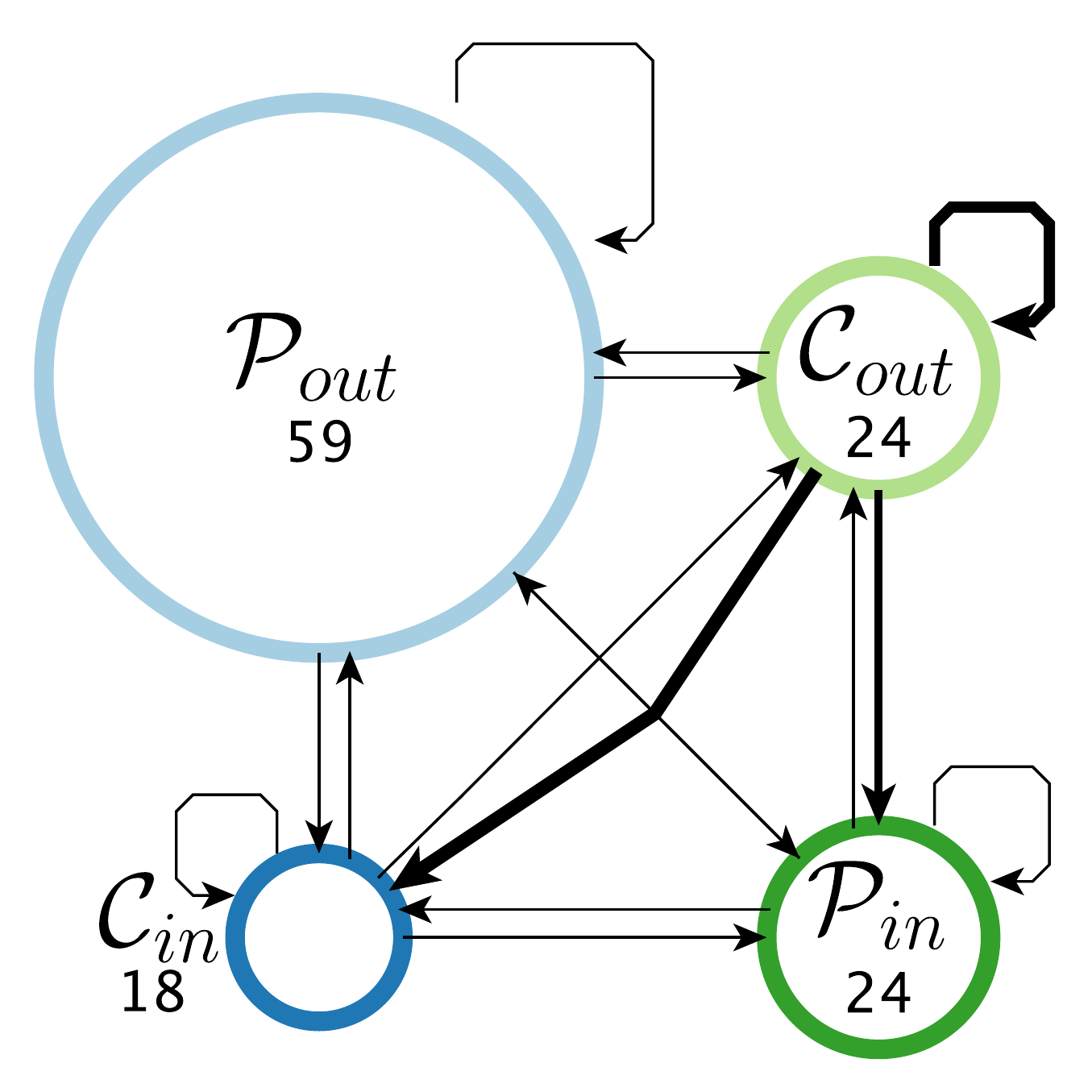}
\hspace{-0.25cm}
\includegraphics[width=0.23\linewidth]{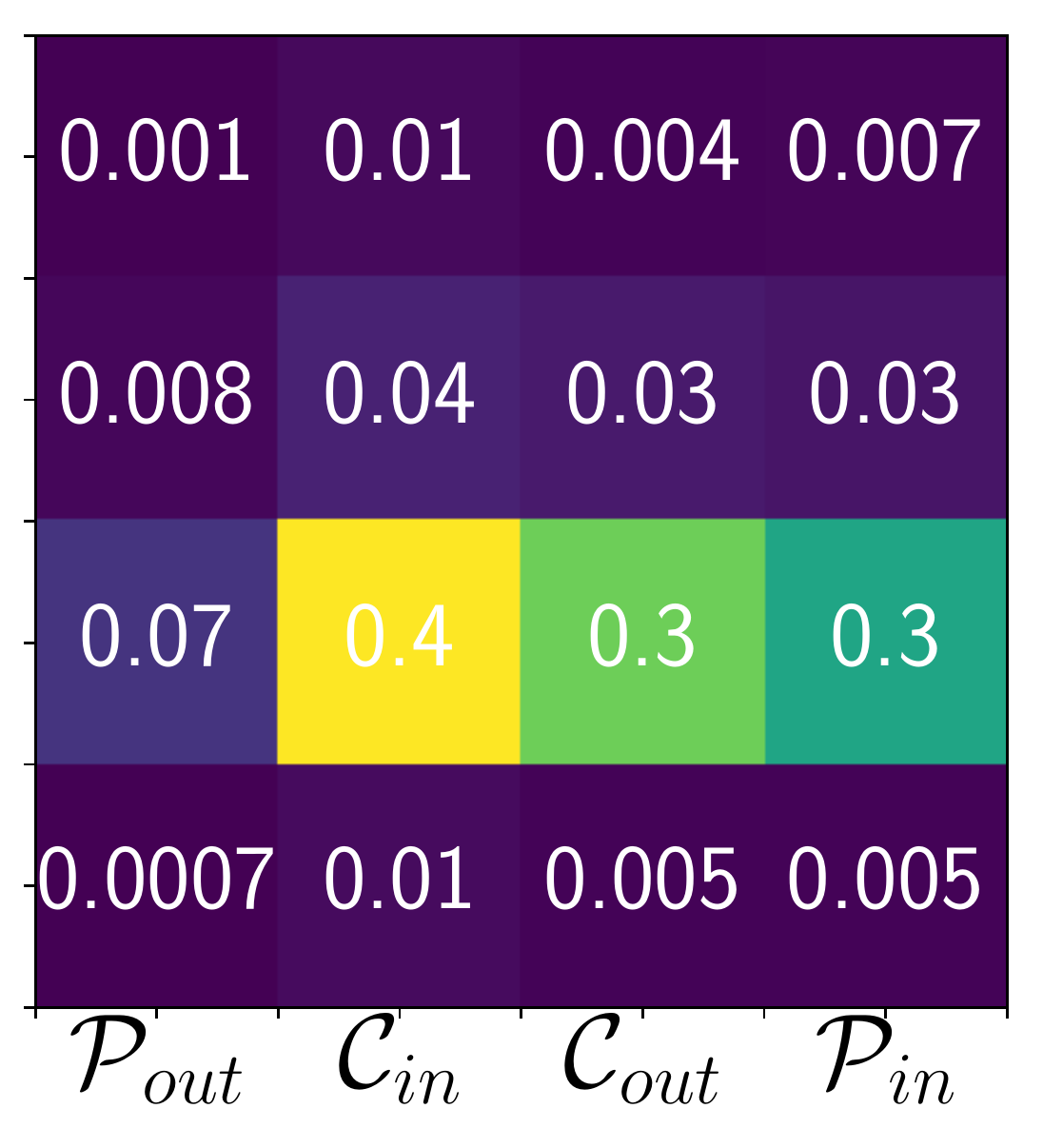}
\hspace{-0.2cm}
\includegraphics[width=0.5\linewidth]{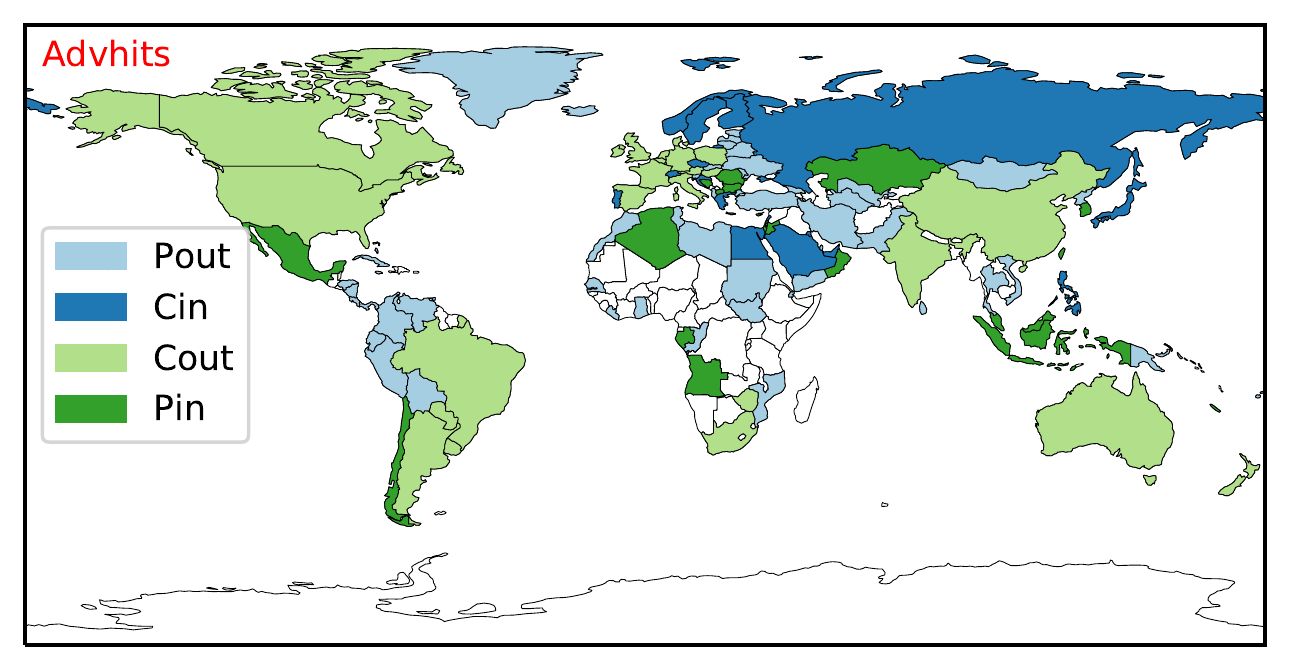}

\vspace{-0.20cm}
\includegraphics[width=0.5cm]{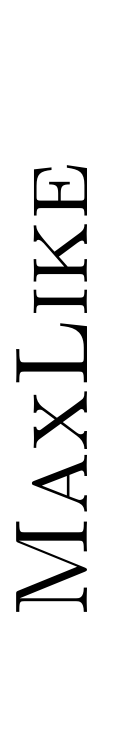}
\hspace{-0.25cm}
\includegraphics[width=0.25\linewidth]{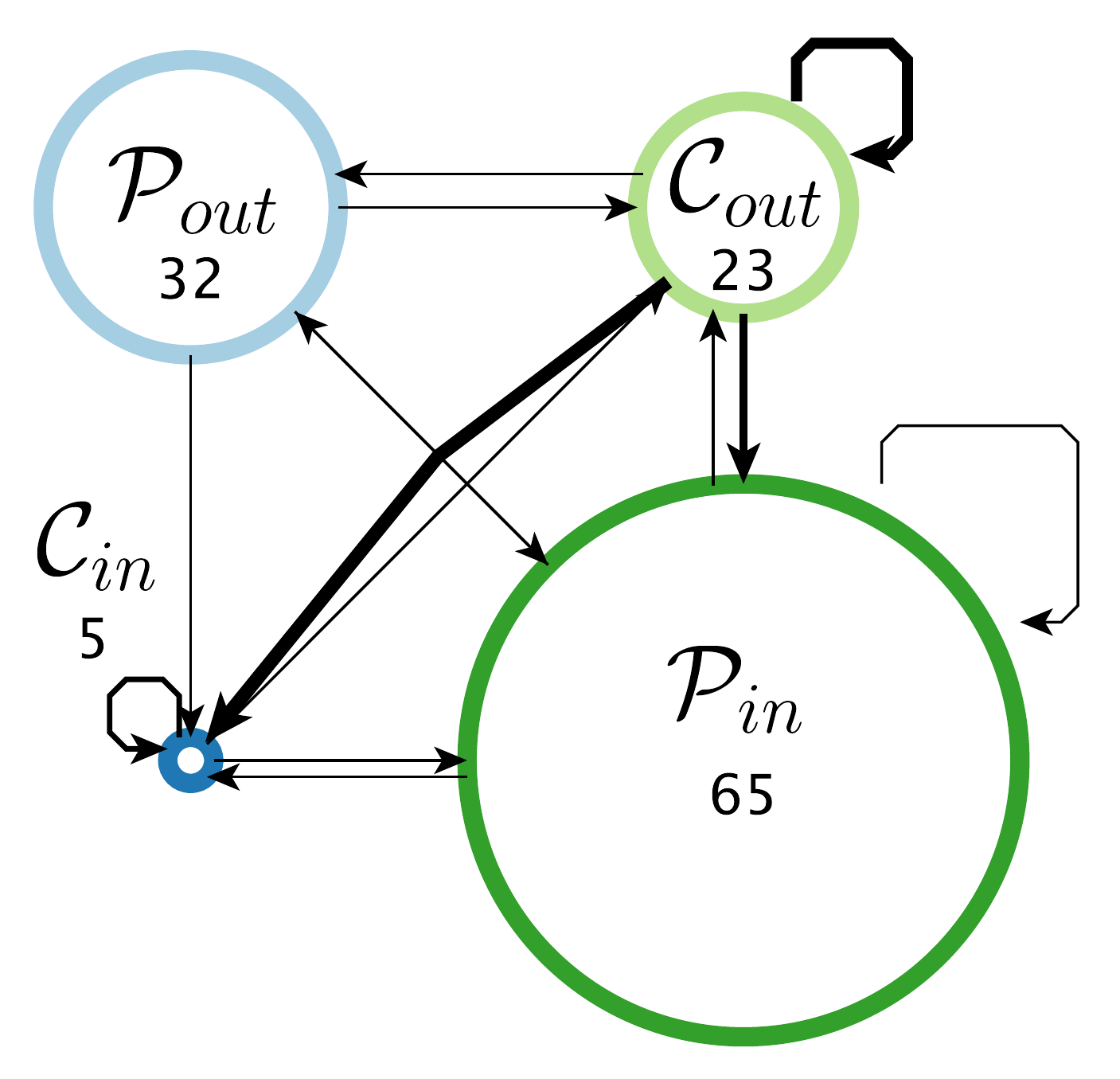}
\hspace{-0.25cm}
\includegraphics[width=0.23\linewidth]{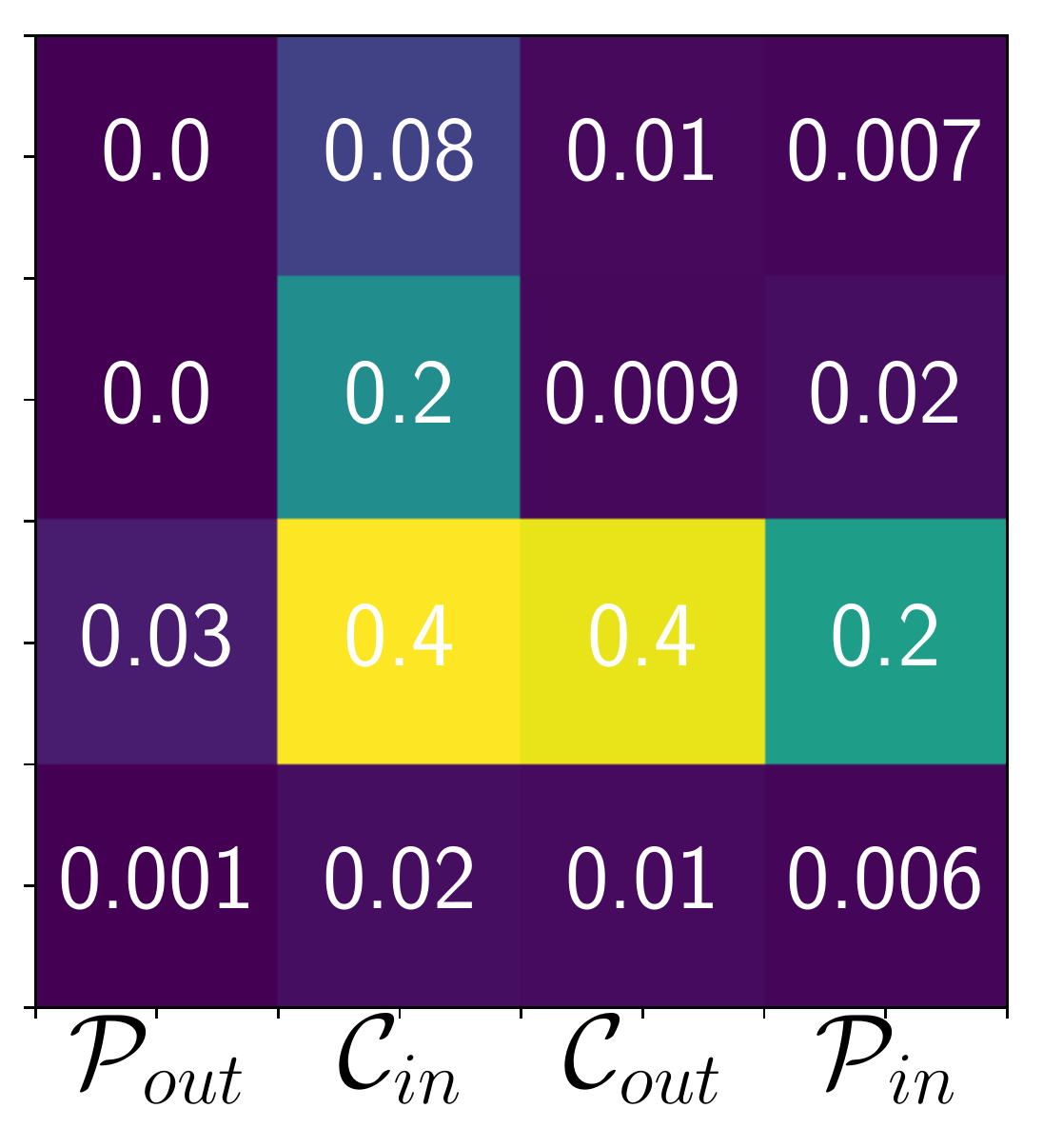}
\hspace{-0.2cm}
\includegraphics[width=0.5\linewidth]{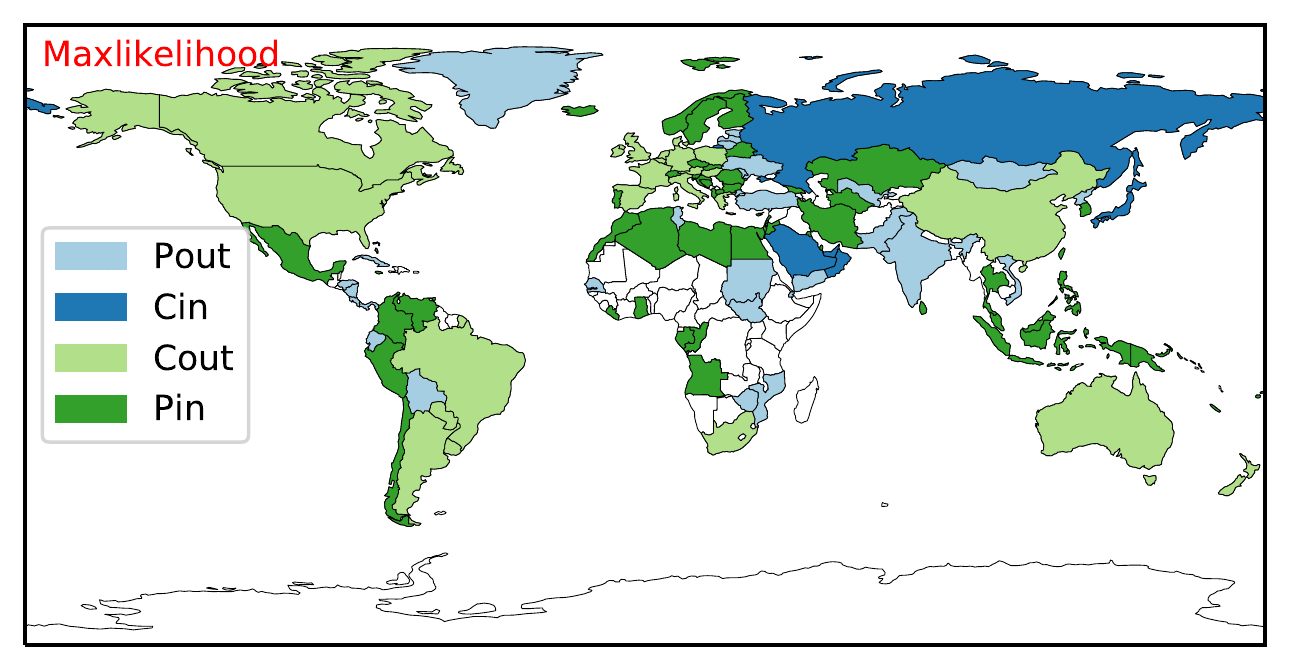}
\caption{Structures in \datasetFmt{WorldTrade}.
In the {\bf top} ({\sc HITS}), 
{\bf middle} ({\sc AdvHits})
and {\bf bottom} ({\sc MaxLike}) panels we show summary network diagrams associated with the uncovered structures. In these summaries the size of each of the vertices is proportional to the number of vertices in each set, the width of 
the lines is given by the percentage of edges that are present between the
sets. The width of lines is designed to highlight the differences and is therefore not comparable between plots. The centre panel displays the percentage of edges between each pair of blocks which allowing easily visualisation of the `L'-structure. The right panel visualises the partition on a World map.
}
\label{fig:tradeNetworks}
\end{figure}

The ARIs of the partitions again show high similarity within method class (data not shown). All methods except {\sc LowRank} show significant deviation from random 
when compared to the directed ER and directed configuration models.
First, we focus on the partition uncovered by {\sc HITS} 
shown in
\cref{fig:tradeNetworks}.
The structure does not have a clearly visible
`L'-shape structure, but in {\sc HITS} the internal `L' density is sufficiently high to pass the significance tests. Inspecting the sets,  $\mathcal{C}_{out}$ has high out-degree and often low in-degree vertices, which corresponds to large beef exporting nations, including Argentina, Australia, and Brazil. The set $\mathcal{C}_{in}$
corresponds to nations that import and export, with strong imports from the
$\mathcal{C}_{out}$ set, including the UK and South Africa, whereas
$\mathcal{P}_{in}$ countries are primarily importers with many vertices with
out-degree $0$, including Russia which strongly imports from
$\mathcal{C}_{out}$.  Finally, $\mathcal{P}_{out}$ corresponds to
mostly low degree vertices.  
{\sc AdvHits} yields a different partition, producing a weak `L'-shape structure. The $\mathcal{C}_{in}$ and $\mathcal{C}_{out}$ sets from the {\sc HITS} division being combined with additional vertices to form the {\sc AdvHits}'  $\mathcal{C}_{out}$ 
set, resulting in a set with large out-degrees and a range of in-degrees. The sets 
$\mathcal{P}_{in}$ and $\mathcal{C}_{in}$ then are enriched for  countries with strong imports from the $\mathcal{C}_{out}$ countries, with $\mathcal{C}_{in}$ having a larger average in-degree  (and the set with the largest out-degree after $\mathcal{C}_{out}$) , and  $\mathcal{P}_{out}$ countries have on average a low in- and out-degree. This structure is close to a classic core--periphery structure, with one core having a high out-degree linking to a large number of peripheral countries. %

Finally, {\sc MaxLike} obtains a full `L' structure; 
\cref{fig:tradeNetworks} shows that the geographic distribution is similar to the
structure from {\sc AdvHits}. There is a strong overlap
between their respective $\mathcal{C}_{out}$ sets ($22$ shared vertices), indicating that it finds a similar core to the {\sc AdvHits}. Similarly, $\mathcal{C}_{in}$ is a subset of {\sc AdvHits}'s $\mathcal{C}_{in}$ (overlap $4$), but in {\sc MaxLike} this focuses on a small set of heavily importing countries with a large in-degree and a small out-degree, which includes Japan, Oman, Russia, Saudi Arabia and United Arab Emirates.
The connection pattern clearly indicates this is a set which behaves differently from the others, and we leave further investigation as future work.
The periphery division is also different;  $\mathcal{P}_{in}$ consists of countries with a large number of imports from $\mathcal{C}_{out}$. In $\mathcal{P}_{out}$, there are certain vertices with an expected structure, e.g. Ukraine, Mongolia, Lithuania and Moldova supplying Russia ($\mathcal{C}_{in}$) and Pakistan, India and Sudan which supply the Arabian Peninsula. However, there are a number of vertices that do not match the pattern and may just be low degree vertices that are difficult to place.

Overall, in our final real-world data set, we demonstrated the power of our method, uncovering three related but different core-periphery structures: a two-core structure from {\sc HITS}, a single-core structure from {\sc AdvHits}, and a two-core, two-periphery structure from  {\sc MaxLike}.

\section{Conclusion and further work}
\label{conc}

We  provide the first comprehensive treatment of 
a directed discrete core--periphery structure which is not a simple extension of bow-tie structure. 
The structure we introduced consists of two core sets and two periphery sets defined in an edge-direction-dependent way, each with a unique connection profile.
To detect this structure algorithmically, we proposed three method
classes
with different
speed-quality trade-offs.
The fastest algorithms exploit the low-rank structure of the network
and cluster the corrected degrees using k-means++, the slower set of algorithms
use an iterative scheme based on the connection structure and are inspired by the HITS algorithm, and the slowest but most accurate approaches directly maximise a likelihood function. 

Using our synthetic model for this novel directed core--periphery structure, we created 
three benchmarks, to 
 compare our specialised approaches to
four directed network clustering methods that have similar
speed-quality trade-offs. 
We found that all of our proposed approaches match or outperform clustering solely using the in- and out-degrees. Furthermore, our fast methods outperform the fast spectral comparisons from the literature, and similarly,  our slow methods outperform existing slow methods.  As our methods are tailored to detect the planted directed core--periphery structure, this finding is reassuring but not surprising. %

Then we explored the existence of our directed core--periphery structure in three real-world data sets, namely a political blog network, a faculty hiring network,  and a world trade
network.%
 In each data set,
we found at least one significant structure, when compared to random ER and configuration model graphs.
In the political blogs data set, our {\sc AdvHits} method, revealed a division of the classically
discovered core into two independent components, a $\mathcal{C}_{in}$ core which we hypothesise
consists of authorities which are 
highly referenced,  and a $\mathcal{C}_{out}$
core which link to a large number of other blogs.
We support this
hypothesis by 
noting that $\mathcal{C}_{in}$ has a much lower percentage of 
`blogspot' sites 
than the other set, and  that $\mathcal{C}_{in}$ 
contains all of the top blogs identified 
by~\cite{politBlogs}.
In the faculty hiring data set, 
the {\sc MaxLike} partition uncovers  a new structure, namely
Canadian universities which have a large number of links with the top US
schools, but also appear to strongly recruit from their
own schools, indicating a complementary structure to the one found in \cite{clauset2015systematic}. 
In the trade data, we found different but related core--periphery structures
using each of our methods. With {\sc HITS} we obtained a two-core structure which
highlights strong exporters as well as those who both import and export; with {\sc AdvHits}
we obtain a structure which is closer to a classical core--periphery structure,
and finally with {\sc MaxLike} we obtain two cores related to strong
exporters and a surprising set of strong importers, a well defined $\mathcal{P}_{in}$ set, and a less well defined but coherent $\mathcal{P}_{out}$ set. 

Here are some avenues for future work. The trade data  highlighted that some vertices simply may not
fit the core--periphery pattern, and thus following the formulation of bow-tie, it would be interesting  to explore modifications to our approaches that would allow us not to place vertices if they do not match the pattern (for example, by introducing a separate set for outlier vertices). Further future work would explore graph regularisation techniques which may increase performance for sparse networks. 
Moreover, as detailed in 
SI~A,
other directed core--periphery patterns are possible. 
Some of our methods could be adapted to also detect such core--periphery patterns. In principle, all possible core--periphery structures could be tested simultaneously, with an appropriate correction for multiple testing.
Such a development should of course be motivated by a suitable data set which allows for interpretation of the results.
More generally, meso-scale structures may change over time, and it would be fruitful to extend our structure and methods to include time series of networks.

\paragraph{Acknowledgements}
This work was funded by EPSRC grant EP/N510129/1 at The Alan Turing Institute and Accenture Plc.
In addition, we acknowledge support from COST Action CA15109.   
We thank Aaron Clauset for useful discussions and the authors of~\cite{deJeude2019} for providing the code for the bow-tie structure.

\bibliographystyle{plain}
\bibliography{AAA_mybibb.bib}

\clearpage
\newpage

\renewcommand{\thesubsection}{(\alph{subsection})}

\appendix
\noindent
{\huge \bf Supplementary Information}
\renewcommand{\theequation}{\Alph{section}.\arabic{equation}}

\renewcommand{\thefigure}{SI \arabic{figure}}
\setcounter{figure}{0}

\section{Alternative Definitions of Directed Core-Periphery}
\label{app:structureVariants}
In the main text, we proposed a novel directed core--periphery structure, introduced methods to detect it, compared the performance of these methods on synthetic benchmarks, and explored this structure in real-world networks. While we find the extension of core--periphery  to the directed case a natural one to consider 
and insightful in real-world data,
there are other possible choices. To better understand  the space of possible directed configurations that still maintain a core--periphery-like structure, we propose a set of criteria, and then enumerate all possible structures which satisfies these criteria.
We start with a partition of the vertex set into four potentially empty sets, which we call Core 1, Core 2, Periphery 1, and Periphery 2 (we suppress the corresponding number when there is only one set of its kind).
For a directed core--periphery structure, we impose the following restrictions 
\begin{enumerate}
  \setlength\itemsep{0.1mm}
\item %
There cannot be connections within or between periphery sets.
\item %
A periphery set must connect to exactly one core set. 
\item %
Any pair of sets designated as core must connect to each other, must be connected internally, and each core set must be connected to at least one periphery set.
\item %
The structure can have no reciprocal edges between any two distinct blocks. 
\end{enumerate}
We detail below the structures that satisfy the above assumptions.
\paragraph{(i) 1 Core set and 1 periphery set}
We first consider the case of one core and one periphery set. As the core must connect internally, and the periphery set must connect to the core, there are two possible design matrices

\begin{equation}
\label{eq:designSize1}
\begin{minipage}{0.35\textwidth}
$
\begin{array}{rccr}
 & \ \ \ \ \text{Core} & \text{Periphery}  \ \ \ \\
 \text{Core} \! \! \! \!\! \! \! \! \! \! & \vline \ \ \ 1  &  1   \ \ \ \vline \\
 \text{Periphery} \! \! \! \!\! \! \! \! \! \! & \vline \ \ \ 0  &  0   \ \ \ \vline \\
\end{array}\quad
$
\end{minipage}
\begin{minipage}{0.25\textwidth}
\resizebox{0.95\textwidth}{!}
{\includegraphics{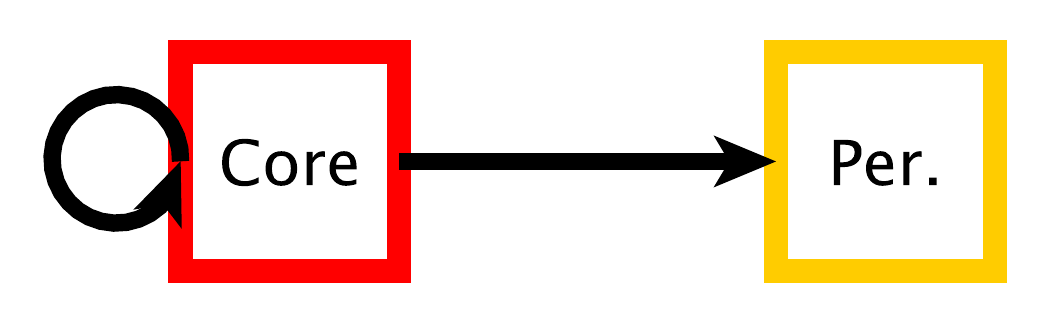}
}
\end{minipage}
\end{equation}
\begin{equation}
\label{eq:designSize2}
\begin{minipage}{0.35\textwidth}
$
\begin{array}{rccr}
 & \ \ \ \ \text{Core} & \text{Periphery}  \ \ \ \\
 \text{Core} \! \! \! \!\! \! \! \! \! \! & \vline \ \ \ 1  &  0   \ \ \ \vline \\
 \text{Periphery} \! \! \! \!\! \! \! \! \! \! & \vline \ \ \ 1  &  0   \ \ \ \vline \\
\end{array}\quad
$
\end{minipage}
\begin{minipage}{0.25\textwidth}
\resizebox{0.95\textwidth}{!}{\includegraphics{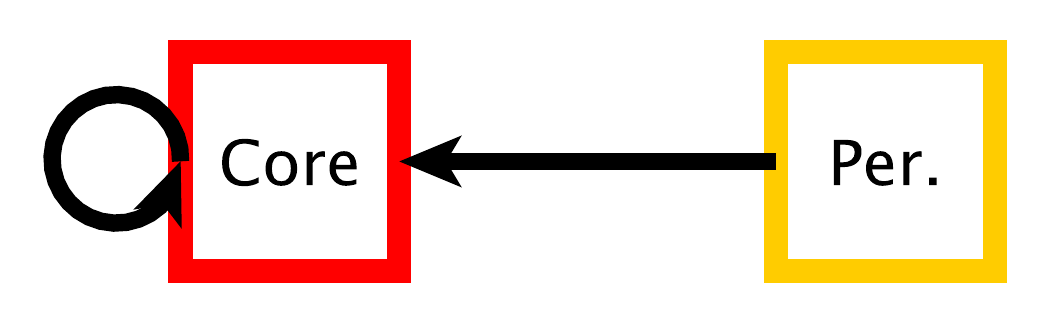}}
\end{minipage}
\end{equation}

Directed core--periphery structure 
\eqref{eq:designSize1}
and 
\eqref{eq:designSize2} represent an
asymmetric flow between the core and the periphery, as a periphery 
either only sends or only receives links. 
This can be a flow of information, such as follower-only accounts in Twitter networks, or a flow of money such as peripheral banks in inter-bank networks who sell their loans to central banks. 

\paragraph{(ii) 2 Core sets and 1 periphery set}
As we restrict the periphery set to connect to only one core, and each core set to connect to at least one periphery, there is no possible design matrices that satisfy these constraints.

\paragraph{(iii)  1 Core set and 2 periphery sets}
In order for this case  %
to make the structure identifiable, we exclude structures that can be obtained by splitting the peripheries of our 1 core and 1 periphery structures. 
Under this restriction,  only one configuration is possible,  
namely having the peripheries connect in different directions.
Below is the resultant design matrix for this structure. 

\begin{equation}
\label{eq:designSize3}
\begin{minipage}{0.35\textwidth}
$
\begin{array}{rccr}
 & \ \ \ \ \text{Core} & \text{Per. 1}  & \text{Per. 2} \ \ \ \\
 \text{Core} \! \! \! \!\! \! \! \! \! \! & \vline \ \ \ 1  &  0  &  1  \ \ \ \vline \\
 \text{Per. 1} \! \! \! \!\! \! \! \! \! \! & \vline \ \ \ 1  &  0  &  0  \ \ \ \vline \\
 \text{Per. 2} \! \! \! \!\! \! \! \! \! \! & \vline \ \ \ 0  &  0  &  0  \ \ \ \vline \\
\end{array}
$
\end{minipage}\quad
\begin{minipage}{0.30\textwidth}
\resizebox{0.95\textwidth}{!}
{\includegraphics{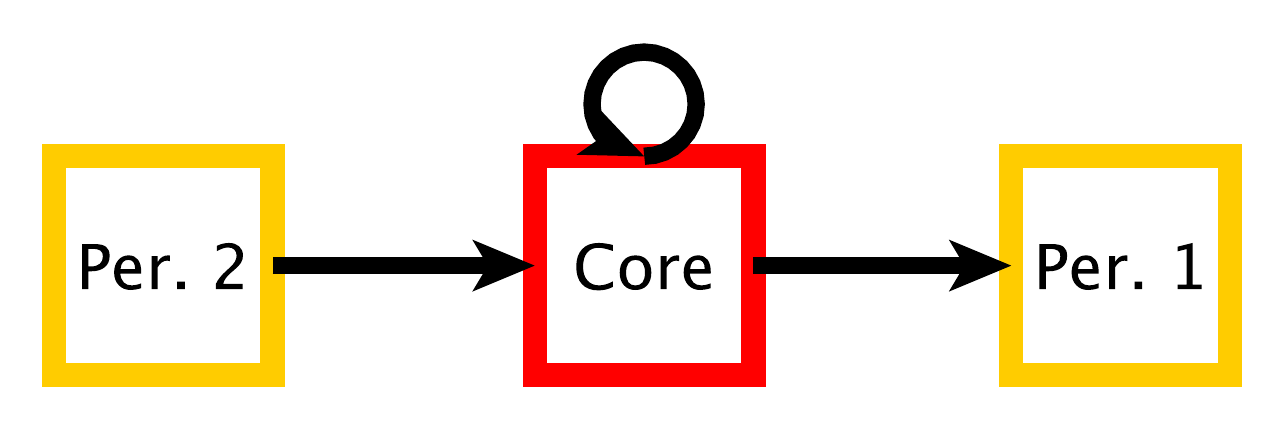}
}
\end{minipage}
\end{equation}

Directed core--periphery structure
\eqref{eq:designSize3}
is  a particular instance of the well-known bow-tie structure, 
essentially having a set of vertices that only send material (content, money,
etc.), and a set of vertices that only receive material;   
there are several known real world examples, namely of the internet in \cite{Broder2000} and in biological networks \cite{ma2003connectivity}. 

\paragraph{(iv) 2 Core sets and 2 periphery sets}
In the case of 2 core and 2 periphery sets, the cores must interconnect, and the link cannot be reciprocated, so that  there is only one possible structure  between the two core sets. By assumption, the periphery vertices only connect to one core set, and each core set must be connected to a periphery. Therefore, there are four possible such structures

\begin{equation}
\label{eq:designSize4a}
\begin{minipage}{0.45\textwidth}
$
\begin{array}{rcccr}
 & \ \ \ \ \text{Core 1} & \text{Core 2} & \text{Per. 1}  & \text{Per. 2} \ \ \ \\
 \text{Core 1} \! \! \! \!\! \! \! \! \! \! & \vline \ \ \ 1  &  0  &  0  &  1  \ \ \ \vline \\
 \text{Core 2} \! \! \! \!\! \! \! \! \! \! & \vline \ \ \ 1  &  0  &  0  &  0  \ \ \ \vline \\
 \text{Per. 1} \! \! \! \!\! \! \! \! \! \! & \vline \ \ \ 0  &  1  &  0  &  0  \ \ \ \vline \\
 \text{Per. 2} \! \! \! \!\! \! \! \! \! \! & \vline \ \ \ 0  &  0  &  0  &  0  \ \ \ \vline \\
\end{array}
$
\end{minipage}\quad
\begin{minipage}{0.25\textwidth}
\resizebox{0.95\textwidth}{!}
{\includegraphics{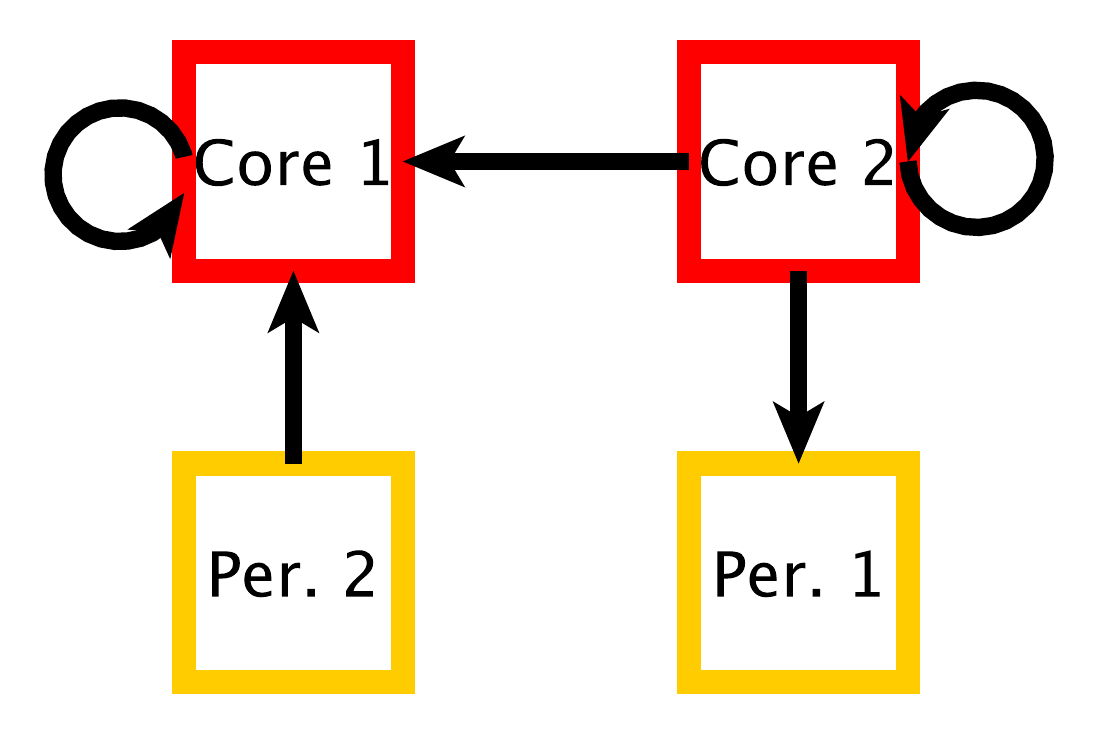}
}
\end{minipage}
\end{equation}

\begin{equation}
\label{eq:designSize4b}
\begin{minipage}{0.45\textwidth}
$
\begin{array}{rcccr}
 & \ \ \ \ \text{Core 1} & \text{Core 2} & \text{Per. 1}  & \text{Per. 2} \ \ \ \\
 \text{Core 1} \! \! \! \!\! \! \! \! \! \! & \vline \ \ \ 1  &  0  &  0  &  0  \ \ \ \vline \\
 \text{Core 2} \! \! \! \!\! \! \! \! \! \! & \vline \ \ \ 1  &  1  &  1  &  0  \ \ \ \vline \\
 \text{Per. 1} \! \! \! \!\! \! \! \! \! \! & \vline \ \ \ 0  &  0  &  0  &  0  \ \ \ \vline \\
 \text{Per. 2} \! \! \! \!\! \! \! \! \! \! & \vline \ \ \ 1  &  0  &  0  &  0  \ \ \ \vline \\
\end{array}
$
\end{minipage}\quad
\begin{minipage}{0.25\textwidth}
\resizebox{0.95\textwidth}{!}
{\includegraphics{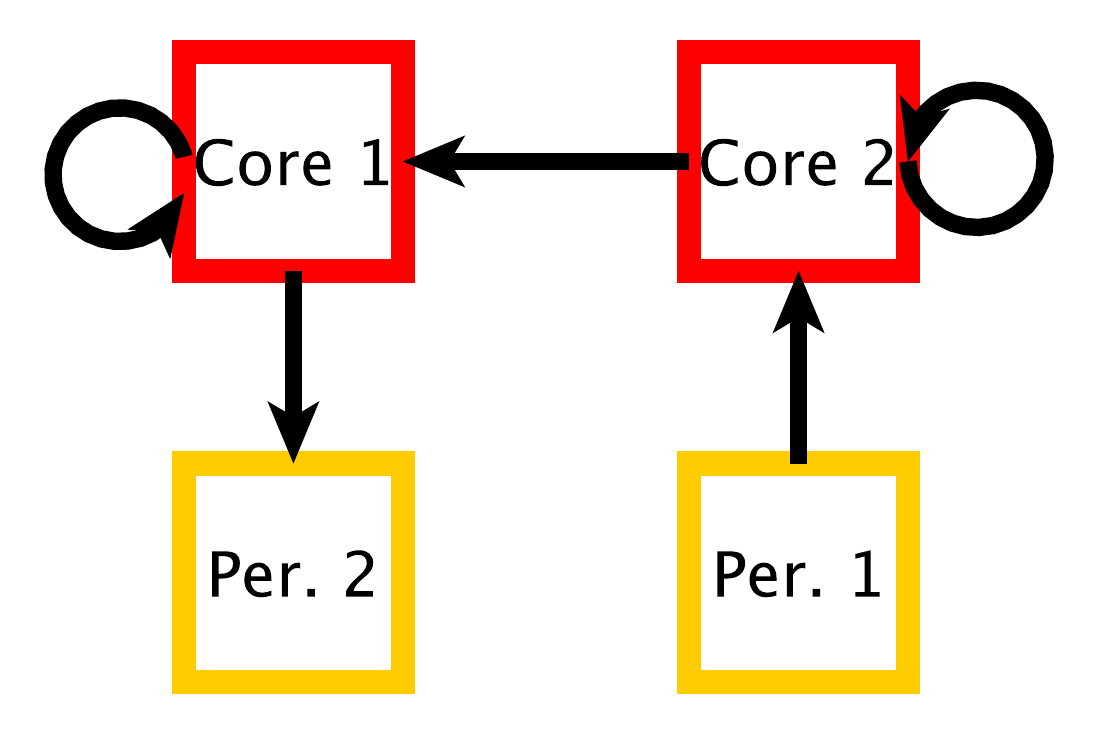}
}
\end{minipage}
\end{equation}

\begin{equation}
\label{eq:designSize4c}
\begin{minipage}{0.45\textwidth}
$
\begin{array}{rcccr}
 & \ \ \ \ \text{Core 1} & \text{Core 2} & \text{Per. 1}  & \text{Per. 2} \ \ \ \\
 \text{Core 1} \! \! \! \!\! \! \! \! \! \! & \vline \ \ \ 1  &  0  &  0  &  0  \ \ \ \vline \\
 \text{Core 2} \! \! \! \!\! \! \! \! \! \! & \vline \ \ \ 1  &  1  &  0  &  0  \ \ \ \vline \\
 \text{Per. 1} \! \! \! \!\! \! \! \! \! \! & \vline \ \ \ 0  &  1  &  0  &  0  \ \ \ \vline \\
 \text{Per. 2} \! \! \! \!\! \! \! \! \! \! & \vline \ \ \ 1  &  0  &  0  &  0  \ \ \ \vline \\
\end{array}
$
\end{minipage}\quad
\begin{minipage}{0.25\textwidth}
\resizebox{0.95\textwidth}{!}
{\includegraphics{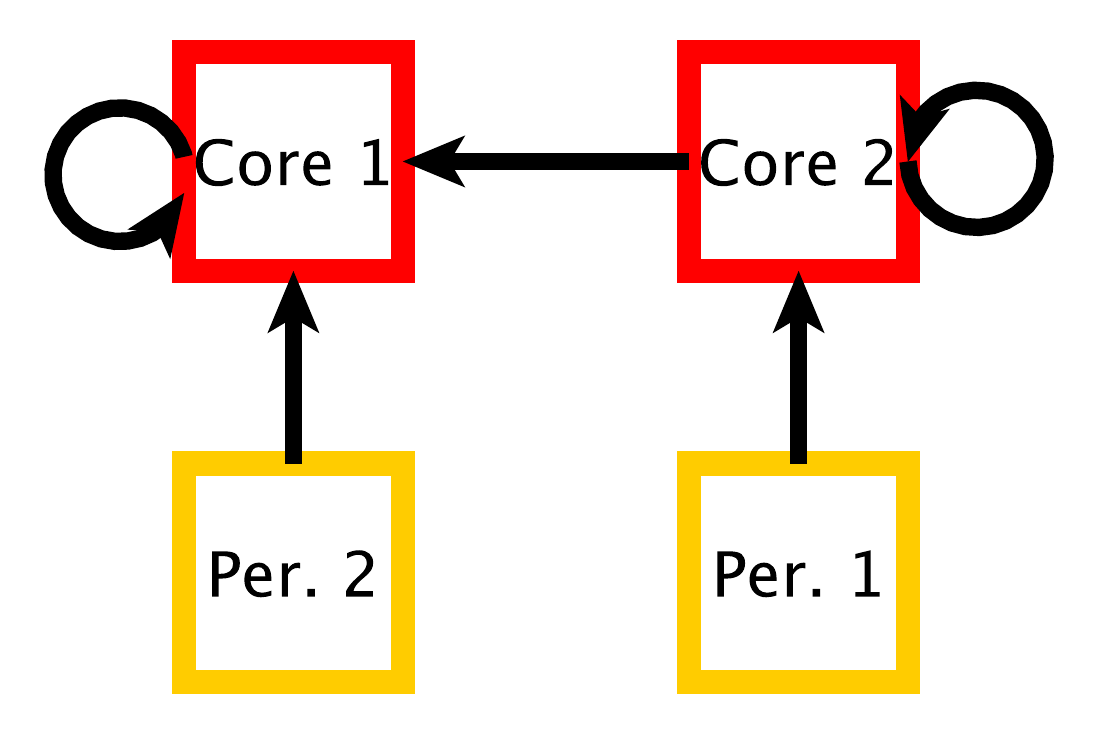}
}
\end{minipage}
\end{equation}

\begin{equation}
\label{eq:designSize4d}
\begin{minipage}{0.45\textwidth}
$
\begin{array}{rcccr}
 & \ \ \ \ \text{Core 1} & \text{Core 2} & \text{Per. 1}  & \text{Per. 2} \ \ \ \\
 \text{Core 1} \! \! \! \!\! \! \! \! \! \! & \vline \ \ \ 1  &  0  &  0  &  1  \ \ \ \vline \\
 \text{Core 2} \! \! \! \!\! \! \! \! \! \! & \vline \ \ \ 1  &  1  &  1  &  0  \ \ \ \vline \\
 \text{Per. 1} \! \! \! \!\! \! \! \! \! \! & \vline \ \ \ 0  &  0  &  0  &  0  \ \ \ \vline \\
 \text{Per. 2} \! \! \! \!\! \! \! \! \! \! & \vline \ \ \ 0  &  0  &  0  &  0  \ \ \ \vline \\
\end{array}
$
\end{minipage}\quad
\begin{minipage}{0.25\textwidth}
\resizebox{0.95\textwidth}{!}
{\includegraphics{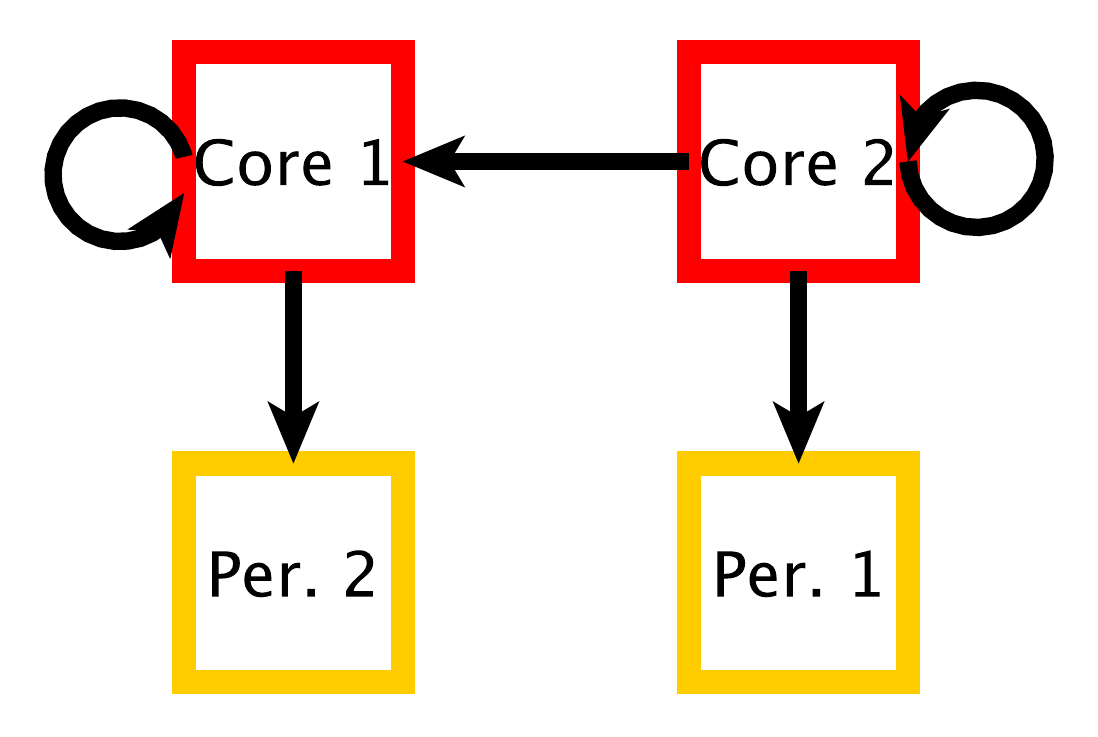}
}
\end{minipage}
\end{equation}

To the best of our knowledge, these structures have not been directly studied before. The directed core--periphery
structure \eqref{eq:designSize4b} is essentially a bow-tie in which the core can be split
into two sets, one which takes the incoming connections of the periphery
vertices, and the remaining core set which sends links to the additional
periphery set. This could be seen as a two-step process version of the
bow-tie structure, in which material is taken in, processed, and then
distributed to a second set that performs a similar action. 
{{In contrast, the structure~\eqref{eq:designSize4a}, which is the one on which the main paper focuses,}} is not a simple 
extension of the bow-tie structure due to the fact that, contrary to bow-tie, the flow in~\eqref{eq:designSize4a} is not uni-directional.

\section{Stochastic Block Model Fitting}
\label{app:SBMmethod}
In this section, we detail the algorithms for fitting a stochastic block model to the proposed structure.

\subsection{{\sc HillClimb} Method}

The first approach is {\sc HillClimb}  from \cite{snijdersSBM}. Our implementation is as follows. 
\begin{enumerate}
  \setlength\itemsep{0.1mm}
\item Repeat the following procedure $10$ times
\begin{enumerate}
  \setlength\itemsep{0.1mm}
\item Assign all vertices to a random set from 
$\mathcal{P}_{in}$, 
$\mathcal{P}_{out}$, 
$\mathcal{C}_{in}$ and  
$\mathcal{C}_{out}$.
\item For a fixed number of repeats ($5000$ for results in the paper):  
\begin{enumerate}
  \setlength\itemsep{0.1mm}
\item Compute the optimal value of $p_1$ and $p_2$ by looking at the density in
both regions of the block structure, enforcing $p_1>p_2$.
\item Consider each vertex in a random order and place it in the set with the highest probability (using the fixed values of $p_1$ and $p_2$). 
\item If there are no changes to the vertex sets then exit the loop. 
\end{enumerate}
\end{enumerate}
\item  Select the partition with the largest likelihood. 
\end{enumerate}
{To enforce the constraint that $p_1>p_2$, we define $p_1 = \text{max}(p_1,p_2)$, and $p_2 = \text{min}(p_1,p_2)$.} 
The algorithm differs slightly from \cite{snijdersSBM},  firstly as we have a fixed number of repeats, and secondly as we fit a likelihood with  fixed set sizes  and only two parameters to estimate, 
$p_1$ and $p_2$.

\subsection{{\sc MaxLike} Method}

We implemented {\sc MaxLike} from \cite{karrer2011} as follows.
\begin{enumerate}
  \setlength\itemsep{0.1mm}
\item Assign all vertices to a random set ({\sc CurrentPartition}). 
\item Calculate the likelihood of the random partition, and store it and the partition itself in {\sc BestPartition}.
\item Make a list {\sc List1} of all vertices. 
\item Find the vertex in {\sc List1} for which the movement to a new set will increase the likelihood the most (or decrease the least). Perform the move and remove the vertex from {\sc List1}. 
\label{algorithmStep4}
\item If the likelihood of the current partition is larger than {\sc BestPartition} then replace {\sc BestPartition} with the current partition.
\item If {\sc List1} is not empty go to {\bf iv}.
\item If {\sc BestPartition} has improved since {\bf 3} then set {\sc CurrentPartition} to {\sc BestPartition} and go to {\bf 3.} 
\item Return {\sc BestPartition}. 
\end{enumerate}

\subsection{Computational Complexity}
\label{app:complexity}
There is a considerable difference in computational cost between the two likelihood
approaches. In a na\"ive implementation, one pass of {\sc HillClimb} takes on the 
order of $O(m)$ to compute the parameters $p_1$ and $p_2$ (each edge must be considered), and
$O(k_i)$ to update each vertex (querying membership of each neighbour) which
overall is again $O(m)$. As there is a fixed number of iterations, this leads to $O(m)$ overall. 
{\sc MaxLike} has a larger
computational cost, with the na\"ive approach to assessing whether to 
change a vertex's
set having a complexity of $O(k_i)$ (as before). %
Further, as we consider moving
each unmoved vertex each time we run step 
\ref{algorithmStep4}, and step \ref{algorithmStep4}
is ran $n$ times,  each iteration considers moving a vertex 
$\binom{n+1}{2}$
times.
If we store the neighbourhood structure of each vertex and update it 
whenever there are changes, we can reduce the complexity of exploring changing a vertices
community to $O(1)$ with the addition of $O(k)$ when changing the structure,
leading to an overall complexity of $O(n^2+m)$.

\section{Selection and Performance of Different Variants of the {\sc LowRank} and the {\sc HITS} Methods} 
\label{sec:methodologicalChoices}

Several methodological choices were made during the development stage of {\sc HITS}, {\sc LowRank}, and {\sc AdvHits}  methods, based on their relative performance on our synthetic benchmarks, as follows.
The {\sc LowRank} method can be broken down into the following steps. 

\begin{enumerate}
  \setlength\itemsep{0.1mm}
\item Construct a rank-$2$ approximation of the observed graph adjacency matrix.  
\item Construct vertex scores from the low-rank approximation. 
\item Cluster the vertices by applying k-means++ to the scores. 
\end{enumerate}
For the low-rank approximation step,  following~\cite{journals/corr/CucuringuRLP14},
we considered adding a threshold to convert the low-rank approximation from a fully connected weighted directed (albeit low-rank) graph to a directed unweighted sparse graph, which may be
computationally advantageous for large graphs. 
For the purpose of choosing whether to apply a threshold or not in the analysis shown in the main text, we compared the non-thresholded case to a threshold of $0.5$. 

For the scoring step described in the main text, we use the scores 
${{\bf C}^{LR}_{in}}$, ${{\bf C}^{LR}_{out}}$, $ {{\bf P}^{LR}_{In}}$ and ${{\bf P}^{LR}_{Out}}$.
This formulation is degree-based and does not leverage other network information such as the score of the neighbours of a vertex. 
In this section, we consider two scoring variants for the periphery sets, the first leveraging 
network information and the second further exploiting the low-rank structure.
For the first alternative formulation, heuristically, 
a vertex $i$ should have a high $\mathcal{P}_{in}$ score if

\begin{enumerate}
  \setlength\itemsep{0.1mm}
	\item it has in-coming edges from vertices that have high $\mathcal{C}_{out}$ scores;
	\item it does not have  high $\mathcal{C}_{out}$, $\mathcal{C}_{in}$ and $\mathcal{P}_{out}$ scores.
\end{enumerate}

\noindent Additionally, a vertex $i$ should have a high $\mathcal{P}_{out}$ score if

\begin{enumerate}
  \setlength\itemsep{0.1mm}
	\item it has out-going edges to vertices that have high $\mathcal{C}_{in}$ scores;
	\item it does not have high $\mathcal{C}_{out}$, $\mathcal{C}_{in}$ and $\mathcal{P}_{in}$ scores.
\end{enumerate}

\noindent  Combing these,  we propose the following alternative scoring scheme
\begin{flalign}
	\label{eq:penalityVar1}
	{{{ P^{LR-Alt}_{In}}}}(i) 
	= &
	\sum\limits_{j=1,j\neq i}^{n} \left(-  \widehat{A}_{ij}
	{{{ C^{LR}_{In}}}}(j)  - \widehat{A}_{ij}
	{{{ C^{LR}_{Out}}}}(j) +  \widehat{A}_{ji}
	{{{ C^{LR}_{In}}}}(j)  - \widehat{A}_{ji}
	{{{ C^{LR}_{Out}}}}(j) \right),
	\\
	\label{eq:penalityVar2}
	{{{ P^{LR-Alt}_{Out}}}}(i) 
	= &
	\sum\limits_{j=1,j\neq i}^{n} \left(  \widehat{A}_{ij}
	{{{ C^{LR}_{In}}}}(j) - \widehat{A}_{ij}
	{{{ C^{LR}_{Out}}}}(j) -  \widehat{A}_{ji}
	{{{ C^{LR}_{In}}}}(j) - \widehat{A}_{ji}
	{{{ C^{LR}_{Out}}}}(j) \right).
\end{flalign}
For the second score variant, we further leverage the low-rank structure. 
Rather than assigning the scores based on degrees, we assign scores from the
left and right singular vectors in the rank-$2$ decomposition, either 
using the values directly or by first re-scaling by the square root of the
singular values.

To decide which approach to use in the main paper, we assessed the variants without thresholds and with a threshold of $0.5$ for the  original and alternative  scoring scheme (thresholding does not impact the direct singular value schemes).
We selected the approach suggested in the main paper as it has  
the best overall 
performance (although there are narrow regions with very differently-sized sets and a high network density where other 
approaches may perform better).
Similarly to the low-rank structure,  we also considered two variants of the {\sc HITS} method, based on the alternative scoring in 
\cref{eq:penalityVar1,eq:penalityVar2}.
The scoring scheme we presented in the main body outperforms the alternatives on all three benchmarks. 

\section{Synthetic Benchmarks}
\label{app:diffPartSimFunc}
In this section, we present additional content and results which relate to our synthetic benchmarks. We first detail the ARI similarity measure used in the main document,  as well as alternative similarity measures one could use (Section~\ref*{app:diffPartSimFunc}\ref{app:DefinitionSim}). 
Next, we present results for alternative network sizes  (Section~\ref*{app:diffPartSimFunc}\ref{app:comparisonTo400}), 
timing results for Benchmark 1 
(Section~\ref*{app:diffPartSimFunc}\ref{sec:TimingSyntheticSec}), and 
finally, results using alternative similarity measures 
(Section~\ref*{app:diffPartSimFunc}\ref{app:additionalResults}).

\subsection{Similarity Measures}
\label{app:DefinitionSim}

\paragraph{Adjusted Rand index} 
The Adjusted Rand Index (ARI) \cite{huberta85} measures the similarity between two partitions of a data set. 
Let $V = (v_1,v_2,\ldots,v_n)$ be a set of vertices we wish to cluster, let $X
= (X_1,X_2,X_3,X_4)$ be a partition of $V$ into four clusters, and let $Y = (Y_1,Y_2,Y_3,Y_4)$ be
the ideal partition. %
We let  
$a$ be the number of vertex pairs that are placed in the same set of $X$ and in the same set of $Y$;
$b$ be the number of vertex  pairs that are placed in the same set of $X$ but in different sets of $Y$;
$c$ be the number of vertex  pairs that are placed in different sets of $X$ but in the same set of $Y$;
$d$ be the number of vertex  pairs that are placed in different sets of $X$ and in different sets of $Y$. 
The Adjusted Rand Index~\cite{huberta85} is  
\[
\text{ARI} = \dfrac{
\binom{n}{2}
(a+d) - [(a+b)(a+c)+(c+d)(b+d)]}{
\binom{n}{2}^2 - [(a+b)(a+c) + (c+d)(b+d)]}.
\]
The Adjusted Rand Index (ARI) is bounded by $\pm$1 with an expected value of 0. An ARI close to 1 indicates an almost perfect match between the two partitions, whereas an ARI close to -1 indicates that the agreements between two partitions is less than what is expected from random labellings, which has an ARI of 0.

\paragraph{Normalised Mutual Information (NMI) and Variation of Information (VOI)} 
Both of these measures are based on the concept of entropy and mutual
information~\cite{elementsOfInformationTheory}. For a random variable $X$ which takes
values in a discrete set $\mathcal{S}_X$, and a random variable $Y$ which take values in a discrete set 
$S_Y$~\cite{elementsOfInformationTheory}, the entropy of $X$ is 
\begin{equation*}
H(X) = -\sum_{a \in S_X}  P(X=a)\log(P(X=a))  
\end{equation*}
and the  mutual information $MI(X,Y)$ of $X$ and $Y$ is 
\begin{equation*}
MI(X,Y) = \sum_{a \in S_X}  \sum_{b \in S_Y}  P(X=a,Y=b)\log\left(\frac{P(X=a,Y=b)}{P(X=a)P(Y=b)}\right) . 
\end{equation*}
The variation of information $VOI(X,Y)$  is defined as
\begin{equation*}
VOI(X,Y) = H(X) + H(Y) - 2MI(X,Y) ,
\end{equation*}
see for example~\cite{MEILA2007873}.
Thus, if $MI(X,Y)=H(X)=H(Y)$, indicating that the information in $X$ is
perfectly captured by the information in $Y$, $VOI(X,Y)$ would have a value of
$0$.
If $MI(X,Y)=0$,
then $VOI(X,Y)$ takes the largest possible value. 
Normalised mutual information $NMI(X,Y)$ is defined in~\cite{kvalseth1987entropy} as
\begin{equation*}
NMI(X,Y) = \frac{2MI(X,Y)}{H(X)+H(Y)} 
\end{equation*}
Again, if $H(X)=H(Y)=MI(X,Y)$, then $NMI(X,Y)=1$, whereas if $MI(X,Y)=0$, then
$NMI(X,Y)=0$.

\subsection{Additional ARI Results}
\label{app:comparisonTo400}
\label{app:model2DifferentThresholds}

Here we show the performance on Benchmark $1$ and Benchmark $2$ of our approaches for $n=400$ in comparison to $n=1000$ (to show the effect of different networks sizes), and additional 
Benchmark $2$ contour plots displayed at ARI=0.9  in contrast to the ARI=0.75 
(to show the effect of different contour thresholds for a given network size).

\begin{figure}%
\centering
\includegraphics[height=9.5cm]{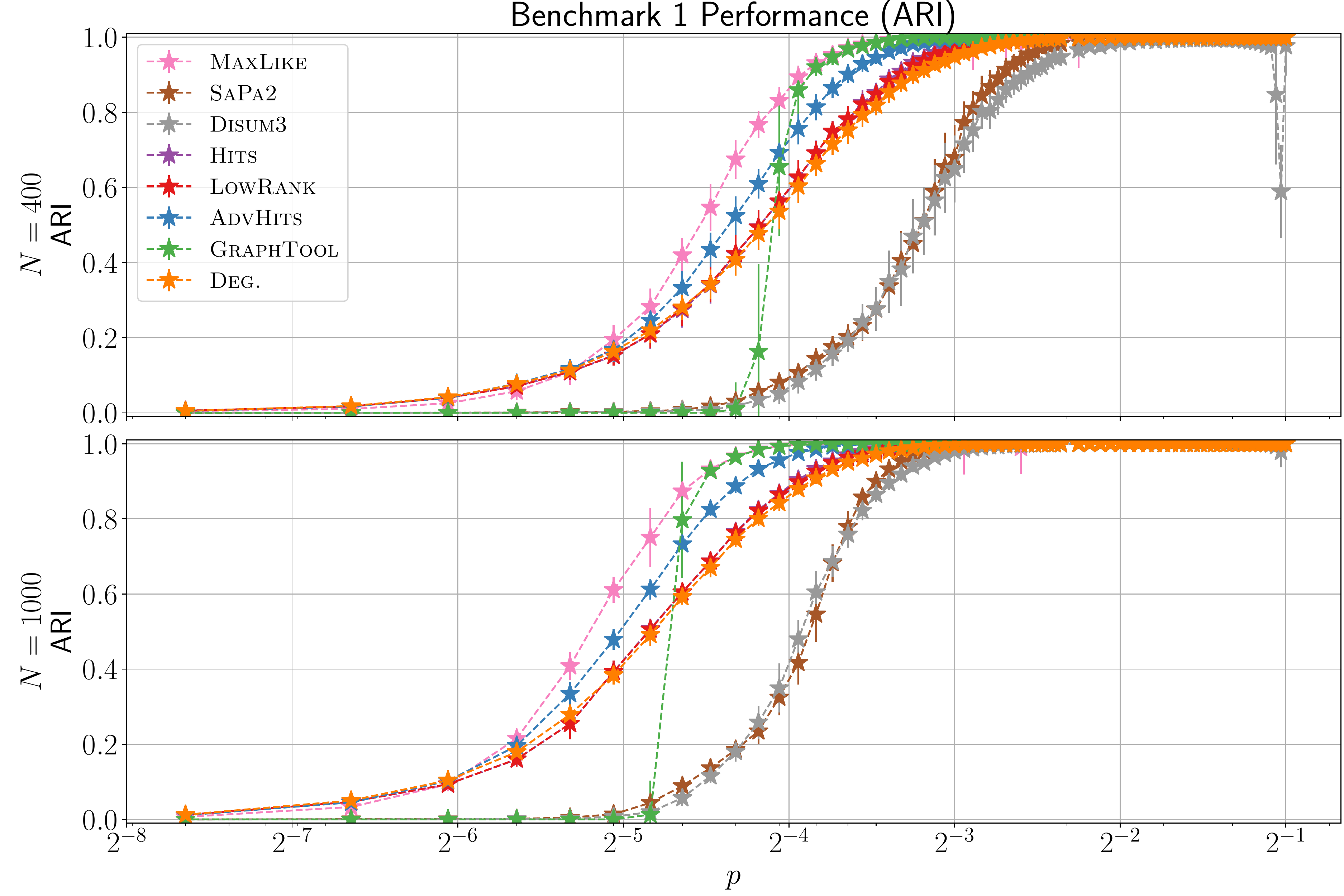}
\caption{Performance on Benchmark $1$. The ARI between the planted
partition of the graph and the partition detected by each method, with the
upper, respectively lower, panel showing results for networks of size 
$n=1000$. On the $x$ axis, we vary the parameter $p$
on a log scale.  
Error bars are one sample standard
deviation.}
\label{fig:model1ande_doubleN}
\end{figure}

\begin{figure}
\centering
\includegraphics[height=5cm]{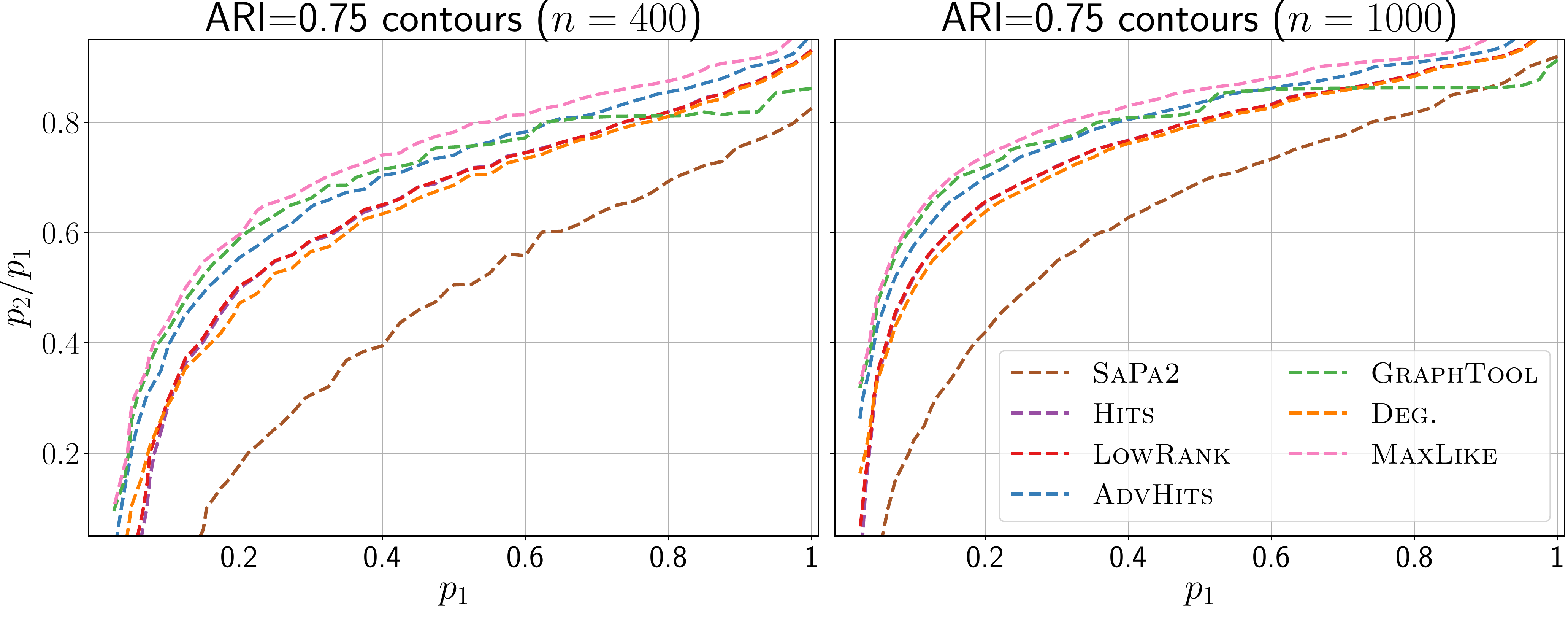}
\includegraphics[height=5cm]{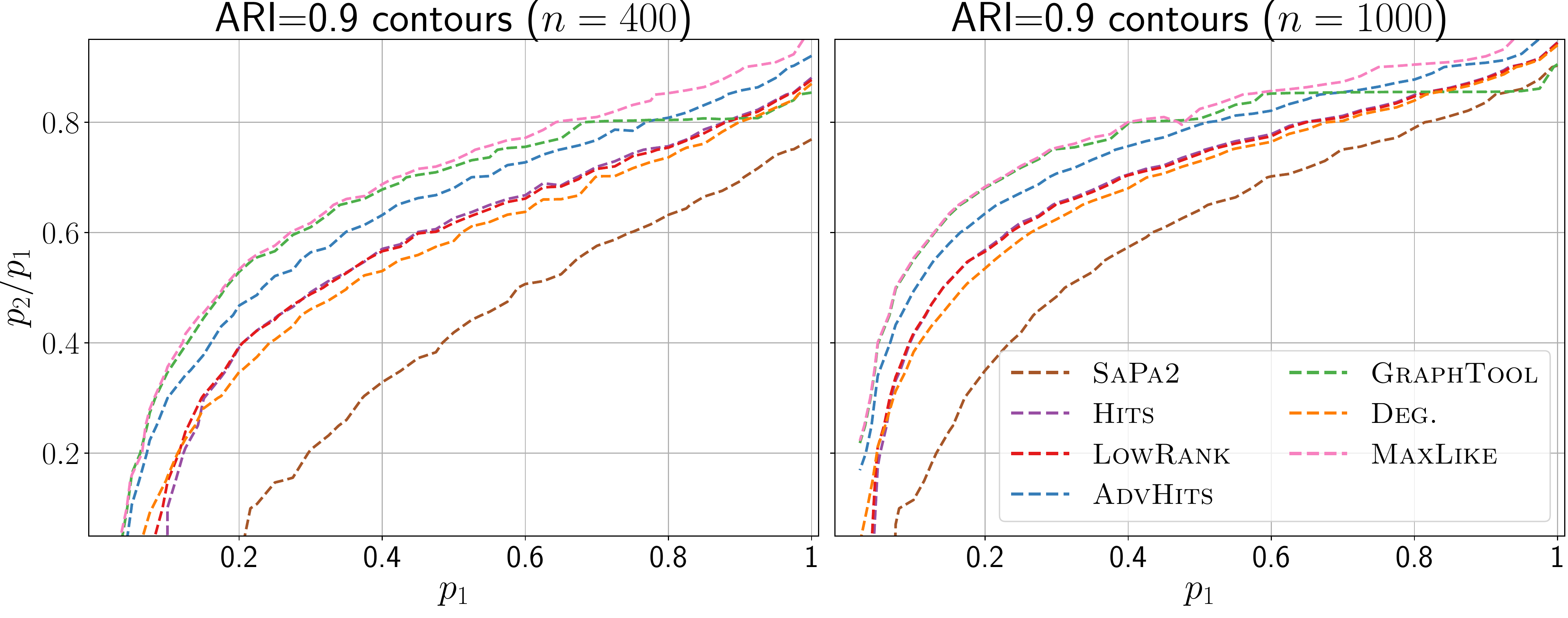}
\caption{Contour plot of ARI=0.75 and ARI=0.9 on Benchmark $2$.  For this
plot, we calculate the average ARI over 10 networks of size $400$ (left panel)
and $1000$ (right panel) varying $p_1 \in \{0.025,0.05,\ldots,1.0\}$ and
$\frac{p_2}{p_1} \in \{0,0.05,0.1,\ldots,0.95\}$, and then display the contours
corresponding to an average ARI of $0.75$. We note that the fast methods {\sc
HITS} and {\sc LowRank} outperform the comparisons {\sc Deg.} and {\sc SaPa2},
although they are hard to distinguish due to similar performance.  {\sc
AdvHits} also outperforms the fast comparisons. {\sc MaxLike } has similar
performance to {\sc GraphTool}, but like {\sc AdvHits} outperforms for larger
values of $p_1$. 
}
\label{fig:model2_mk2_Full}
\end{figure}

\begin{figure}
\centering
\includegraphics[height=9.5cm]{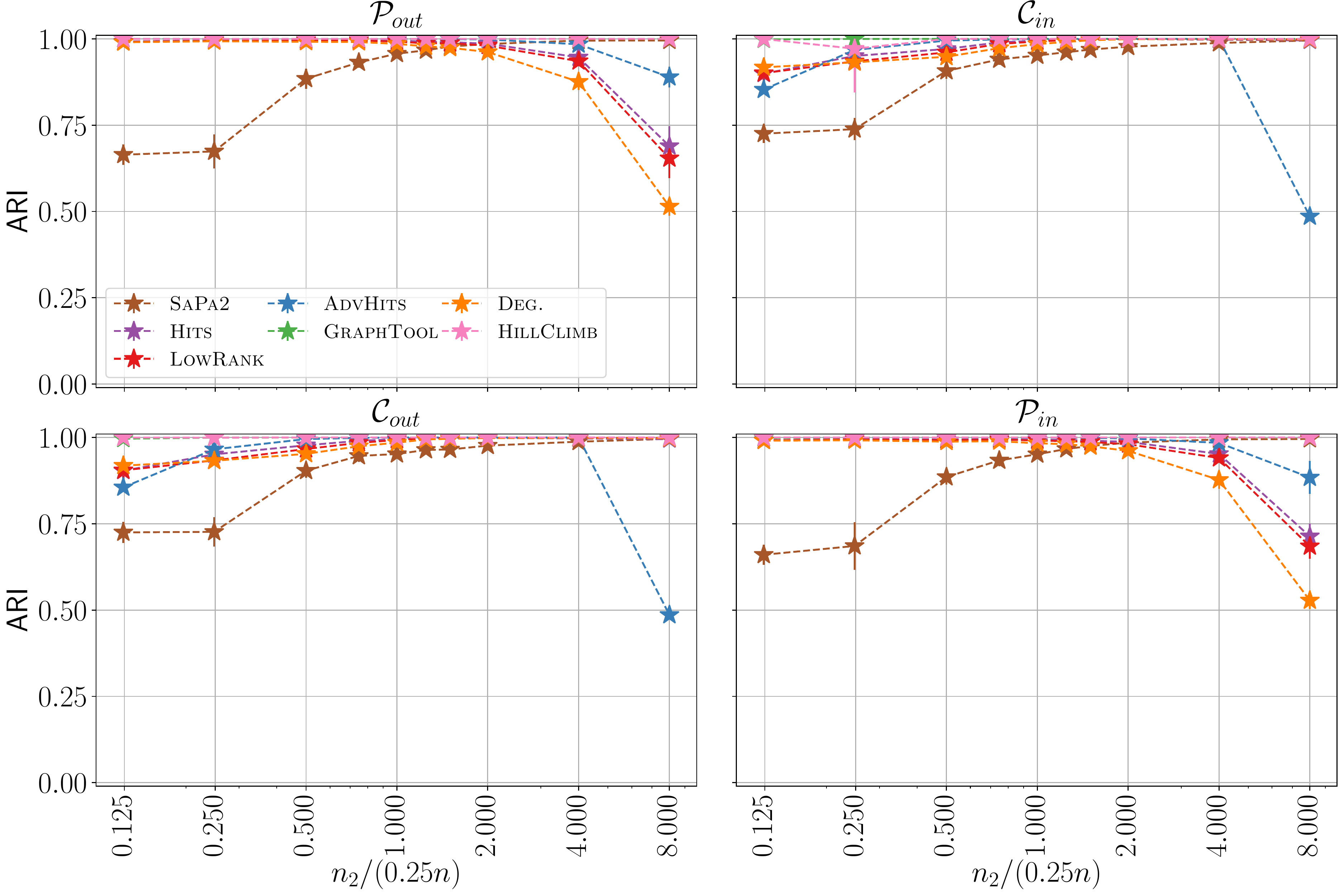}
\caption{Performance of all methods on Benchmark $3$ with sets of
different sizes using ARI, for $n=1000$. We
fix $p=0.1$ and the size of 3 sets at $n/4$, and then measure the ability
of each method to detect the sets when the size of the final set is
varied.}
\label{fig:model3n1000}
\end{figure}

The comparison  for graphs of different size is shown 
in Fig.~\ref{fig:model1ande_doubleN}
(Benchmark 1); 
Fig.~\ref{fig:model2_mk2_Full}
(Benchmark 2)
and 
Fig.~\ref{fig:model3n1000}
(Benchmark 3). 
In Benchmark 1 and 2 the results are qualitatively similar between the graph sizes with methods failing in the same order and in the same manner, although the range of values where this failure occurs shifts. This is expected as there is additional information present in larger graphs, so we would expect smaller values of $p$ to have potentially larger effects on performance (up to detectability limits). We observe a similar effect in Benchmark 3 for $n=1000$, with the increased network size (and fixed value of $p=0.1$) many methods perform better over the range of block sizes, although we note that we do observe a similar decay in the performance of {\sc AdvHits}. 
The $0.75$ and $0.9$ contours for ARI are shown in 
(\cref{fig:model2_mk2_Full}).
The results are broadly consistent,
with similar
performance for many regimes in the $0.9$ contour but a much larger
performance difference in the $0.75$ contour.

\subsection{Run Time On Synthetic Networks}
\label{sec:TimingSyntheticSec}
\begin{figure}
\centering
\includegraphics[height=9.5cm]{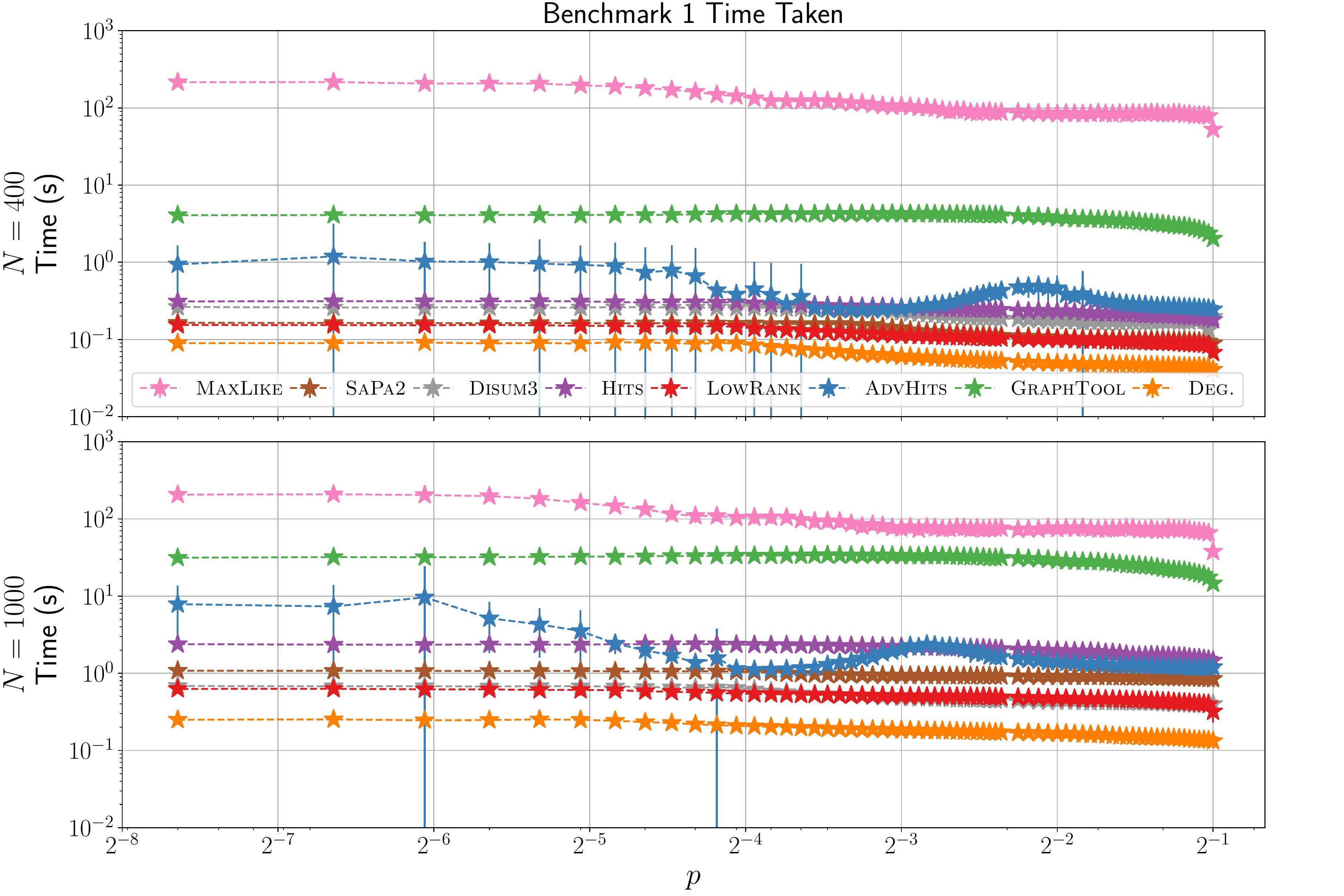}
\caption{Time taken by all methods on Benchmark $1$ with the upper, respectively lower, panel showing results for
networks of size $n=400$,  respectively $n=1000$. On the $x$ axis, we vary the
parameter $p$, with values close to $0$ more difficult to detect than
higher values of $p$. 
The timing for {\sc MaxLike} for $n=1000$ was ran under PyPy, a faster Python
interpreter, to highlight the possible speed up.
}
\label{fig:model1andeTime}
\end{figure}

The two panels of \cref{fig:model1andeTime} show the run time of each
of the methods. We note that the time comparison is advisory as this calculation was
performed on a Azure Virtual Machine. 
To mitigate the effects of hyperthreads
we used half of the available vCPUs 
but we cannot control for other users present on the same physical hardware. 
Further, different
approaches are implemented in different languages, with the matrix based
methods ({\sc HITS}, {\sc AdvHits}, {\sc SaPa}, {\sc DiSum}, {\sc HITS}, and {\sc LowRank}) being
implemented in Python using NumPy and SciPy vectorised matrix operations, {\sc GraphTool} in
C++ with Python wrapping\footnote{The {\sc GraphTool} timings also contains the C++ graph
construction, and the matrix approaches timings contain the matrix
construction}, and the likelihood methods are implemented in pure Python
ran under cPython with the exception of {\sc MaxLike} for $n=1000$ which
was ran under PyPy~\cite{pypy}. We note that when testing these methods under PyPy
we obtain a $\approx 10$x speed up on the cPython times. 
Finally,  we note  that the implementation of  {\sc SaPa}, {\sc DiSum}, {\sc HITS}, {\sc LowRank}  are relatively na\"ive, and could be improved, for example, by exploiting the specialised matrix construction code we use in {\sc AdvHits}. 

We observe that the performance of many of the methods is relatively constant
with respect to $p$, 
with an increase in run time when $p$ is small. 
This behaviour is especially pronounced in {\sc AdvHits} with comparatively fast times for
the easier graphs, and slower times for the harder graphs, which is likely to be because the 
method falls back to the single vertex update scheme for small values of $p$.  This is also
likely to be related to the error bars (one sample standard deviation),
which 
increase for low $p$ in {\sc AdvHits} for $n=400$. 
We note that {\sc AdvHits} would have similar time complexity to the
PyPy version of {\sc HillClimb} (with $10$ repeats) but have substantially
better time complexity than {\sc MaxLike}, highlighting that
the {\sc AdvHits} approach is a medium speed approach in contrast to the fast
approaches. 

\subsection{Benchmark Results for VOI and NMI}
\label{app:additionalResults}
\begin{figure}
\centering
\includegraphics[height=6.5cm]{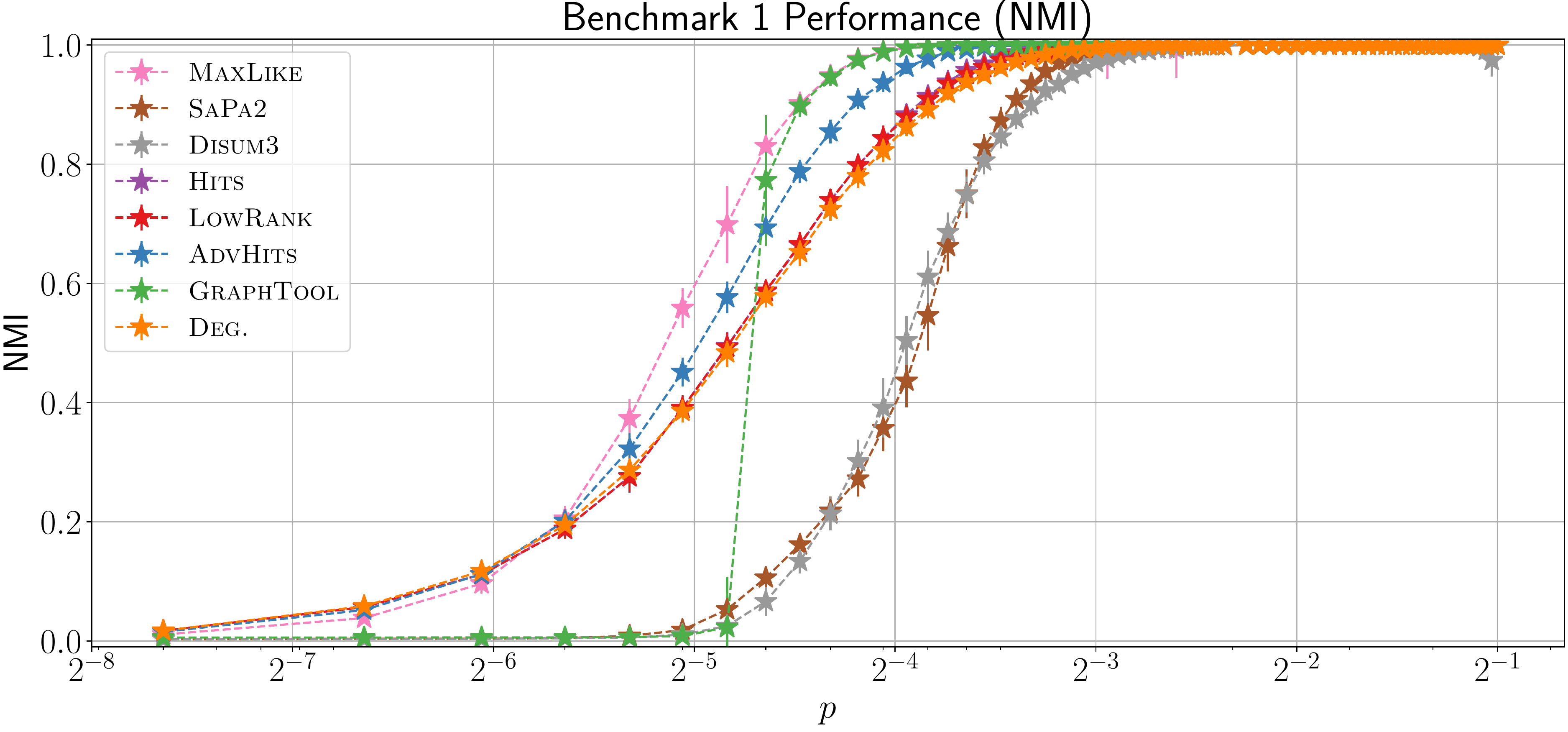}

\includegraphics[height=6.5cm]{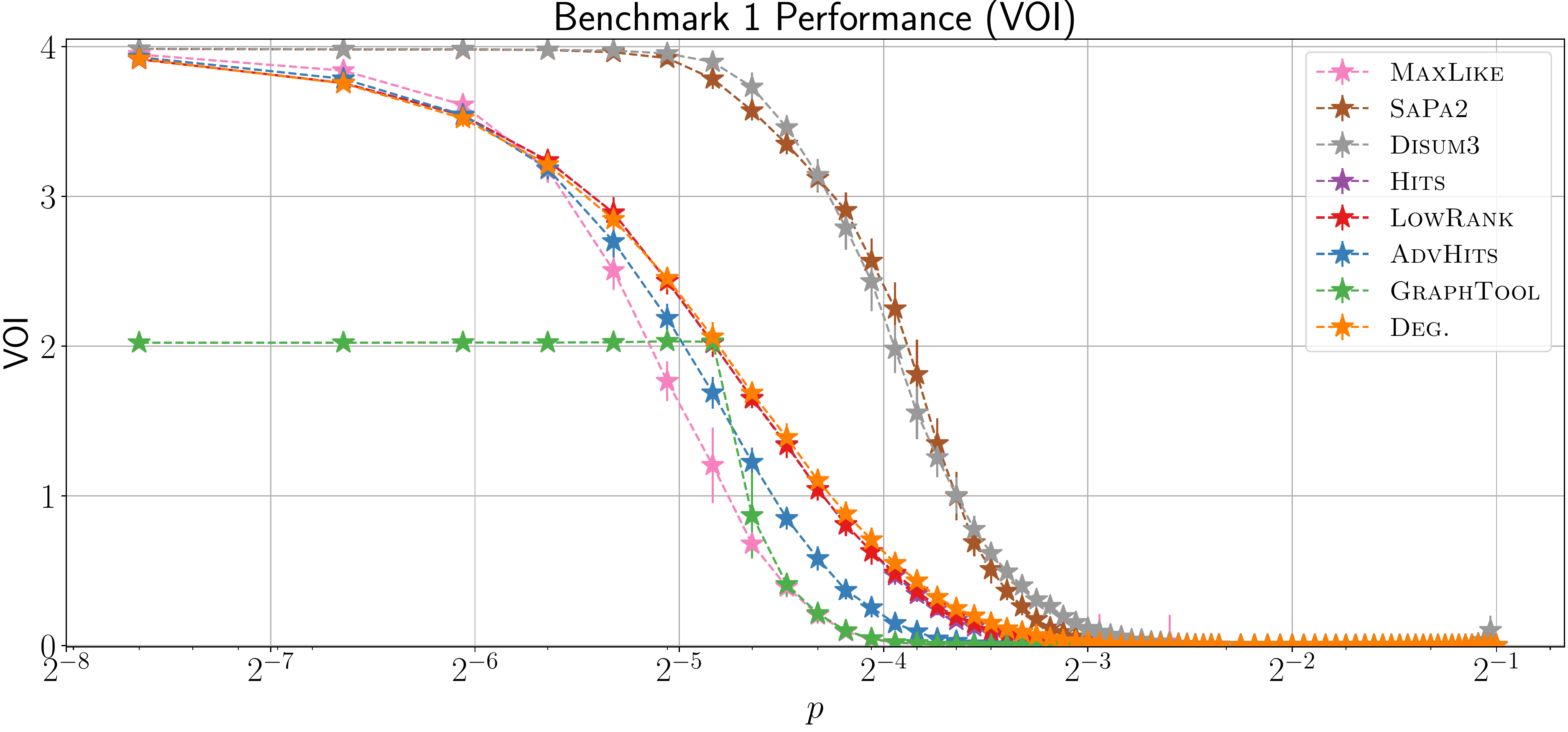}
\caption{Performance on Benchmark $1$. The NMI  ({\bf Top Panel}) and VOI  ({\bf Bottom Panel}) between the
underlying partition of the graph and the detected partition for $n=1000$.
On the $x$ axis, we vary the parameter $p$, with values close to $0$ more
difficult to detect than higher values of $p$. Error bars are one sample
standard deviation.}
\label{fig:model1NMI}
\end{figure}

\paragraph{Benchmark $1$}
The NMI and VOI results for Benchmark $1$ can be found in \cref{fig:model1NMI}. Broadly, the results are qualitative similar to the ARI results. 
For both NMI and VOI, {\sc SaPa2} and {\sc DiSum}, are outperformed by
all other methods, which is consistent with the results of ARI. Similarly {\sc
AdvHits} and {\sc MaxLike}  outperform clustering based on degree in all
cases, with the {\sc HITS} and {\sc LowRank} methods performing at least as
well as clustering based on degree. We also observe an equivalent result performance result for {\sc AdvHits}, with
a performance improvement over {\sc LowRank} and {\sc HITS} methods, but
outperformed by {\sc MaxLike}. 
We only show the results for $n=1000$; the results for $n=400$ are similar.

\begin{figure}
\centering
\includegraphics[height=5.75cm]{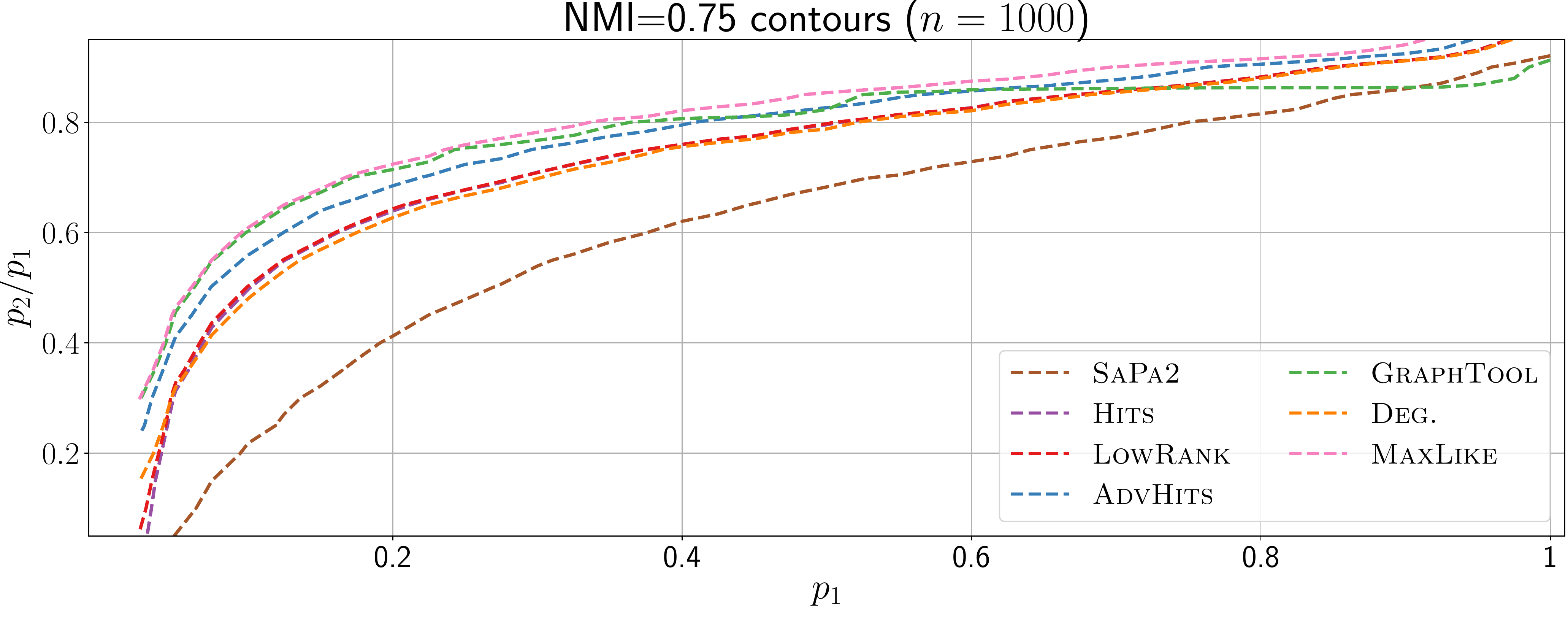}
\caption{Contour plot of NMI=0.75 on the Benchmark $2$.  For this
plot, we calculate the average NMI over $10$ networks of size 
$1000$ 
varying $p_1 \in \{0.025,0.05,\ldots,1.0\}$ and
$\frac{p_2}{p_1} \in \{0,0.05,0.1,\ldots,0.95\}$, and then display the contours
corresponding to an average NMI of $0.75$ and $0.9$. 
}
\label{fig:model2_mk2_FullNMI}
\end{figure}

We highlight the VOI results for {\sc GraphTool}. It
outperforms the other methods for $p<2^{-6}$ whereas ARI and NMI show precisely the opposite performance. This discrepancy may be related to the effect that placing almost all vertices into one set has on the similarity measures.  
Indeed, \cref{tab:avLargestGroupSize}  shows the average size of
the largest set in {\sc LowRank}, {\sc AdvHits} and {\sc GraphTool}. For {\sc LowRank} and {\sc AdvHits}  the average largest set 
size  is close to a equal division into $4$ equally-sized sets, whereas {\sc
GraphTool} incorrectly  places most of the vertices in a single set for $p<0.04   \approx 2^{-4.64}$.  
With most vertices in the {\sc GraphTool} partition in one set, the entropy will be close to $0$, as
$-\log(1-\epsilon)\approx 0$, thus, as mutual information is by definition less
than the entropy, the VOI becomes close to the entropy of the almost constant sequence,   
which is equal to $-\log_2 (0.25) = 2$,  a relatively high value. 
The behaviour of NMI on {\sc GraphTool} can also be similarly explained, as
mutual information is close to $0$, NMI then also becomes close to $0$.

\begin{table}
\center
\begin{tabular}{|cc||c|c|c|}
\hline 
$n$ & $p$ &  {\sc LowRank} & {\sc AdvHits} & {\sc GraphTool}
\\
\hline
\hline
 1000 & 0.050 &  260.8 & 264.42 & 252.62 \\ \hline
 1000 & 0.045 & 264.82 & 270.92 & 255.62 \\ \hline
 1000 & 0.040 & 270.74 &  279.6 & 298.68 \\ \hline
 1000 & 0.035 &  274.9 &  288.5 & 980.24 \\ \hline
 1000 & 0.030 & 280.62 & 295.52 & 995.42 \\ \hline
 1000 & 0.025 & 286.64 & 298.72 & 996.56 \\ \hline
 1000 & 0.020 & 290.64 & 292.18 & 996.88 \\ \hline
 1000 & 0.015 &  290.4 & 282.16 & 996.86 \\ \hline
 1000 & 0.010 &  296.2 & 282.18 &  997.0 \\ \hline
 1000 & 0.005 & 298.54 &  283.1 &  997.0 \\ \hline
\end{tabular}
\caption{Average (mean) size of the largest set in the detected partition in Benchmark $1$. Note that the maximum achievable set size for a division into $4$ non empty sets is $1000 - 1 - 1 - 1 = 997$.} 
\label{tab:avLargestGroupSize}
\end{table}

\medskip 
\paragraph{Benchmark $2$}
For brevity, in the main text we considered contour plots for a fixed value of ARI.  In this section, we consider alternative similarity measures.  We again only show the results for $n=1000$; the results for $n=400$ are similar.

\begin{figure}
\centering
\includegraphics[height=5.75cm]{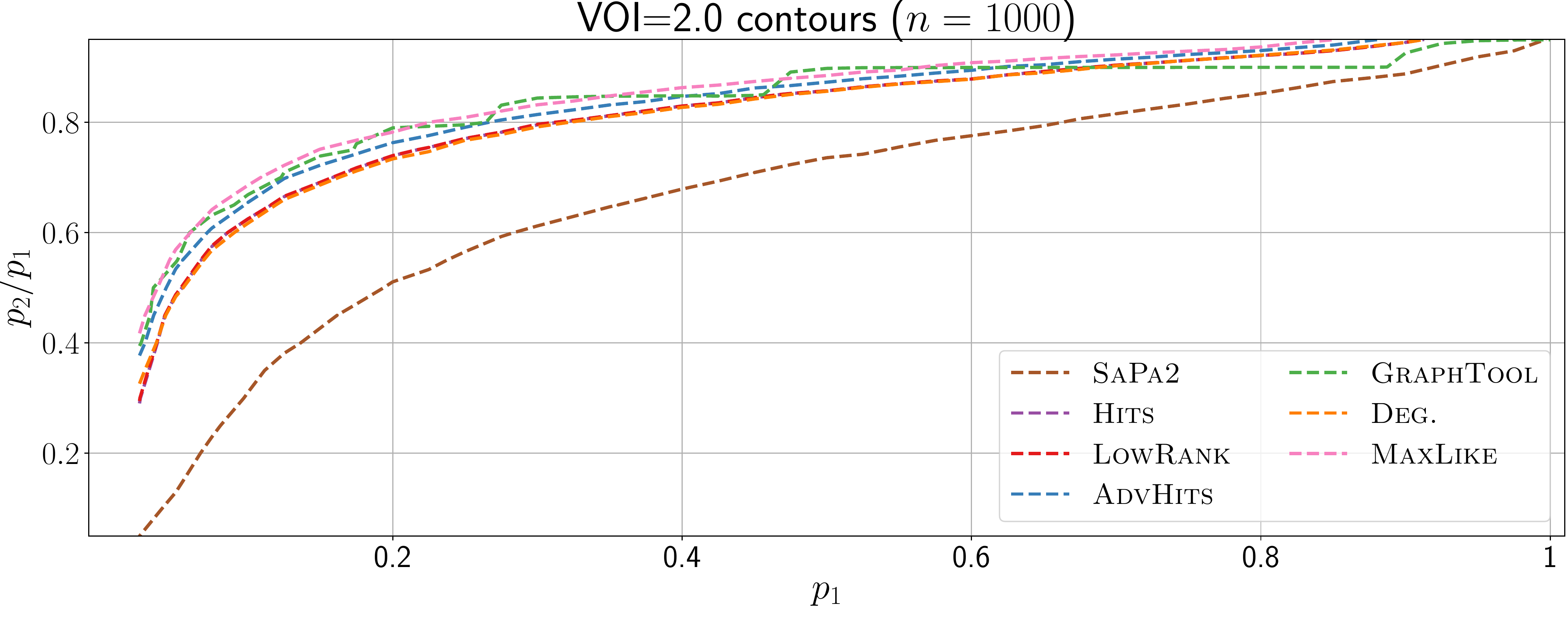}
\caption{Contour plot of VOI=2.00 a on the Benchmark $2$.  For this
plot, we calculate the average VOI over 10 networks of size 
$1000$ 
varying $p_1 \in \{0.025,0.05,\ldots,1.0\}$ and
$\frac{p_2}{p_1} \in \{0,0.05,0.1,\ldots,0.95\}$, and then display the contours
corresponding to an average VOI of  $2.0$. 
}
\label{fig:model2_mk2VOI}
\end{figure}

The NMI contours, displayed in \cref{fig:model2_mk2_FullNMI}, tell a broadly
similar story to the ARI results (\cref{fig:model2_mk2_Full})
with the methods having the same ordering and performance as we
observe  ARI.
The VOI contours, displayed in \cref{fig:model2_mk2VOI}, show broadly
consistent findings but with a few small deviations. 

\begin{figure}
\centering
\includegraphics[height=8.5cm]{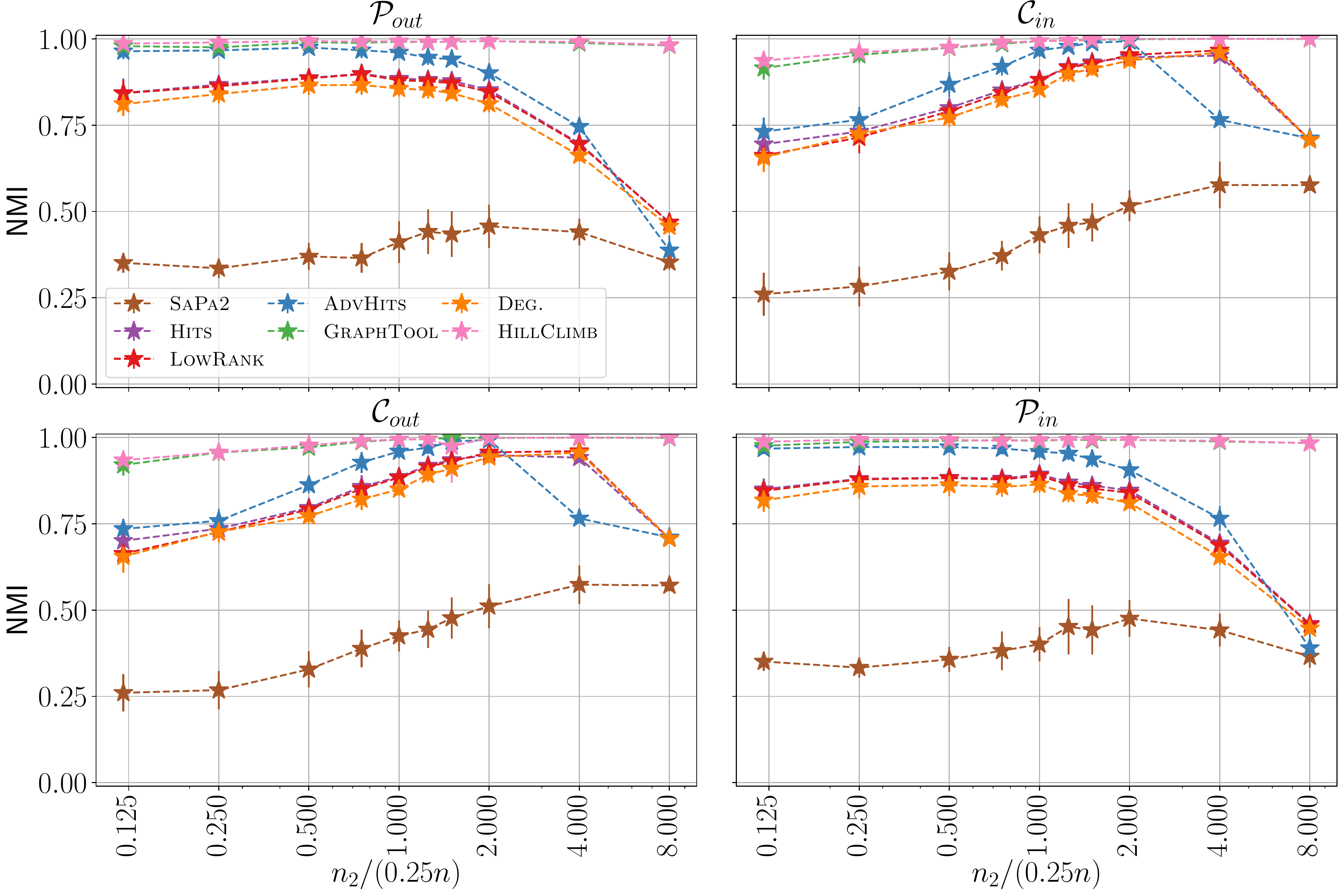}

\includegraphics[height=8.5cm]{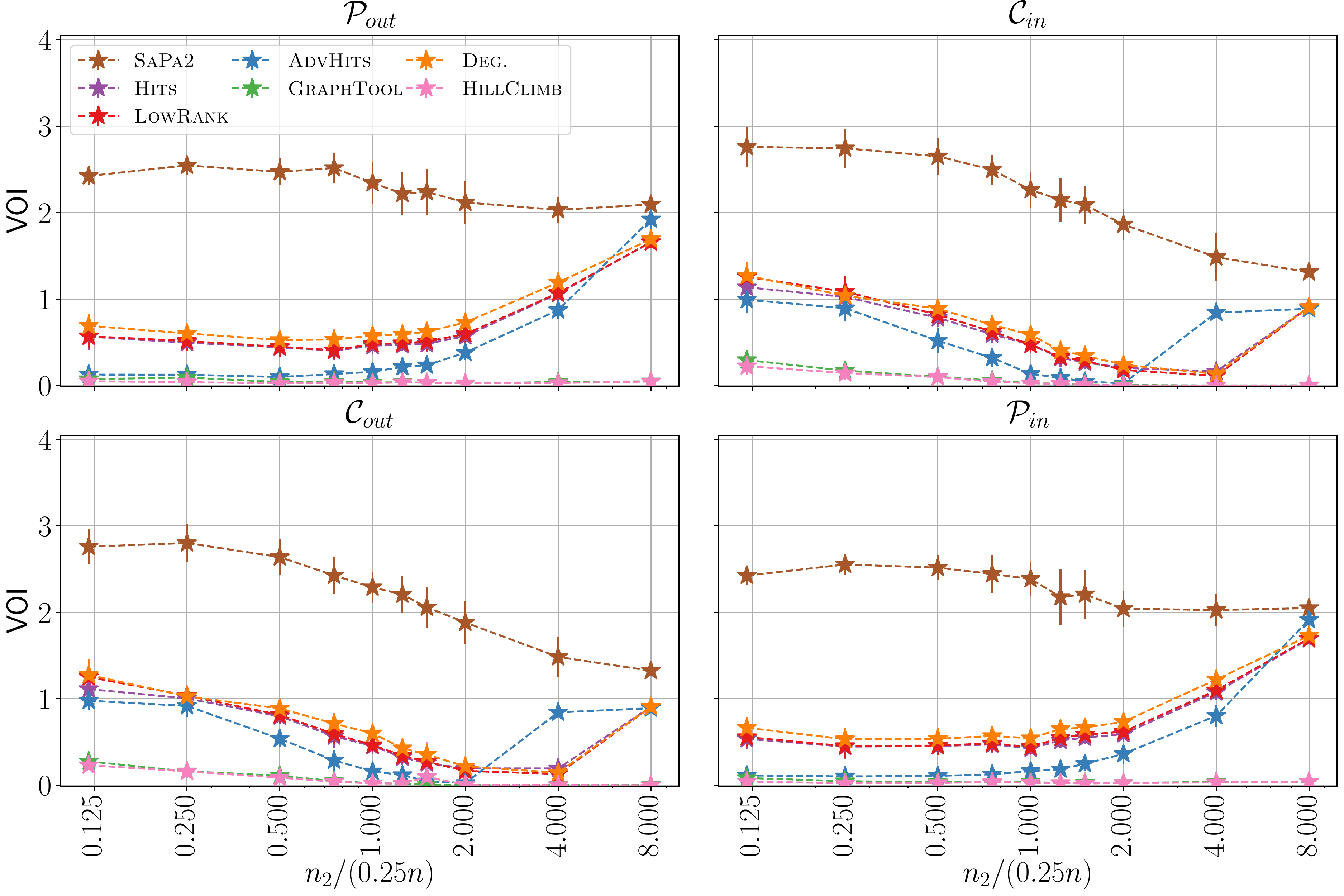}
\caption{Performance of all methods on Benchmark $3$ with sets of
different sizes using NMI ({\bf Top Panel}), and VOI ({\bf Bottom Panel}). We
fix $p=0.1$ and the size of 3 sets at $n/4$, and then measure the ability
of each method to detect the sets when the size of the final set is
varied.}
\label{fig:model3andeNMI_VOI}
\end{figure}

\medskip 
\paragraph{Benchmark $3$}
The results for Benchmark $3$
be seen in \cref{fig:model3andeNMI_VOI}. 
They  are broadly consistent across each of the similarity measures, although there is a small deviation with
the performance of {\sc AdvHits} for $\frac{n_2}{n}>2$.  
The information theoretic measures (NMI and VOI) yield a slightly higher performance for {\sc AdvHits} (in comparison to the controls) than the ARI.
In both cases, the performance of {\sc AdvHits} collapses in this region.

\subsection{Performance of {\sc DiSum} and {\sc SaPa} variants}
\label{spectralNullVariants}
As alluded to in the main text we consider a small number of variants of our spectral comparison methods and only display the best performing method. In this section we discuss the other variants we considered on each of our benchmarks.

\paragraph{{\sc SaPa} Variants}
From~\cite{satuluri2011sym}, {\sc SaPa} 
employ a similarity measure using two different types of symmetrisations  for directed (asymmetric) adjacency matrices,  namely bibiometric symmetrisation and degree discounted symmetrisation. We use them here to construct two variants of {\sc Sapa}, {\sc SaPa1} and {\sc SaPa2}, which we then cluster. They are defined as follows:
\begin{enumerate}
  \setlength\itemsep{0.1mm}
\item {\sc SaPa1 } - Bibiometric symmetrisation: $ A_IA_I^T  + A_I^TA_I $
\item {\sc SaPa2 } - Degree discounted:
$
D_o^{-0.5}
A
D_i^{-0.5}
A^T
D_o^{-0.5}
+
D_i^{-0.5}
A
D_o^{-0.5}
A^T
D_i^{-0.5}
$
\end{enumerate}

\noindent where  $D_i$ is a diagonal matrix with the in-degrees on the diagonal, 
$D_o$ is an diagonal matrix with the out-degrees on the diagonal, 
$A_I = A + I$, where $I$ is the identity matrix. In the original formulation of {\sc SaPa}, these similarity matrices were then clustered via various clustering algorithms. Instead, we use the same kmeans++ pipeline that we leverage for other results. 

\paragraph{{\sc DiSum} Variants}
The second fast approach, {\sc DiSum} from~\cite{Rohe12679},
clusters the graph using the
left and right singular vectors of the regularised %
and degree normalised  adjacency matrix $((k^{out}_i+\frac{m}{n}) (k^{in}_i+\frac{m}{n}) )^{-1/2} A_{ij}$.
The original approach in \cite{Rohe12679}, performed a column and row clustering, as we are comparing to a full network clustering. We consider to $4$ different variants of this approach:

\begin{enumerate}
  \setlength\itemsep{0.1mm}
\item {\sc DiSum1} A row clustering into $4$ sets,
\item {\sc DiSum2} A column clustering into $4$ sets, 
\item {\sc DiSum3} A combined row and column clustering into $4$ sets (using the concatenation of the left and right singular vectors), 
\item {\sc DiSum4} Combining a row clustering into $2$
sets with a column clustering into $2$ sets to obtain $4$ resultant sets (with each set defined by the intersection of a pair of sets from the different clusterings).   Note, we use a separate row and column clustering rather than the combined clustering in \cite{Rohe12679}.
\end{enumerate}

\noindent
We again cluster the resulting vectors using our standard pipeline i.e. k-means++.
\paragraph{Results}
Unlike in previous cases as this is a comparison method rather than computing an independent set of networks we display the method that performs best on our benchmark networks (results not shown). 

We observe that {\sc SaPa2} and {\sc DiSum3} outperform there counterparts
on Benchmark $1$;
Benchmark $2$ (where the other approaches are below the threshold). 
In Benchmark $3$ {\sc DiSum4} outperforms {\sc DiSum3},  
for large differences in group sizes. 
 However, given that the {\sc DiSum3} 
performs reasonably 
and outperforms {\sc DiSum4} considerably in the Benchmark $1$ and Benchmark $2$, we select this as our overall {\sc DiSum} method.

\section{Political Blogs: Additional Results}
\label{app:realWorld}

In this section,  we present additional results on the political blogs data set. We first explore the relationship between the sets we uncover and the political division of the network. 
Second we relate it to undirected core-periphery structure.%

\begin{figure}
\centering
\includegraphics[height=5.5cm]{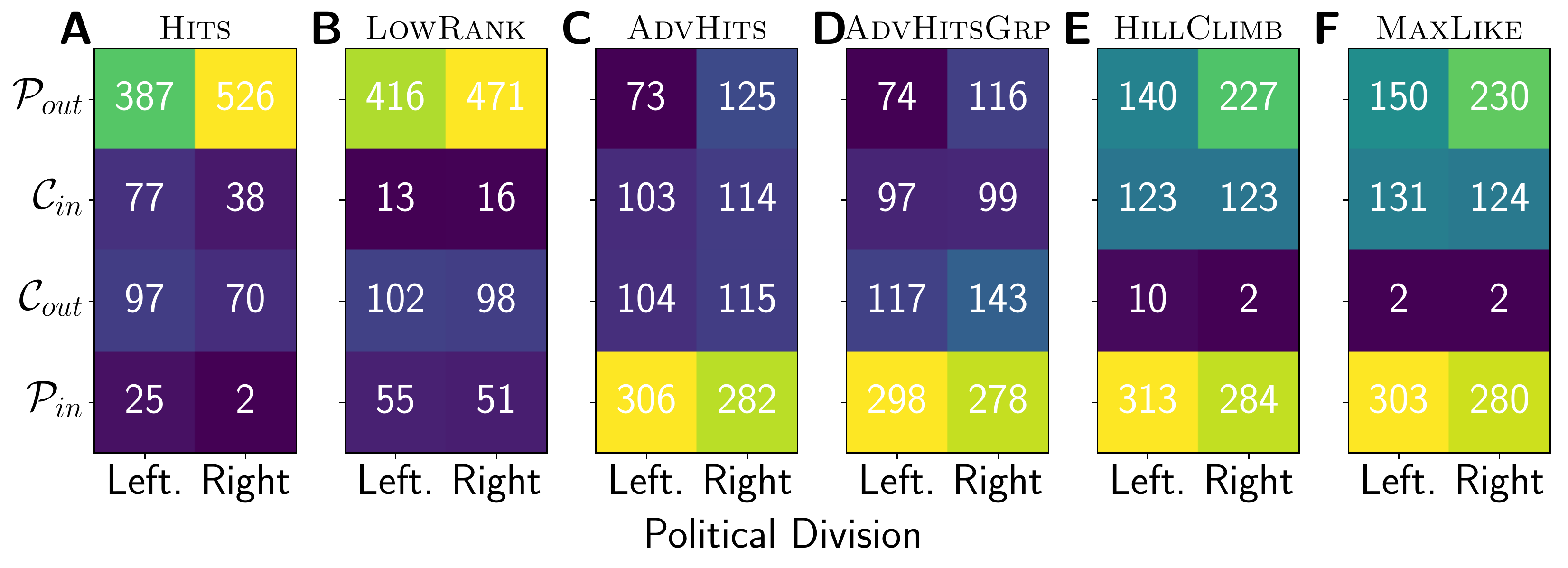}
\caption{Each panel shows the confusion table of the detected sets by each method and the division into left and right leaning political blogs. 
Colour scales are different in each plots and are designed to show the differences within a panel.
}
\label{fig:politBlogPoliticalComparison}
\end{figure}

In the main paper we state that the partitions from {\sc AdvHits} and {\sc MaxLike} are related 
to the political division into liberals (Left) and conservative (Right) blogs, but are not purely a function
of it.  Figure~\ref{fig:politBlogPoliticalComparison} gives the confusion matrices between each of
our divisions and the political division. 
We observe that with the exception of {\sc HITS}, 
the sets 
divide into two roughly similarly sized sets. These sets are not in 1-1 relationship with the Left/Right division of the blogs. Hence there is evidence that  %
the sets also relate to structural properties of the
vertices.

\begin{figure}[!htbp]
\centering
\includegraphics[height=5.5cm]{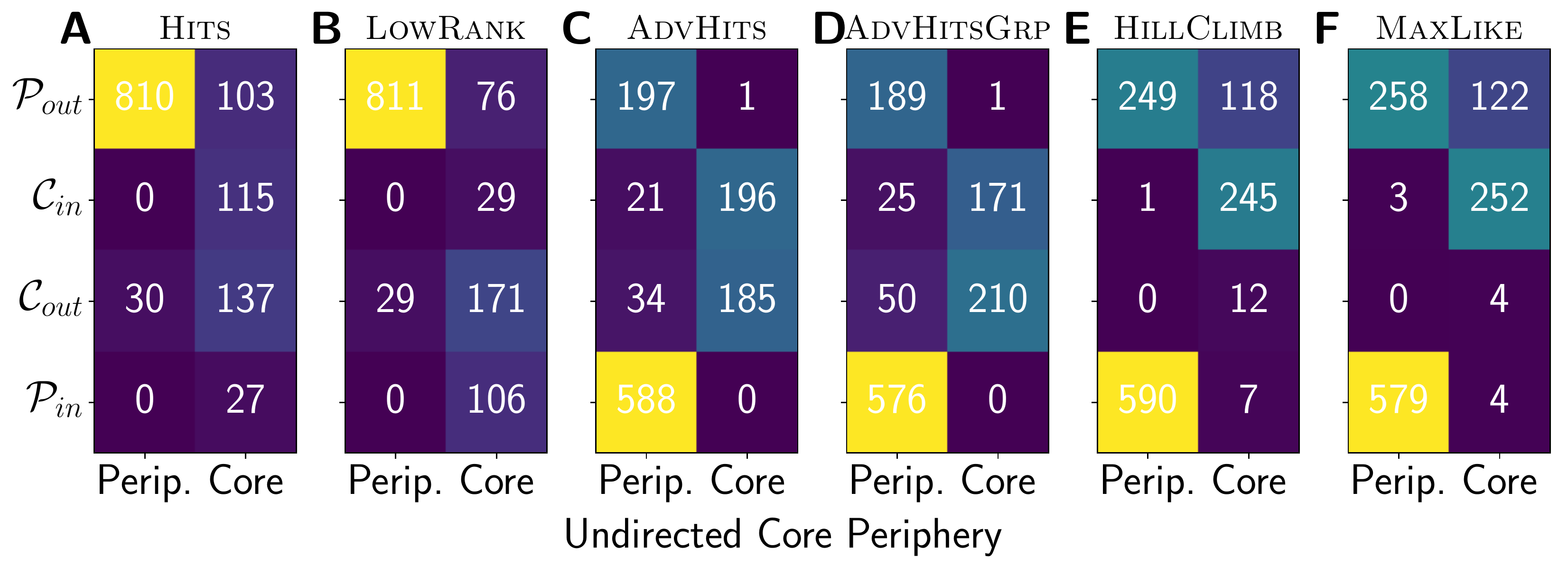}
\caption{Each panel shows the confusion table of the detected sets by each
method and the division into the undirected core--periphery structure.  Colour
scales are different in each plot, and are designed to show the differences
within a panel.}
\label{fig:politBlogUndirCPComparison}
\end{figure}

Next we compare the division obtained using our approaches with the
division found using the classical core--periphery approach.  The results can
be seen in \cref{fig:politBlogUndirCPComparison}.
We note that the split of the undirected core into $\mathcal{C}_{in}$ and $\mathcal{C}_{out}$ 
and the split of the undirected periphery 
into 
$\mathcal{P}_{in}$ and $\mathcal{P}_{out}$
is present in {\sc AdvHits} and {\sc
AdvHitsGrp}.

\end{document}